\documentclass[12pt]{article}

\usepackage{csvsimple}
\usepackage{comment}

\usepackage[english]{babel}
\usepackage[T1]{fontenc}
\usepackage[utf8]{inputenc}

\usepackage{beton}
\usepackage{euler}

\usepackage{amsmath, amssymb, amsfonts, amsthm}
\usepackage{mathtools}
\usepackage{bbm}
\usepackage{mathrsfs}
\usepackage{dsfont}           
\DeclareMathOperator*{\argmin}{arg\,min}
\newcommand{\indep}{\perp\!\!\!\!\perp}

\newtheoremstyle{boldexample}
  {\topsep}{\topsep}
  {\itshape}{}{\bfseries}{.}{ }
  {\thmname{#1}\thmnumber{ #2}\thmnote{ (#3)}}

\theoremstyle{boldexample}
\newtheorem{theorem}{Theorem}[section]

\newtheorem{lemma}[theorem]{Lemma}
\newtheorem{proposition}[theorem]{Proposition}
\newtheorem{example}[theorem]{Example}

\theoremstyle{definition}
\newtheorem{definition}[theorem]{Definition}

\newtheorem{assumption}[theorem]{Assumption}

\theoremstyle{remark}
\newtheorem{remark}[theorem]{Remark}

\numberwithin{equation}{section}

\usepackage{titlesec}

\titleclass{\subsubsubsection}{straight}[\subsubsection]

\newcounter{subsubsubsection}[subsubsection]
\renewcommand\thesubsubsubsection{%
  \thesubsubsection.\arabic{subsubsubsection}}

\titleformat{\subsubsubsection}
  {\normalfont\normalsize\bfseries}
  {\thesubsubsubsection}{1em}{}

\titlespacing*{\subsubsubsection}
  {0pt}{3.25ex plus 1ex minus .2ex}{1.5ex plus .2ex}

\usepackage[margin=1in]{geometry}
\usepackage{setspace}
\usepackage{indentfirst}

\usepackage{graphicx}
\graphicspath{{images/}}
\usepackage{caption}
\usepackage{subcaption}
\usepackage{threeparttable}
\usepackage{adjustbox}
\usepackage{booktabs}
\usepackage{longtable}
\usepackage{rotating}
\usepackage{array}
\newcolumntype{H}{>{\setbox0=\hbox\bgroup}c<{\egroup}@{}}
\usepackage{nicematrix}

\usepackage{authblk}
\usepackage{hyperref}
\hypersetup{
    colorlinks=true,
    linkcolor=blue,
    citecolor=blue,
    urlcolor=blue
}
\usepackage{enumerate}
\usepackage{textcomp}
\usepackage{xcolor}
\usepackage{soul}

\usepackage[longnamesfirst]{natbib}%

\usepackage[babel]{csquotes}

\begin{document}
\title{Testing the Solvability of Systems of Linear Inequalities\thanks{This paper draws on research supported by the Social Sciences and Humanities Research Council of Canada. We thank 
Federico Bugni, Ivan Canay, Gregory Cox, Kirill Ponomarev, Jonathan Roth and Alex Torgovitsky for useful discussions. We thank John Pepper for assistance with replication data. We thank Alex Lu for excellent research assistance. This paper was previously circulated with the title ``Inference on the value of a linear program''.}}
\date{}
\author[1]{Leonard Goff}
\author[2]{Eric D. Mbakop}
\affil[1]{Department of Economics, University of Calgary}
\affil[2]{Department of Economics, The Ohio State University}
\maketitle

\begin{abstract}
This paper studies the problem of testing whether a system of linear equality and inequality constraints admits a solution when the coefficients of that system may have to be estimated. We show that a wide range of inferential questions in partially identified models can be formulated as hypotheses of this form. Our approach exploits an alternative characterization of the hypothesis based on whether the value of a certain linear program is equal to zero. Building on this characterization, we develop bootstrap-based testing procedures and establish their uniform validity over large classes of data-generating processes. Simulation results demonstrate good finite-sample performance, even for moderate sample sizes. We illustrate the usefulness of the approach in two empirical applications.
\end{abstract}

\section{Introduction}
Many economic models imply restrictions that can be formulated as linear equalities or inequalities involving estimable quantities. Such restrictions can be used for testing the model, or to provide identifying information about parameters of interest under weaker assumptions than are required to point identify them. 

With either goal in mind, the inferential task can often be reformulated as a test of the hypothesis
\begin{equation} \label{eqnARP}
H_0: Q \in \mathcal{Q}_0 \quad \text{versus} \quad H_1: Q \in \mathcal{Q} \setminus \mathcal{Q}_0,
\end{equation}
where 
\begin{equation*}
\mathcal{Q}_0 = \left\{ Q \in \mathcal{Q} \;\middle|\; 
\left\{ \eta \in \mathbb{R}^d \;\middle|\; b(Q) \leq A(Q)\eta \right\} \neq \emptyset 
\right\}
\end{equation*}
where \(A(Q) \in \mathbb{R}^{p \times d}\) and \(b(Q) \in \mathbb{R}^p\). Since any linear equality can be expressed as a pair of inequalities, our hypothesis can be interpreted as determining whether a system of linear equality and inequality constraints is solvable, i.e., whether it admits a solution.

In this paper, we develop an approach to testing \eqref{eqnARP} when entries of $b$ and/or $A$ must be estimated from data, and $\mathcal{Q}$ is a large class of distributions that we describe below. 
We show in Section \ref{secApplications} how special cases of \eqref{eqnARP} arise in a range of settings involving partial identification, including tests of moment inequalities, tests of the hypothesis that there exists a solution to a system of linear equality restrictions, and subvector inference in the presence of nuisance parameters that enter into moment inequalities/equalities linearly. We detail a series of specific examples from the treatment effects literature where hypotheses of this type are of empirical interest. 

Despite being of broad interest in applied work, developing tests of \eqref{eqnARP} that are uniformly valid over large classes of DGPs poses substantial technical challenges. Natural test statistics for this problem have null asymptotic distributions that are discontinuous in the DGP and depend on difficult-to-estimate nuisance features, such as which inequalities bind as equalities and the structure of the solution set to the system. As a consequence, standard procedures such as the naive bootstrap are generally inconsistent. Moreover, this discontinuity implies that inference procedures based on asymptotic approximations can perform poorly in certain regions of the null, regardless of sample size. These challenges motivate the need for alternative approaches.

We note that in principle, one could test hypothesis \eqref{eqnARP} using procedures designed for moment inequalities, whether the matrix $A$ is estimated or known. For a given value of $\eta$, the quantity $A(Q)\eta-b(Q)$ is estimable and a test for  $A(Q)\eta-b(Q) \ge 0$ can be repeated across values of $\eta$. Procedures like \cite{DAGS} would be uniformly valid here under weak assumptions. In practice however, discretizing and iterating over the possible values of $\eta$  would be computationally prohibitive if the dimension of $\eta$ is not small.\footnote{For a given $A$ and $b$, this approach involves testing the null hypothesis at each point on a grid in $\mathbb{R}^{d_{\eta}}$.} Moreover, as noted in \cite{CBS}, \cite{HKFMJS} and \cite{CCT} projection-based tests can be quite conservative. The objective of our approach is to propose an inference procedure that is computationally feasible and not unduly conservative, without compromising on uniformity.

Our inference procedure is bootstrap-based and controls size uniformly over large classes $\mathcal{Q}_0$ of data-generating processes (DGPs). Moreover our method is not unnecessarily conservative, potentially obtaining asymptotically exact size. We give two versions of the procedure. First, we consider the simpler case in which $A$ is known and only $b$ is estimated. In this setting, our method gives an alternative to procedures previously proposed by \cite{ARP}, \cite{CS} and \cite{FSST} (see also \cite{KS}). Second, we consider the more general case in which both $A$ and $b$ are potentially estimated. Computationally, both procedures involve generating a test statistic by solving a linear program (LP) in the original sample, and then computing a critical value though a bootstrap procedure in which an additional LP is solved within each bootstrap iteration. In the case where A is estimated, a (convex) quadratic program must also be solved once.

In the case of a known matrix \(A\), the classes of DGPs for which we establish uniform validity of our procedure are comparable to those considered in \citet{FSST} (FSST). But in some ways our conditions are more general. In particular, our results are established in a ``semi--high-dimensional'' regime in which the number of estimated components of \(b\) is bounded, while the number of its deterministic components, as well as the number of columns of $A$, are unrestricted and can vary arbitrarily with the sample size (see Assumption \ref{assAVKAM0} below). Thus, while FSST allow the dimension \(p\) (the number of rows of \(A\)) to grow at a controlled rate, with the dimension \(d\) (the number of columns of \(A\)) unrestricted, our approach permits unrestricted growth in both \(p\) and \(d\), provided that the number of estimated components of \(b\) remains bounded. In this sense, our framework accommodates a higher-dimensional regime than those considered in \citet{BSS}, \citet{ARP}, and \citet{CS}, which all study testing problem \eqref{eqnARP} under a regime where the dimensions of \(A\) remain bounded as the sample size increases.

Compared with the well-studied setting in which \(A\) is known, the more general case in which the coefficient matrix \(A\) must be estimated from data remains relatively underexplored. One exception is \citet{CSS} (CSS), who, concurrently with this work, also propose an inference procedure for the case in which both \(A\) and \(b\) are estimated. The CSS procedure computes a test statistic by solving a single convex quadratic program, with critical values taken from a chi-squared distribution whose degrees of freedom are determined by the rank of the inequality constraints active at the solution. Compared with our approach, their method is computationally inexpensive and requires no tuning parameters; however, these advantages come at a cost. First, any linear dependencies in the estimation error across the rows of \(A\) must be known and must match corresponding dependencies in \(b\). This requirement excludes some interesting configurations and may require careful implementation by the researcher. In addition, the CSS approach relies on a ``rank stability'' assumption, which ensures that the ranks of certain subsets of rows of \(A\) remain unchanged by estimation, with probability approaching one as the sample size grows. We show that this assumption can exclude some economically relevant \(A\) matrices. Consequently, whereas our procedure is guaranteed to be pointwise valid for any DGP (i.e., any \((A,b)\) pair), the approach of \citet{CSS} fails to be pointwise valid for certain DGPs (see Section \ref{sectionComp} for details).

In comparison to the assumptions under which we establish uniformity when the coefficient matrix is known, the case of an estimated \(A\) matrix requires an additional assumption (Assumption \ref{assAVEAM4} below). This assumption, in part, requires that a ``condition-like'' quantity, \(\Lambda(A)\), which measures how close the rows of \(A\) are to being collinear (see Definition \ref{deflambda}), is bounded.\footnote{\citet{AV} also assumes the boundedness of a similar condition-like quantity when establishing the uniform validity of the inference procedure.} This condition is not needed when the coefficient matrix is known. However, we show that if one only maintains assumptions similar to those used for uniformity with a known \(A\), there may not exist a uniformly valid test of hypothesis \eqref{eqnARP} with nontrivial power (see Proposition \ref{propImp}). This illustrates the necessity of additional assumptions in the setting with an estimated \(A\) matrix. Although \(\Lambda(A)\) is finite for any fixed matrix \(A\), it may be unbounded in the neighborhood of certain matrices, precluding the uniform validity of our procedure in such regions. Nevertheless, for the applications we consider, we show that this limitation is not restrictive.

We demonstrate the empirical relevance of our method with an estimated coefficient matrix $A$ to two problems involving inference on the average treatment effect (ATE) in a setting where the ATE is partially identified. In our first application, we build on the work of \cite{MST} who show that instrumental variables generally yield systems of linear equalities that relate causal parameters to moments involving the outcome variable. These systems can be augmented to capture auxiliary assumptions, which often appear as linear inequalities. Existing inference procedures such as FSST cannot always be applied in such settings, as both the outcome moments and the coefficient matrix can depend on the DGP. We design a Monte Carlo simulation around the empirical setting of \cite{Dupas2014}, in which valid instruments are provided by an experiment that randomized the price of an antimalarial bed net. Our procedure demonstrates desirable power properties, even with a relatively small sample size typical of field experiments. In our second application, we consider the ``monotone instrumental variables'' (MIV) framework introduced by \citet{MP}. We reframe the MIV assumption as a set of linear inequalities. We again demonstrate good power of our method in simulations. We apply the procedure empirically to estimating the average returns to college.

The structure of the paper is as follows. In Section \ref{secApplications}, we show how the problem \eqref{eqnARP} relates to inference problems from the literature, and gives some motivating empirical examples. Section \ref{sectionTP} introduces the test itself, describing how it would be applied by a practitioner. Section \ref{sectionAV} presents our main results establishing the theoretical properties of the test. Section \ref{secApplication} revisits our motivating examples and presents a simulation study. All proofs are deferred to the appendix.

\subsection{Notation}  
Let \( x \) be a vector in Euclidean space and \( A \) a matrix. Unless stated otherwise, we use \( \|x\| \) to denote the Euclidean norm of \( x \), \(\|x\|_p\) ($1\leq p \leq \infty$) for its general $l_p$-norm, and  \( \|A\| \) to denote the spectral norm of \( A \). We use $ \mathbb{R}_{+}^{p}$ to denote the set of elements of $ \mathbb{R}^{p}$ with non-negative entries. We use $\mathbbm{1}$ to denote a vector with all entries equal to one. For a given vector \( x \), we define \( (x)_+ \) and \( (x)_- \) as the positive and negative parts of \( x \), respectively, where the \( i^{\text{th}} \) entry of \( (x)_+ \) is given by \( \max\{x_i, 0\} \) and the \( i^{\text{th}} \) entry of \( (x)_- \) is given by \( \max\{-x_i, 0\} \). Given a point \( x \) and a subset \( A \) of some Euclidean space, we denote the Euclidean distance from \( x \) to \( A \) by \( d(x, A) \).  For two sets \( A \) and \( B \), we define \( \overset{\rightarrow}{d}_H(A,B) \) and \( d_H(A,B) \) as the directed (one-sided) Hausdorff distance from \( A \) to \( B \) and the Hausdorff distance between \( A \) and \( B \), respectively. These are given by $\overset{\rightarrow}{d}_H(A,B) = \sup_{x \in A} d(x, B)$, $d_H(A,B) = \max\{\overset{\rightarrow}{d}_H(A,B), \overset{\rightarrow}{d}_H(B,A)\}$. Given a matrix \( A \in \mathbb{R}^{m \times n} \) and a subset of row indices \( I \subset [m] \), where for $m\in \mathbb{N}$ we have $[m]:=\{1,\cdots,m\}$, we denote by \( A_I \) the submatrix obtained by retaining only the rows of \( A \) indexed by \( I \) and deleting all other rows. Given a vector \( b \in \mathbb{R}^m \) and a subset \( I \subset [m] \), we define \( b_I \) analogously. We use \( \Rightarrow \), \( \xrightarrow{p} \), and \( \xRightarrow{p} \) to denote convergence in distribution, convergence in probability, and convergence in distribution in probability, respectively. We use $d_{\mathrm{pr}}(\cdot,\cdot)$ to denote the Prohorov distance.

\section{Motivating examples} \label{secApplications}
Below, we show how several broad classes of problems can be written in the form of Eq. \eqref{eqnARP}. We begin with cases in which $A$ is known, and then move to the general case where both $A$ and $b$ are estimated. While our testing procedure differs depending on whether $A$ is known or is estimated,
both testing procedures make use of a linear programming characterization of the null hypothesis \eqref{eqnARP}. In particular, using Farkas' Lemma, hypothesis \eqref{eqnARP} is equivalent to 
\begin{equation} \label{eqnLPARP}
H_0: v(Q)=0 \quad \text{versus} \quad H_1: v(Q)>0,
\end{equation}
where
\begin{equation*}
v(Q) = \max \left\{ b(Q)^\top \lambda \mid \lambda \in \mathbb{R}_{+}^{p}, \ A(Q)^\top \lambda = 0 \right\}.
\end{equation*}
We end this section by illustrating the practical relevance of the case in which both $A$ and $b$ contain estimated components by presenting two specific examples.

\medskip

\subsection{Known coefficient matrix} \label{exampMIARP}
Many papers in the literature have considered hypothesis \eqref{eqnARP} with a known $A$. One case where it arises is in the context of subvector inference in moment inequality models in which nuisance parameters enter linearly. Specifically, consider a model in which a true parameter $\beta_0 = (\theta_0, \eta_0)$ is assumed to satisfy the moment inequality
\begin{equation} \label{eqnMIARP}
    E_{F_0}[m(W, \beta_0)] = E_{F_0}[h(W, \theta_0) - M(\theta_0)\eta_0] \leq 0,
\end{equation}
where the function $h$ and the coefficient matrix $M(\theta)$ $(\in \mathbbm{R}^{p \times d})$ are known given $\theta$. We are interested in inference on the component $\theta_0$ of $\beta_0$, treating $\eta_0$ as a nuisance parameter. A confidence interval for $\theta_0$ can be obtained by inverting a test of the hypothesis (for a fixed $\theta$)
\begin{equation} \label{eqH0Ab}
    H_0: \quad \exists \eta \in \mathbb{R}^d \quad \text{ s. t. } \quad A(\theta) \eta -b(\theta) \ge 0
\end{equation}
with $A(\theta)=M(\theta)$ and $b(\theta)=E_{F_0}[h(W, \theta)]$.\footnote{This involves conducting the test at a set of points $\theta$ on a grid in $\mathbb{R}^{d_{\theta}}$, where $d_\theta$ is the dimension of the parameter of interest $\theta_0$.} Procedures of this type for subvector inference on $\theta_0$ have been proposed by \cite{ARP} and \cite{CS}.\footnote{Although a different hypothesis is initially posed, the one ultimately tested in \cite{ARP} and \cite{CS} coincides with ours, as their inference is conducted conditional on the realized values of the instruments.} Note that for the special case in which $A=0$, \eqref{eqnMIARP} reduces to the moment inequality setting studied by \cite{DAGS}.\footnote{Moment equalities can be equivalently expressed as pairs of moment inequalities, so this setting also accommodates moment equalities.} 

Equivalently, \cite{FSST} consider the problem of testing hypotheses of the form:
\begin{equation}
\label{eqnMIFSST}
H_0:  \exists \eta \geq 0 \quad \text{such that} \quad A\eta = b,
\end{equation}
where the coefficient matrix $A\in \mathbb{R}^{p\times d}$ is known, and the vector $b$ is to be estimated (here again both $A$ and $b$ may depend on a parameter of interest $\theta_0$). The form of \eqref{eqnARP} nests that of \eqref{eqnMIFSST} and vice versa.

\citet{FSST}, \citet{ARP} and \cite{CS} give several examples of empirical applications where hypotheses of the form \eqref{eqnARP} or \eqref{eqnMIFSST} (with a known $A$ matrix) are relevant. We give two more examples below.
\medskip


\begin{example} \label{exDTE}
Consider an experiment in which a binary treatment $D$ is randomized, and there is perfect compliance. An outcome variable $Y$ with $J$ support points is available, with potential outcomes $Y(0),Y(1)$. For simplicity, label these points by integers $Y \in \{1,2, \dots J\}$. An important causal parameter in this setting is the share harmed by treatment, i.e. $\theta_0 = P(Y(1)-Y(0) \le 0) = P(Y(1)\le Y(0))$. The parameter $\theta_0$ is partially identified from the marginal distributions of $Y(0)$ and $Y(1)$, which are in turn identified given randomization of the treatment \citep{FP}. Let $\eta_{j,k} = P(Y(0) = j, Y(1)=k)$, $\theta = \sum_{j,k} 1[j \ge k] \cdot \eta_{jk}$. We can form a confidence interval for $\theta_0$ by inverting a test along a grid of values $\theta$, where the hypothesis takes the form of Eq. \eqref{eqnMIFSST} with:
$$A=\begin{bmatrix} M \\ N \\ (1 \dots 1) \\ c^T \end{bmatrix}, \quad \quad b(\theta,Q)=\begin{bmatrix} q^0(Q) \\ q^1(Q) \\ 1 \\ \theta \end{bmatrix},$$
where $M$ is a known $J \times J^2$ matrix with entries $M_{j',jk} = 1[j=j']$ while $q_0$ is a $J-$component vector with components $q^0_j(Q) = P_Q(Y=j|D=0)$, $N$ is a $J \times J^2$ matrix with entries $N_{k',jk} = 1[k=k']$ while $q_0$ is a $J-$component vector with components $q^1_k(Q) = P_Q(Y=k|D=1)$, and $c$ is a $J^2$-component vector with component $c_{jk}=1[j \ge k]$. Appendix \ref{secApShareHarmed} discuss further assumptions that can be layered on to this example to narrow the bounds on the parameter $\theta$. 
\end{example}

\begin{example} \label{exDefiers}
In the analysis of instrumental variables with heterogeneous treatment effects and a bounded discrete outcome, the monotonicity assumption of \citet{IA94}---while ensuring point identification of the average treatment effect among compliers---does not reduce the identified set for the overall average treatment effect (ATE) \citep{BP93,BHMSV}. However, alternative assumptions may be useful in narrowing bounds on the ATE. For example, \citet{dC} introduces the assumption that there are ``more compliers than defiers'' (MCTD), among each sub-population that shares a value of the treatment effect: i.e. $P(G=C|Y(1)-Y(0)) \ge P(G=D|Y_i(1)-Y_i(0))$, where $G$ indicates group membership in the compliers $(C)$, defiers $(D)$, always-takers or never-takers.

Let $\eta_{g,y,y'} = P(G=g,Y(1)=y,Y(0)=y')$. As in Example \ref{exDTE}, we can combine model restrictions, identifying information from the data, and a hypothesis about the ATE into a single hypothesis of the form $A \eta \ge b(Q)$ for some vector $\eta$. For example, with binary $Y$, we have a matrix with $4\times 2 \times 2 = 16$ columns, and 13 rows (encoding both equality and inequality restrictions). 4 equality restrictions come from moments of the form $E[Y1[D=d]|Z=z]$ for $z,d \in \{0,1\}$, one further equality restriction capturing the hypothesis about the ATE, $J$ positivity restrictions for the $\eta_{g,y,y'}$, and one normalization equality. The MCTD condition implies $3$ inequality restrictions. For all restrictions, the coefficients in $A$ are known.
\end{example}

\subsection{Estimated coefficient matrix} \label{exampMILSUC}

In many settings, hypothesis of the form \eqref{eqnARP} arise where both the matrix \( A \) and the vector $b$ must be estimated. For instance, moment inequalities that are linear in nuisance parameters \eqref{eqnMIARP} may feature a random matrix $M(W,\theta)$ that depends on the data $W$. In this case, $A(\theta)=E_{F_0}[M(W,\theta)]$ in  hypothesis \eqref{eqH0Ab} has to be estimated. 

Below we provide two examples of hypotheses of the form \eqref{eqnARP} with an estimated $A$ from the treatment effects literature. Further details on empirical applications that motivate the examples below are provided in Section \ref{secApplication}. We refer the reader to \cite{CSS} for further examples in which hypothesis \eqref{eqnARP} arises with an estimated coefficient matrix $A$.

\begin{example} \label{exMST}
    In settings with instrumental variables (IV) and heterogeneous treatment effects, interesting parameters like the average treatment effect (ATE) are often only partially identified by the instrument variation in the data. For example, \citet{Dupas2014} randomizes prices $Z$ for antimalarial bed nets. With treatment $D$ indicating purchase of a bed net and the outcome $Y$ and indicator for usage of the net, we consider in Section \ref{secApplication} inference on the ATE $E[Y(1)-Y(0)]$. As described by \cite{MST}, the identified set for the ATE can be narrowed considerably if the researcher is willing to parameterize the marginal treatment response curves $m_d(u)=E[Y(d)|U=u]=0$ for $d \in \{0,1\}$, where $U$ is a latent variable determining selection into treatment. Motivated by the application of \citet{Dupas2014}, we assume $m_1$ takes the form of a decreasing quadratic function $m_1(u)  = a + \beta + \gamma u^2$ and $m_0(u)=0$. With a binary instrument $Z$ (e.g. low vs. high price), the hypothesis that the ATE takes value $\theta$ leads to the following linear system of equalities and inequalities 
    \begin{equation} \label{eqMST}
        \begin{bmatrix}
    p(0) & p(0)^2/2 & p(0)^3/3\\
    p(1) & p(1)^2/2 & p(1)^3/3\\
    -1 & 0 & 0\\
    1 & 1 & 1\\
    0 & -1 & 0\\
    0 & -1 & -2\\
    1 & 1/2 & 1/3
    \end{bmatrix}\begin{pmatrix} a \\ \beta \\ \gamma
    \end{pmatrix} \quad  \begin{matrix} = \\ = \\ \ge \\ \ge \\ \ge \\ \ge \\ =
    \end{matrix} \quad \begin{pmatrix}
    	E[Y \cdot D|Z=0]\\E[Y \cdot D|Z=1]\\
        -1\\ 0 \\
        0 \\ 0\\ \theta
    \end{pmatrix}
    \end{equation}
    where $p(z) :=E[D|Z=z]$ is the propensity score function, which must be estimated. The first two rows capture the identifying content of the instrument variation in the data. The four inequalities that follow encode monotonicity of $m_1(u)$ along with the constraint that $m_1(u) \in [0,1]$ which follows from the binary outcome $Y$. The  final row tests a given hypothesis $\theta$ for the ATE, which in this model is equal to $a + \beta/2 + \gamma/3$.
\end{example}

\begin{example} \label{exMP}
    In settings where a valid IV is not available, \citet{MP} introduce the notion of a ``monotone IV'' $Z$ that satisfies $E[Y(t)|Z=z'] \ge E[Y(t)|Z=z]$ when $z' \ge z$, for each treatment value $t$. This weakens the assumption $E[Y(t)|Z=z'] = E[Y(t)|Z=z]$ for a valid instrument $Z$. In an application, we take $Y$ to be a measure of wages, treatment $T$ an indicator for going to college, and let $Z$ be a test score. Consider the special case of $\mathcal{T}=\mathcal{Z}=\{0,1\}$ and let the parameter of interest be $\theta_0 = E[Y(1)-Y(0)]$, the average treatment effect of college on wages. In Section \ref{secMonotoneIV}, we show that we can write the hypothesis that $\theta_0=\theta$ and that the MIV restriction is satisfied as a system of linear equalities and inequalities, where $\eta$ is an 8-component vector with entries $\eta_{tt'z}:=E[Y(t)|T=t',Z=z]$.  In particular, let $P(t,z):=P_Q(T=t,Z=z)$ and $P(t|z):=P_Q(T=t|Z=z)$. The system is:
    \[
    \resizebox{\textwidth}{!}{$
    \begin{bNiceArray}[first-row]{cccccccc}
    (000) & (001) & (010) & (011) & (100) & (101) & (110) & (111) \\
    1 & 0 & 0 & 0 & 0 & 0 & 0 & 0\\
    0 & 1 & 0 & 0 & 0 & 0 & 0 & 0\\
    0 & 0 & 0 & 0 & 0 & 0 & 1 & 0\\
    0 & 0 & 0 & 0 & 0 & 0 & 0 & 1\\
    - P(0|0) &  P(0|1) & -P(1|0) & P(1|1) & 0 & 0 & 0 & 0\\
    0 & 0 & 0 & 0 & - P(0|0) & P(0|1) & -P(1|0) & P(1|1)\\
    - P(0,0) & - P(0,1) & - P(1,0) & - P(1,1) & P(0,0) & P(0,1) & P(1,0) & P(1,1)
    \end{bNiceArray} \eta
    \qquad
    \begin{NiceArray}[first-row]{c}
        =\\=\\=\\=\\\ge \\\ge\\=
    \end{NiceArray} \qquad
    \begin{pNiceArray}[first-row]{c}\\
    E_Q[Y|T=0,Z=0]\\
    E_Q[Y|T=0,Z=1]\\
    E_Q[Y|T=1,Z=0]\\
    E_Q[Y|T=1,Z=1]\\
    0\\
    0\\
    \theta
    \end{pNiceArray}
    $}
    \]
    In the matrix above, the tuples $(tt'z)$ indicate the ordering of the components of $\eta$ for legibility. The two inequalities above impose the MIV assumption, the first four equalities reflect the identifying content of the data, and the final equality imposes the hypothesis that $\theta_0=\theta$.    
\end{example}

\begin{remark} \label{remark:Ceta}
In many cases, the researcher is interested in inferring \( \theta_0 = C \eta_0 \), for some known matrix \( C \), where the identified set for \( \eta_0 \) is defined by the linear system of inequalities $b \le A\eta$, and both the matrix \( A \) and the vector \( b \) must be estimated. Following the suggestion of \cite{FSST}, a confidence interval for the subvector \( \tau_0 \) can be obtained by inverting a test of the following hypothesis for various candidate values of \( \tau \):
\begin{equation}
\label{eqnLSUC}
H_0: \exists\, \theta \text{ such that } A\theta \ge b \quad \text{and} \quad C\theta = \tau.
\end{equation}
The latter hypothesis is of the form \eqref{eqnARP}. Examples \ref{exDTE}, \ref{exDefiers}, \ref{exMST} and \ref{exMP} are all of this form with $\theta_0$ a scalar.
\end{remark}

\begin{remark}
    A related problem to the one described in Remark \ref{remark:Ceta} is to conduct inference on $\theta_0$ defined as a maximizer of a linear function subject to linear inequality constrraints, i.e.:
    \[
    \theta_0 \in \arg \max \{ c^\top \theta \mid M\theta \le d \},
    \]
    We refer the reader to \cite{HSS22} for economic applications that motivate this problem.
    
    \cite{HSS22} give a method for this problem that constructs confidence regions via a profile test statistic. In Appendix \ref{exampHSS}, we describe how this procedure may become impractical when the number of constraints in the linear program becomes large. We show there that this problem can be recast into a hypothesis of the form \ref{eqnARP}, and our approach thus provides an alternative that remains computationally tractable in high-dimensional settings.
\end{remark}


\section{Testing Procedures}
\label{sectionTP}

In this section, we describe two inference procedures for the testing problem \ref{eqnARP}. We provide details on the computation of the test statistics and critical values, and we discuss the statistical properties of these procedures in Section \ref{sectionAV} below. 

The first procedure concerns the setting in which the coefficient matrix is known and offers an alternative to the testing procedures proposed in \cite{CS}, \cite{ARP}, and \cite{FSST}. The second procedure considers the case in which the matrix $A$ must be estimated.

\subsection{Known coefficient matrix}
\label{subsectionKAM}

In this section, we propose a testing procedure for hypotheses of the form \ref{eqnARP}, where the matrix $A$ is known while some components of $b(Q)$ are unknown and must be estimated from the data.

Our proposed test statistic is based on the LP characterization of the hypothesis in \ref{eqnLPARP} and is given by
\begin{equation}
\label{eqnKAM2}
\hat{T}_N = \max\{\lambda^\top \hat{b} \mid \lambda \in \mathbb{R}_+^p,\ A^\top \lambda = 0,\ \mathbbm{1}^\top \hat{D}\lambda \leq 1\}.
\end{equation}

Here, as in Assumption~\ref{assAVKAM1} below, it is assumed that $\hat{b}$ is a $\sqrt{N}$-consistent estimator of $b$ and that 
$\hat{D} = \operatorname{diag}(\hat{\Sigma})^{1/2}$, where $\hat{\Sigma}$ is a consistent estimator of the asymptotic variance of $\sqrt{N}(\hat{b}-b)$.

For instance, if $b = E[m(W,\theta)]$ for a given value of $\theta$, then given an iid sample $\mathbb{W}:=\{W_i\}_{i=1}^N$, $\hat{b}$ and $\hat{D}$ can be given by
\begin{equation}
\label{eqnKAM21}
\hat{b} = \bar{m}(\theta) = \frac{1}{N} \sum_{i=1}^N m(W_i, \theta), 
\quad \text{and} \quad 
\hat{D} = \operatorname{diag}(\hat{\Sigma})^{1/2},
\end{equation}
where
\[
\hat{\Sigma} = \frac{1}{N} \sum_{i=1}^N 
\left( m(W_i, \theta) - \bar{m}(\theta) \right)
\left( m(W_i, \theta) - \bar{m}(\theta) \right)^\top.
\]

Here, the constraint in the LP \ref{eqnKAM2} involving $\hat{D}$ is a normalization that serves two purposes. 
First, it ensures that the components of $\lambda$ corresponding to the estimated elements 
of $b$ remain bounded over the feasible region of the LP. Second, it renders the test statistic 
scale-invariant by accounting for heteroskedasticity through $\hat{D}$.

Given a significance level $\alpha \in (0, 1/2)$, we reject the null hypothesis in favor of the alternative if and only if $\sqrt{N}\hat{T}_N > \hat{c}_N(1 - \alpha)$, where $\hat{c}_N(1 - \alpha)$ is defined by
\begin{equation}
\label{eqnKAM3}
\hat{c}_N(1 - \alpha) := \inf\{ u \mid \hat{H}_N(u) \geq 1 - \alpha \},
\end{equation}
and $\hat{H}_N(\cdot):=\mathcal{L}(v^*|\mathbb{W})$ is the conditional (on $\mathbb{W}$) distribution of $v^*$, where $v^*$ is the value of the following LP:
\begin{equation}
\label{eqnKAM4}
v^*=\max_{\lambda\in \hat{\Delta}}\lambda^\top \zeta^* \quad \text{where} \quad \hat{\Delta}:=\left\{ \lambda \in \mathbb{R}_+^p  \;\middle\vert\; \ A^\top \lambda= 0, \ \|\hat{D} \lambda\|_1 \leq 1, \ \hat{b}^\top \lambda \geq \hat{T}_N - \kappa_N/\sqrt{N} \right\},
\end{equation}
and where $\kappa_N=o(\sqrt{N})$, $\kappa_N\to \infty$. Here, $\zeta^*$ is a simulated variable whose distribution approximates that of $\sqrt{N}(\hat{b}-b)$ (see Assumption~\ref{assAVKAM3} below). In the setting of equation \ref{eqnKAM21} for instance, we can take $\hat{\zeta^*}$ to be the bootstrap statistic
\[
\zeta_N^*=\frac{1}{\sqrt{N}}\sum_{i=1}^N \bigl(m(W_i^*,\theta)-\bar{m}(\theta)\bigr), \quad \text{where}\quad W_i^* \mid \mathbb{W}_N\overset{i.i.d.}{\sim} \hat{F}_N,
\]
where $\hat{F}_N$ denotes the empirical distribution of the data $\mathbb{W}_N$.

\begin{remark}
\label{remKAM}
Our test differs from that proposed in \cite{ARP} in that we normalize using a weak inequality constraint ($\|\hat{D}\lambda\|_1 \leq 1$), whereas they use a strict equality ($\|\hat{D}\lambda\|_1 = 1$). We prefer the weak inequality formulation for two reasons. First, it simplifies the analysis. Our derivation relies crucially on the fact that, under this normalization, the value of the linear program is zero under $H_0$ and remains nonnegative under small perturbations. That is, under $H_0$, the LP value functional is locally minimized at zero. Second, it guarantees that the feasible region is always nonempty (as it always contains the origin), which may fail to hold under the strict normalization $\|\hat{D}\lambda\|_1 = 1$.\footnote{However, as noted in \cite{ARP}, the feasible region is empty under this normalization if and only if the matrix $A$ has full range, in which case the null holds for any $b$.}
\end{remark}

\subsection{Estimated coefficient matrix}
\label{subsectionEAM}
In this section, we propose a testing procedure for hypotheses of the form \ref{eqnARP},  now allowing for the possibility that some entries of the matrix $A(Q)$,  as well as some components of $b(Q)$, are unknown and must be estimated from the data.

Let $\hat{c} = \operatorname{vec}([\hat{A}\ \hat{b}])$ be a $\sqrt{N}$-consistent estimator of 
$d = \operatorname{vec}([A\ b])$, with asymptotic variance $\Sigma$ and asymptotic standard 
deviation matrix $D = \operatorname{diag}(\Sigma)^{1/2}$ (see Assumption~\ref{assAVEAM1} below). 
Let $\hat{\Sigma}$ and $\hat{D}$ denote consistent estimators of $\Sigma$ and $D$, respectively.

Let $\sigma_{A_{ij}}$ and $\sigma_{b_i}$ denote the asymptotic standard deviations of the estimators 
of the $(i,j)$-th entry of the matrix $A$ and the $i$-th entry of the vector $b$, respectively. 
These correspond to diagonal entries of $D$.

 Let $\hat{\sigma}_{A_{ij}}$ and $\hat{\sigma}_{b_i}$ be the analogous entries from the estimator $\hat{D}$. Define the diagonal matrices $\Omega$ and $\hat{\Omega}$ (both in $\mathbb{R}^{p \times p}$) by:
\begin{equation}
\label{eqnEAM1}
\Omega_{ii} = \sigma_{b_i} \vee \max_{j \in [d]} \sigma_{A_{ij}}, \quad \quad 
\hat{\Omega}_{ii} = \hat{\sigma}_{b_i} \vee \max_{j \in [d]} \hat{\sigma}_{A_{ij}}.
\end{equation}
The matrix serves the same normalizing role as $\hat{D}$ in Section \ref{subsectionKAM}, and is used to ensure that the test statistic that we propose is scale invariant. 

Following the linear programming characterization \ref{eqnLPARP}, our test statistic is given by:
\begin{equation}
\label{eqnEAM2}
\hat{T}_N:=\max\bigl\{\hat{b}^\top \lambda \mid \lambda\geq 0,\  \hat{A}^\top \lambda=0,\ \mathbbm{1}^\top \hat{\Omega}\lambda \leq 1\bigr\}.
\end{equation}
We reject the null hypothesis in favor of the alternative whenever $\sqrt{N}\hat{T}_N$ is large. Specifically, given a significance level $\alpha \in (0, 1/2)$, we reject the null hypothesis in favor of the alternative if and only if
 \[
\sqrt{N}\hat{T}_N > \hat{c}_N(1 - \alpha),
\]
 where $\hat{c}_N(1 - \alpha)$ is defined by
\begin{equation}
\label{eqnEAM3}
\hat{c}_N(1 - \alpha) := \inf\{ u \mid \hat{H}_N(u) \geq 1 - \alpha \},
\end{equation}
and $\hat{H}_N(\cdot):=\mathcal{L}(v^*|\mathbb{W})$ is the conditional (on $\mathbb{W}$) distribution of $v^*$, where $v^*$ is the value of the following LP:
\begin{equation}
\label{eqnEAM4}
v^* = \max_{\lambda \in \hat{\Delta}} \left\langle \zeta_b^* - \zeta_A^* \hat{\eta}, \lambda \right\rangle
\end{equation}
where
\begin{equation}
\label{eqnEAM5}
 \hat{\Delta}:=\left\{ \lambda \in \mathbb{R}^p_+  \;\middle\vert\; \  -\frac{\kappa_{1N}}{\sqrt{N}}\mathbbm{1}\leq \hat{A}^\top \lambda\leq \frac{\kappa_{1N}}{\sqrt{N}}\mathbbm{1}, \ \|\hat{\Omega} \lambda\|_1 \leq 1, \ \hat{b}^\top \lambda \geq \hat{T}_N - \kappa_{2N}/\sqrt{N} \right\},
\end{equation}
and $\hat{\eta}$ is given by the (unique) solution of the quadratic program\footnote{We could replace the quadratic program with a linear program by minimizing the $\ell_1$ norm instead. However, in that case, Assumption~\ref{assAVEAM5} would need to hold for all covariance matrices $V(Q)$, where $\eta_Q$ is now any element of the set $\arg\min\{\|\eta\|_1 \mid b - A\eta \leq 0\}$, which may now contain multiple elements.}
\begin{equation}
\label{eqnEAM6}
\hat{\eta}=\argmin\Bigl\{\|\eta\|^2 \  \big| \  \hat{b}-\hat{A}\eta \leq \bigl(\hat{T}_N+\frac{\kappa_{3N}}{\sqrt{N}}\bigr)\hat{\Omega}\mathbbm{1}\Bigr\}.
\end{equation}
Here, $\zeta_b^*$ and $\zeta_A^*$ are the $b$ and $A$ components of $\zeta_N^*$, the bootstrap-like estimator of the asymptotic distribution of $\hat{b}$ (see Assumption \ref{assAVEAM3} below). The tuning parameters $\kappa_{1N}, \ \kappa_{1N}$, and $\kappa_{3N}$ are all required to satisfy $\kappa_N=o(\sqrt{N})$ and $\kappa_N\to \infty$. 

\section{Uniform asymptotic validity}
\label{sectionAV}

In this section, we present assumptions under which the testing procedures discussed in 
Section \ref{sectionTP} are uniformly asymptotically valid. Following the structure of Section 
\ref{sectionTP}, we divide the exposition into two subsections. Section 
\ref{subsectionAVKAM} considers the case in which the coefficient matrix is known, 
whereas Section \ref{subsectionAVEAM} considers the case in which the coefficient matrix 
must be estimated from the data. In each subsection, we also discuss how our assumptions 
compare with those required for the uniform validity of alternative procedures proposed 
in the literature.


\subsection{Known coefficient matrix}
\label{subsectionAVKAM}

In this subsection, we formally state the restrictions on the class of null DGPs ${\cal Q}_0$ 
under which we establish the uniform asymptotic validity of the testing procedure introduced 
in Section \ref{subsectionKAM}. Throughout, we focus on a semi–high-dimensional setting in which the dimension of the unknown components  of $b$ remains bounded, while the dimension of the deterministic component of $b$, as well as the dimension of the nuisance parameter, are allowed to grow arbitrarily with the sample size, as stated in in Assumption \ref{assAVKAM0} below.

Let \( b(Q) = (b^u(Q), b^k(Q)) \in \mathbb{R}^p \), where \( b^k(Q) \in \mathbb{R}^{p_k} \) represents the deterministic (known) component of \( b \), and \( b^u(Q) \in \mathbb{R}^{p_u} \) represents the unknown component of \( b(Q) \).

\begin{assumption}
\label{assAVKAM0}
We assume that \( p_u \) remains bounded as \( N \) increases, with all other dimensions of the problem allowed to vary arbitrarily with \( N \).\footnote{That is, we allow \( p_k \) to vary arbitrarily with \( N \), and, if \( A \in \mathbb{R}^{p \times d} \), then \( d \) is also allowed to vary arbitrarily with \( N \); however, it is without loss of generality to assume \( d \leq p \), since when A is known, the hypothesis~\ref{eqnARP} remains unchanged if \( A \) is replaced by a subset of its columns that spans its column space.}
\end{assumption}

\medskip

As in Section~\ref{subsectionKAM}, we assume that estimators $\hat{b}$ and $\hat{D}$ of 
$b$ and $D$ (the matrix of asymptotic standard deviations) are available and satisfy 
Assumptions~\ref{assAVKAM1} and~\ref{assAVKAM2} below. Assumption~\ref{assAVKAM1} requires 
that $\hat{b}$ is asymptotically Gaussian and converges weakly to its Gaussian limit 
uniformly over ${\cal Q}_0$. Assumption~\ref{assAVKAM2}, in part, requires that the 
diagonal entries of the asymptotic standard deviation matrix corresponding to the 
unknown component $b^u$ of $b$ are uniformly bounded away from zero and infinity. 

\begin{assumption}
\label{assAVKAM1}
For each $Q \in \mathcal{Q}_0$, there exists a symmetric matrix $\Sigma(Q)$, representing the asymptotic variance of $\hat{b}$, such that for all $\epsilon > 0$,
\[
\overline{\lim}_{N \to \infty} \sup_{Q \in \mathcal{Q}_0} d_{pr} \left( \sqrt{N}(\hat{b} - b(Q)), \mathcal{N}(0, \Sigma(Q)) \right) = 0.
\]
\end{assumption}

\begin{assumption}
\label{assAVKAM2}
\begin{enumerate}[i)]
\item There exist positive constants $\underline{\sigma}$ and $\bar{\sigma}$, with $\underline{\sigma} < \bar{\sigma}$, such that for all $Q \in \mathcal{Q}$, all diagonal elements of $D(Q) (= \operatorname{diag}(\Sigma(Q))^{1/2})$ that correspond to the estimated component of $b$ lie in the interval $[\underline{\sigma}, \bar{\sigma}]$.

\item For all $\epsilon > 0$,
\[
\overline{\lim}_{N \to \infty} \sup_{Q \in \mathcal{Q}_0} Q \left( \|\hat{D} - D(Q)\| > \epsilon \right) = 0.
\]
\end{enumerate}
\end{assumption}
\medskip

\begin{remark}
\label{remAVKAM1}
Consider the setting of equation \ref{eqnKAM21}, where for a given value of $\theta$, \( \hat{b}\) (=$\hat{b}(\theta)$) and \( \hat{D} \) (=$\hat{D}(\theta)$) are given by
\[
\hat{b} = \bar{m}(\theta) = \frac{1}{N} \sum_{i=1}^N m(W_i, \theta), \quad \text{and} \quad 
\hat{D} = \operatorname{diag}(\hat{\Sigma})^{1/2}, 
\]
with
\[
\hat{\Sigma} = \frac{1}{N} \sum_{i=1}^N \left( m(W_i, \theta) - \bar{m}(\theta) \right)\left( m(W_i, \theta) - \bar{m}(\theta) \right)^\top.
\]

Then Assumptions~\ref{assAVKAM1} and~\ref{assAVKAM2} (part (ii)) are satisfied if, for some \( \delta > 0 \), the following “\( 2 + \delta \)” condition holds\footnote{For details, see Section 11.4.2 of \cite{ELJR}}:
\[
\sup_{Q \in \mathcal{Q}} \mathbb{E}_F\left[ \|m(W, \theta)\|^{2 + \delta} \right] < \infty.
\]
\end{remark}
\medskip

We assume that the bootstrap-like estimator $\zeta_N^*$ of the asymptotic distribution 
of $\sqrt{N}(\hat{b} - b)$, used to compute the critical values, satisfies the following 
assumption.

\begin{assumption}
\label{assAVKAM3}
We have available a random variable $\zeta_N^*$ whose distribution is entirely determined by the realized sample $\mathbb{W}$ and which satisfies: for all $\epsilon > 0$,
\[
\overline{\lim}_{N \to \infty} \sup_{Q \in \mathcal{Q}} Q\Biggl(d_{\text{Pr}}\Bigl(\mathcal{L}(\zeta_N^* \mid \mathbb{W}), \mathcal{N}(0, \Sigma(Q))\Bigr) > \epsilon\Biggr) = 0.
\]
\end{assumption}
\medskip 
\begin{remark}
\label{remAVKAM2} Continuing with Remark \ref{remAVKAM1}, the arguments in Section 11.4.2 of \cite{ELJR} (see also Theorem 2.4 in \cite{JRAS}) can be adapted to show that the same $2+\delta$ condition implies that Assumption \ref{assAVKAM3} holds for the bootstrap statistic
\[
\zeta_N^*=\frac{1}{\sqrt{N}}\sum_{i=1}^N \bigl(m(W_i^*,\theta)-\bar{m}(\theta)\bigr), \quad \text{where}\quad W_i^* \mid \mathbb{W}_N\overset{i.i.d.}{\sim} \hat{F}_N,
\]
and $\hat{F}_N$ denotes the empirical distribution of the data $\mathbb{W}_N$.
\end{remark}
\medskip

We consider additionally an assumption which puts restrictions on the relationship between the coefficient matrix $A$ and the asymptotic variance matrix $\Sigma(Q)$ for DGPs in the null family $\mathcal{Q}_0$. Assumption \ref{assAVKAM4} below ensures that the asymptotic distribution of our test statistic does not ``concentrate too much'' along arbitrary sequences of DGPs $Q_N \in \mathcal{Q}_0$, which could otherwise undermine the uniform validity of our procedure. It is possible however to show that whithout this assumption, our method remains valid if we modify the rejection rule and instead reject iff $\sqrt{N}\hat{T}_N > \hat{c}_N(1 - \alpha)+\tau$, for some fixed arbitrary $\tau>0$.

Before stating the assumption, we introduce some additional notation. Let the population counterpart of the LP in \ref{eqnKAM2} be given by 
\begin{equation}
\label{eqnAVKAM0}
 \max\{\lambda^\top b(Q) \mid \lambda \in \mathbb{R}_+^p,\ A(Q)^\top \lambda = 0,\ \mathbbm{1}^\top D(Q)\lambda \leq 1\}.
\end{equation}
Let $extr(\mathcal{D}(Q))$ denote the set of extreme points of the feasible region of the LP~\ref{eqnAVKAM0}, defined by
\[
\mathcal{D}(Q) := \left\{ \lambda \in \mathbb{R}_+^p \,\middle|\, A(Q)^\top \lambda = 0,\ \mathbbm{1}^\top D(Q)\lambda \leq 1 \right\}.
\]
Let $\Delta_0(Q)$ denote the subset of ${\cal D}(Q)$, consisting of the optimal solutions to the LP~\ref{eqnAVKAM0}.\footnote{Note that $extr({\cal D}(Q))$ is non-empty as it always contains zero.}

\begin{assumption}
\label{assAVKAM4}
There exists a constant $\rho > 0$ such that for all $Q \in \mathcal{Q}_0$:
\begin{enumerate}[i)]
    \item If $\Delta_0(Q) \neq \{0\}$, then there exists $\lambda \in \Delta_0(Q)$ such that $\lambda^\top \Sigma(Q) \lambda \geq \rho$.
    \item If $\Delta_0(Q) = \{0\}$ and $extr(\mathcal{D}(Q)) \setminus \{0\}\neq \emptyset$, then for all $\lambda \in extr(\mathcal{D}(Q)) \setminus \{0\}$, we have $\lambda^\top \Sigma(Q) \lambda \geq \rho$.
\end{enumerate}
\end{assumption}

\medskip

\begin{remark}
\label{remAVKAM3}
Assumption~\ref{assAVKAM4} is trivially satisfied if the variance matrices $\Sigma(Q)$ are non-singular, with their eigenvalues uniformly bounded away from zero and infinity.\footnote{In that case, the nonzero extreme points are uniformly bounded away from zero as they satisfy $\|D\lambda\|_1=1$, and thus $\|\lambda\|_1\geq 1/\underline{\sigma}$.} While Assumption~\ref{assAVKAM4} is analogous to Assumption 4 in \cite{ARP}, it is weaker.\footnote{ As in \cite{ARP}, in part ii) of the assumption, it suffices to only consider vertices that occur with positive probability as optimal solutions of the sample LP, when $N\to \infty$.} In particular, our assumption allows for the possibility that asymptotically, with positive probability, two distinct vertices arise as optimal solutions of the sample LP \ref{eqnKAM2}, a posibility that is ruled out by Assumption 4 in \cite{ARP}, and which can arise when $\Sigma(Q)$ is singular. The approach in \cite{CS} does not impose this type of assumption. Instead, as in \cite{ARP}, it assumes that in the case of degeneracy (when $\Sigma(Q)$ is singular), the structure of the degeneracy is \emph{known}. Specifically, if $b(Q) = \mathbb{E}[m(W, \theta)]$ and $\Sigma(Q) = \operatorname{Var}(m(W, \theta))$ is singular, then there exists a \emph{known} matrix $B$ and a function $g$ such that $b(Q) = B \mathbb{E}[g(W, \theta)]$, where $\operatorname{Var}(g(W, \theta))$ is non-singular, with eigenvalues uniformly bounded away from zero and infinity. By contrast, in the degenerate case, our framework does not require the structure of the linear dependence among the components of $b$ to be known. 
\end{remark}

\medskip

We now state the main result of this section. The proof is in Section~\ref{pthmAVKAM} of the appendix.

\begin{theorem}
\label{thmAVKAM}
Suppose that Assumptions~\ref{assAVKAM0}--\ref{assAVKAM4} hold, and let $\hat{T}_N$ and $\hat{c}_N(\cdot)$ be defined as in equations~\ref{eqnKAM2} and~\ref{eqnKAM3}, respectively. Then, for all $\alpha \in (0, 1/2)$, we have
\begin{equation}
\label{eqnAVKAM1}
\overline{\lim}_{N \to \infty} \sup_{Q \in \mathcal{Q}_0} Q\left(\sqrt{N}\hat{T}_N > \hat{c}_N(1 - \alpha)\right) \leq \alpha.
\end{equation}
Moreover, for each $Q \in \mathcal{Q}_0$ and any $\epsilon > 0$, we have
\begin{equation}
\label{eqnAVKAM2}
\lim_{N \to \infty} Q\left( \left\{ \mathbb{W}_N \;\middle|\; d_{\mathrm{Pr}}\bigl( \mathcal{L}(\sqrt{N} \hat{T}_N), \hat{H}_N \bigr) > \epsilon \right\} \right) = 0,
\end{equation}
where $\hat{H}_N(\cdot)$ is the conditional distribution (given $\mathbb{W}_N$) of $v^*$ as defined in equation~\eqref{eqnKAM4}.
\end{theorem}

\begin{remark}
\label{remAVKAM4} 
The second part of Theorem~\ref{thmAVKAM} (equation~\ref{eqnAVKAM2}) implies that the inequality in equation~\ref{eqnAVKAM1} can be replaced by an equality if, for some $Q \in \mathcal{Q}_0$, the asymptotic distribution of $\sqrt{N} \hat{T}_N$ is nondegenerate. The asymptotic distribution is derived in Lemma~\ref{lemAD}, and it is nondegenerate if and only if the LP~\ref{eqnAVKAM0} has at least one optimal solution $\lambda \in \Delta_0(Q)$ such that its component $\lambda^u$ that corresponds to the estimated part of $b$ is not zero. Thus our approach yields pointwise asymptotic exact size, for all DGPs on the "boundary" of the null region ${\cal Q}_0$. This shows, in particular, that our test is not unnecessarily conservative. By contrast, the LF (least favorable) and hybrid tests of \cite{ARP} have exact pointwise asymptotic size only for DGPs where all feasible $\lambda$ are optimal (i.e., $\Delta_0(Q)={\cal D}(Q)$)\footnote{Equivalently, this corresponds to DGPs for which there exists a value of the nuisance parameter $\eta$ such that $b(Q) - A\eta =0$.}, which (By Farkas' Lemma) is not possible in settings where for each $Q\in {\cal Q}_0$, the (reverse) inequality $b(Q)-A\eta\geq 0$ does not have a solution $\eta$.\footnote{In the moment inequality setting (where $A=0$) this is for instance the case if all moment inequalities cannot bind simulataneously, i.e., the inequality $b(Q)\leq 0$, never holds as an equality for all $Q \in {\cal Q}_0$.} We refer the reader to \cite{ARP} for details.
\end{remark}
\begin{remark}
Theorem \ref{thmAVKAM} and Assumption \ref{assAVKAM0} imply that our method remains uniformly valid regardless of the dimensions of the matrix $A$ and the vector $b$, provided that the number of estimated components of $b$ remains bounded as the sample size grows. This indicates that the number of deterministic restrictions in the system $b \leq A\eta$ does not complicate the inferential task. While \cite{FSST} establish uniform validity in a setting where $p = o(N)$ and $d$ is unrestricted, to our knowledge, our paper is the first to emphasize that the key factor is the number of estimated components of $b$ and their growth rate relative to $N$.
\end{remark}

\subsection{Estimated coefficient matrix}
\label{subsectionAVEAM}

In this subsection, we formally state the restrictions on the class of null DGPs ${\cal Q}_0$ under which we establish the uniform asymptotic validity of the testing procedure introduced in Section \ref{subsectionEAM}. Throughout, we focus on a semi–high-dimensional setting in which the number of estimated components of $A$ and $b$ remains bounded, while the number of the deterministic components of $A$ and $b$ is allowed to grow arbitrarily with the sample size, as summarized in Assumption \ref{assAVEAM0} below.

To simplify the analysis, we assume that the number of columns of $A$ is fixed, while the number of rows may grow with the sample size. Let $K \cup U = [p]$ denote a partition of $[p]$, such that $i\in K$ iff $A_{ij}$, for $j\in[d]$, and $b_i$ are known and not subject to estimation. Thus $U$ denotes the set of indices $i$ such that either $b_i$ is unknown or $A_{ij}$ is unknown for some $j\in [d]$. Let $p^u:=|U|$ and $p^k=|K|$. We impose the following assumption.\footnote{It is possible to allow the number of columns of $A$ to grow with the sample size; what matters is that the number of estimated components of $A$ remains uniformly bounded.} We let $\Pi^u(\lambda)(\cdot)$ denote the projection on on the $U$-coordinates defined by
\[
[\Pi^u(\lambda)]_i=\lambda_i \quad \forall i\in U \quad\quad \text{and}\quad \quad [\Pi^u(\lambda)]_i=0\quad \forall i\in K.
\]

\begin{assumption}
\label{assAVEAM0}
We assume that $d$ and $p^u$ are uniformly bounded in $N$, while $p^k$ may grow arbitrarily with $N$.
\end{assumption}

Let $c(Q) = \operatorname{vec}\bigl([A(Q)\ \  b(Q)]\bigr)$. Analogously to Assumption~\ref{assAVKAM1}, we assume access to uniformly consistent estimators of the matrix $A$ and vector $b$, denoted by $\hat{c}$, which satisfy a uniform asymptotic normality property, as formalized in the following assumption.

\begin{assumption}
\label{assAVEAM1}
For each $Q \in \mathcal{Q}$, there exists a symmetric matrix $\Sigma(Q)$, representing the asymptotic variance of $\hat{c}$, such that for all $\epsilon > 0$,
\[
\overline{\lim}_{N \to \infty} \sup_{Q \in \mathcal{Q}} d_{pr} \left( \sqrt{N}(\hat{c} - c(Q)), \mathcal{N}(0, \Sigma(Q)) \right) = 0.
\]
\end{assumption}
We note that although the dimension of $c$ may vary with $N$, Assumption~\ref{assAVEAM1} is not restrictive in light of Assumption~\ref{assAVEAM0}, which implies that all but a uniformly bounded number of components of $\sqrt{N}(\hat{c} - c)$ are zero.

Let $D(Q) = \operatorname{diag}(\Sigma(Q))^{1/2}$ denote the asymptotic standard deviation matrix of $\hat{c}$, and let $\hat{D}$ be an estimator of $D$. Analogously to Assumption~\ref{assAVKAM2}, the following assumption requires that $\hat{D}$ is a uniformly consistent estimator of $D$. We also assume that the asymptotic standard deviations of the estimated components of $c$ are uniformly bounded away from zero and infinity over all $Q \in \mathcal{Q}$.

\begin{assumption}
\label{assAVEAM2}
\begin{enumerate}[i)]
\item There exist positive constants $\underline{\sigma}$ and $\bar{\sigma}$, with $\underline{\sigma} < \bar{\sigma}$, such that for all $Q \in \mathcal{Q}$, all diagonal elements of $D(Q)$ that correspond to the estimated component of $c$ lie in the interval $[\underline{\sigma}, \bar{\sigma}]$.
\item For all $\epsilon > 0$,
\[
\overline{\lim}_{N \to \infty} \sup_{Q \in \mathcal{Q}} Q \left( \|\hat{D} - D(Q)\| > \epsilon \right) = 0.
\]
\end{enumerate}
\end{assumption}
\medskip 

The next assumption requires the availability of a uniformly consistent bootstrap-like estimate of the asymptotic distributions $\{\mathcal{N}(0, \Sigma(Q)) \mid Q \in \mathcal{Q}\}$.

\begin{assumption}
\label{assAVEAM3}
There exists a random variable $\zeta_N^*$, whose distribution is entirely determined by the realized sample $\mathbb{W}$, such that for all $\epsilon > 0$,
\[
\overline{\lim}_{N \to \infty} \ \sup_{Q \in \mathcal{Q}} \ Q\left( d_{\mathrm{Pr}}\left( \mathcal{L}(\zeta_N^* \mid \mathbb{W}), \mathcal{N}(0, \Sigma(Q)) \right) > \epsilon \right) = 0.
\]
\end{assumption}

\medskip

To describe the structural condition we impose on the coefficient matrices $A$, we introduce the following definition, which serves as an analogue of a condition number for systems of inequalities. The quantity $\Lambda(A)$ considers subsets of rows of $A$ that form a basis for its row space and measures how close such subsets are from being linearly dependent. It thus serves as a measure of multicollinearity, becoming large when some basis of the row space are nearly collinear.

\begin{definition}
\label{deflambda}
Given a symmetric matrix $B$, let $\lambda_{\min}(B)$ denote its smallest eigenvalue. For an arbitrary non-zero matrix $A \in \mathbb{R}^{m \times n}$ with $\operatorname{rank}(A) = r$, define the (strictly positive) quantity $\Lambda(A)$ by
\begin{equation}
\Lambda(A)^{-1} := \min\bigl\{\lambda_{\min}(A_F A_F^\top) \mid F \subseteq [m],\ |F| = r,\ \operatorname{rank}(A_F) = r \bigr\},
\end{equation}
where $A_F$ denotes the submatrix formed by the rows indexed by $F$. When $A=0$, set $\Lambda(A)=0$.
\end{definition}

As shown in Lemma~\ref{lemhoff} in the appendix, $\Lambda(A)$ provides an upper bound on the Hoffman constant, which characterizes the local behavior of the polyhedron $\{x \mid Ax \leq b\}$ under perturbations of $A$ and $b$. Although sharper bounds for the Hoffman constant exist in the literature—such as the Robinson constant (see~\cite{SMR3})—we employ $\Lambda(\cdot)$ because of its interpretability. Accordingly, the condition in part (i) of Assumption~\ref{assAVEAM4} below could alternatively be replaced with the less restrictive requirement that the Hoffman constant is uniformly bounded.

\begin{assumption}
\label{assAVEAM4}
The coefficient matrices $A(Q)$ and vectors $b(Q)$ satisfy:
\begin{enumerate}[i)]
\item $\sup_{Q\in {\cal Q}_0}\Lambda(A(Q))<\infty$.
\item For each $Q\in {\cal Q}_0$, let 
\[
\eta_Q:=\argmin\{\|\eta\|\mid b(Q)-A(Q)\eta\le 0\}.
\]
We assume that $\sup_{Q\in {\cal Q}_0}\|\eta_Q\|<\infty$.
\end{enumerate}
\end{assumption}

Part~(i) of Assumption~\ref{assAVEAM4} is mild, as it imposes only an upper bound on $\Lambda(A)$, which is finite for any fixed matrix $A$. By choosing the upper bound sufficiently large, any specific $A$ matrix can be included in the set of matrices over which the uniformity holds. In particular, our procedure is pointwise asymptotically valid for any fixed $A$ matrix. This contrasts with the procedure of \cite{CSS}, where some $A$ matrices are excluded by the rank stability condition, and their result does not guarantee pointwise validity for such matrices (see Section~\ref{sectionComp} below). Note, however, that $\Lambda(\cdot)$ is not continuous. Consequently, the condition 
$\sup_{A \in \mathcal{A}} \Lambda(A) < \infty$ excludes any set $\mathcal{A}$ that 
\emph{contains a neighborhood} of a matrix with linearly dependent rows.\footnote{If a 
matrix $A \in \mathbb{R}^{m \times n}$ has a subset $F \subseteq [m]$ of rows, with 
$|F| \le n$, that are linearly dependent, then arbitrarily small perturbations 
$\tilde{A}$ of $A$ can produce $\tilde{A}_F$ with rows that are nearly collinear 
but linearly independent. This can cause $\Lambda(\tilde{A}_F)$—and hence 
$\Lambda(\tilde{A})$—to become arbitrarily large.} \footnote{However, in Section~\ref{secWeakenSSLSLUC4} of the appendix, we provide an example of such a neighborhood ${\cal A}$ on which $\Lambda(\cdot)$ is unbounded while the Hoffman constant remains uniformly bounded, and thus our uniform validity result still applies.} Part~(ii) of the assumption ensures the existence of a constant $M>0$ such that all feasible regions $\{\eta \mid b(Q) - A(Q)\eta \le 0\}$, with $Q\in {\cal Q}_0$, have non-empty intersection with the Euclidean ball $B(0,M)$.

In settings where $A$ is known, the results of Section~\ref{subsectionAVKAM}, as well as those in \cite{ARP, CS, FSST}, do not require an assumption such as Assumption~\ref{assAVEAM4}. However, as we show in Proposition~\ref{propImp} in the appendix, without Assumption~\ref{assAVEAM4} there may not exist non-trivial uniformly valid testing procedures for the hypothesis \ref{eqnARP}; that is, any procedure that uniformly controls size may have trivial power against all alternatives. In some sense, this justifies the necessity of an assumption like \ref{assAVEAM4} (see Remark~\ref{remAlt} below).\footnote{We are grateful to the authors of \cite{BKSSMT} for sharing a more general version of the impossibility result from their working paper. We obtained a related result independently, although their result predates ours.}

The following assumptions is needed to guarantee that, on the event where our test rejects, the distributions of our test statistic (for $Q\in {\cal Q}_0$) do not asymptotically concentrate. We can do without this assumption, if we are willing to modify the rejection rule to: reject iff $\sqrt{N}\hat{T}_N > \hat{c}_N(1 - \alpha)+\tau$, where $\tau>0$ is a fixed  small constant.

 Let the random elements $(\zeta_A,\zeta_b)$, with $\zeta_A\in \mathbb{R}^{p\times d}$ and $\zeta_b\in \mathbb{R}^p$, be such that $\mathcal{L}\Bigl(vec([\zeta_A \ \ \zeta_b]) \big| Q\Bigr)\sim \mathcal{N}(0,\Sigma(Q))$. For $Q\in {\cal Q}_0$, let $V(Q)\in \mathbb{R}^{p\times p}$ be defined by 
\[
\zeta_b-\zeta_A\eta_Q\sim \mathcal{N}(0,V(Q))
\]
where $\eta_Q$ is as in Assumption \ref{assAVEAM4}. Let ${\cal D}(Q)$ denote the feasible region of the LP
\begin{equation}
\label{eqnAVEAM0}
\max\bigl\{b(Q)^\top \lambda \mid \lambda\geq 0,\  A(Q)^\top \lambda=0,\ \mathbbm{1}^\top \Omega(Q)\lambda \leq 1\bigr\},
\end{equation}
and let $\Delta_0(Q)$ denote the subset of ${\cal D}(Q)$ consisting of its optimal solutions.
\begin{assumption}
\label{assAVEAM5}
There exists a constant $\rho > 0$ such that for all $Q \in \mathcal{Q}_0$:
\begin{enumerate}[i)]
    \item If $\Delta_0(Q) \neq \{0\}$, then there exists $\lambda \in \Delta_0(Q)$ such that $\lambda^\top V(Q) \lambda \geq \rho$.
    \item If $\Delta_0(Q) = \{0\}$ and $extr(\mathcal{D}(Q)) \setminus \{0\}\neq \emptyset$, then for all $\lambda \in extr(\mathcal{D}(Q)) \setminus \{0\}$, we have $\lambda^\top V(Q) \lambda \geq \rho$.
\end{enumerate}
\end{assumption}

We now state the main result of this section. Its proof is provided in Section~\ref{pthmAVEAM} of the appendix.

\begin{theorem}
\label{thmAVEAM}
Suppose that Assumptions~\ref{assAVEAM0} through \ref{assAVEAM5} hold, and let $\hat{T}_N$ be as defined in equation~\eqref{eqnEAM2}. Then, for all $\alpha \in (0, 1/2)$,
\begin{equation}
\label{eqnAVEAM1}
\overline{\lim}_{N \to \infty} \sup_{Q \in \mathcal{Q}_0} Q\left(\sqrt{N}\hat{T}_N > \hat{c}_N(1 - \alpha)\right) \leq \alpha,
\end{equation}
where $\hat{c}_N(1 - \alpha)$ is defined as in equation~\eqref{eqnEAM3}.

Moreover, for any $Q \in \mathcal{Q}_0$ such that the system $b(Q) - A(Q)\eta \leq 0$ has $\eta_Q$ as its unique solution, we have
\begin{equation}
\label{eqnAVEAM2}
\lim_{N \to \infty} Q\left( \left\{ \mathbb{W}_N \;\middle|\; d_{\mathrm{Pr}}\bigl( \mathcal{L}(\sqrt{N} \hat{T}_N), \hat{H}_N \bigr) > \epsilon \right\} \right) = 0,
\end{equation}
for any $\epsilon>0$. Here $\hat{H}_N(\cdot)$ denotes the conditional distribution (given $\mathbb{W}_N$) of $v^*$ as defined in equation~\eqref{eqnEAM4}.
\end{theorem}

\begin{remark}
\label{remAVEAM1}
Equation~\ref{eqnAVEAM2} implies that the inequality in Equation~\ref{eqnAVEAM1} can be replaced by an equality if, for some $Q \in \mathcal{Q}_0$, the inequality system $b(Q) - A(Q)\eta \leq 0$ has $\eta_Q$ as its unique solution, and the asymptotic distribution of $\sqrt{N} \hat{T}_N$ is nondegenerate. In Lemma~\ref{lem1AD}, we derive the asymptotic distribution of $\sqrt{N} \hat{T}_N$ under such DGPs and show that it is nondegenerate if and only if the optimal solution set $\Delta_0(Q)$ of the LP~\ref{eqnAVEAM0} strictly contains the set $\{0\}$. As a consequence, our test is not unduly conservative when the family $\mathcal{Q}_0$ includes such DGPs. For DGPs $Q \in \mathcal{Q}_0$ where the inequality system $b(Q) - A(Q)\eta \leq 0$ has multiple solutions and $\Delta_0(Q)$ strictly contains $\{0\}$, it can be shown that the pointwise (i.e., for fixed $Q$) asymptotic limit of $\hat{H}_N(\cdot)$ stochastically dominates the pointwise asymptotic distribution of $\sqrt{N} \hat{T}_N$ in the first-order stochastic dominance sense, with both distributions being nondegenerate. In such cases, our testing procedure is expected to be conservative, with the pointwise asymptotic size of the test potentially much smaller than the nominal level $\alpha$.
\end{remark}
\begin{remark}
\label{remtup}
The procedures in Sections~\ref{subsectionKAM} and \ref{subsectionEAM} require tuning parameters $\kappa_N$ satisfying $\kappa_N \to \infty$ and $\kappa_N = o(\sqrt{N})$. Simulation evidence indicates that these choices materially affect performance. Large values of the tuning parameters $\kappa_{1N}$ and $\kappa_{2N}$ in equations \ref{eqnEAM5} lead to more conservative tests, whereas smaller values may fail to achieve proper size control. The method is not highly sensitive to the choice of $\kappa_{3N}$. Intuitively, the tuning parameters should scale with the standard deviations of the estimated components of $(A, b)$, although a formal analysis is beyond the scope of this paper. In our applications and simulations, we set $\kappa_N = \sqrt{\log N}$, which seems to perform reasonably well in practice. 
\end{remark}

\begin{remark}\label{remAlt}
Proposition \ref{propImp} shows that Assumption \ref{assAVEAM4} is necessary for the existence of nontrivial uniformly valid tests. Since this assumption has two parts—neither of which is required in the known $A$ case (Theorem \ref{thmAVKAM})—it is natural to ask whether one part alone may suffice. 

Indeed, maintaining the other assumptions and replacing Assumption \ref{assAVEAM4} with the boundedness condition $\sup_{Q\in {\cal Q}_0}\|\eta_Q\|\le M$ (for known $M$), one can obtain a uniformly valid test by applying the procedure of \cite{DAGS} over a grid of $\|\eta\| \le M$ and rejecting \ref{eqnARP} if we reject the hypothesis $b - A\eta \le 0$ at all values of $\eta$. However, this approach may be computationally infeasible and overly conservative. Moreover, a close inspection of the proof of Proposition \ref{propImp} reveals that the boundedness condition $\sup_{Q\in {\cal Q}_0}\|\eta_Q\|\le M$ is necessary for pointwise consistency: without it, any uniformly valid test has trivial power against some fixed alternatives. Thus, part (ii) of Assumption \ref{assAVEAM4}, with an explicit bound, is sufficient for the existence of a nontrivial procedure. While our method can be adapted to exploit this bound without part (i) of Assumption~\ref{assAVEAM4}, the resulting test is typically conservative.
\end{remark}

\subsubsection{Comparison to the approach of \cite{CSS}}
\label{sectionComp}
\cite{CSS} propose an alternative testing procedure for the hypothesis in \ref{eqnARP} in settings where both \(A\) and \(b\) are estimated. A key advantage of their approach is that it is tuning-parameter free and computationally inexpensive. By contrast, our procedure requires weaker structural assumptions and achieves uniform validity over a broader class of data-generating processes (DGPs). In fact, their method may fail to be even pointwise valid for certain DGPs (i.e. for some estimated $A$ matrices) that are admissible under our framework.

Their approach is based on the quasi-likelihood ratio statistic
\begin{equation}
\label{eqnCSS}
\hat{T}=\min_{\eta,\, b:\, b\leq \hat{A}\eta}N(\hat{b}-b)^\top \hat{S}^{-1}(\hat{b}-b),
\end{equation}
where \(\hat{S}\) is an estimator of an asymptotic variance matrix (see \cite{CSS} for details). Critical values are for the test statistic are given by the quantiles of a chi-squared distribution, with degrees of freedom determined by the rank of the set of inequalities that bind at the solution of the quadratic programming problem \eqref{eqnCSS}. This feature eliminates the need for bootstrap procedures to compute critical values, as we do, and contributes to the computational simplicity of their method.

These advantages, however, come at the cost of stronger assumptions. Assumption 2 in \cite{CSS}—the stable rank condition—imposes further restrictions on the coefficient matrix \(A\). Let the polyhedron be defined as
\[
\mathcal{D}(Q) := \left\{ \eta \;\middle|\; b(Q) \leq A(Q)\eta \right\},
\]
and define
\[
\mathcal{F}(Q) := \left\{ F \subseteq [p] \;\middle|\; \exists\, \eta^* \text{ such that } b_F = A_F \eta^* \text{ and } b_{F^c} < A_{F^c} \eta^* \right\}.
\]
Here, for any index set \(I \subseteq [p]\), \(A_I\) denotes the submatrix of \(A\) obtained by retaining only the rows indexed by \(I\). The inequality \(b_{F^c} < A_{F^c}\eta^*\) is interpreted to hold componentwise. Each \(F \in \mathcal{F}(Q)\) is associated with the set
\[
\left\{ \eta \;\middle|\; b_F = A_F \eta \text{ and } b_{F^c} \leq A_{F^c} \eta \right\},
\]
which defines a face of the polyhedron \(\mathcal{D}(Q)\).The stable rank condition requires, in part, that
\[
\lim_{N \to \infty} Q\bigl( \mathrm{rank}(\hat{A}_F) = \mathrm{rank}(A_F) \;\; \forall F \in \mathcal{F}(Q) \bigr) = 1,
\]
so that, with probability approaching one as \(N \to \infty\), the set of inequalities that is active at each face of \(\mathcal{D}(Q)\) preserves its rank under estimation. 
This requirement may fail, for instance, in settings where the rows of \(A\) corresponding to true equality constraints—i.e., inequalities in \(\mathcal{D}(Q)\) that are always active—are themselves rank-deficient and must be estimated. To illustrate, consider a simple example of a (one-sided) "union bound" testing problem as studied in \cite{Bei25}. Let \(\{W_i\}_{i=1}^N \overset{i.i.d.}{\sim} \mathcal{N}(\mu, I_2)\), where \(\mu=(\mu_1,\mu_2)^\top\) is unknown. We wish to test \footnote{This hypothesis is of interest, for instance, if we wish to establish that both means (representing for instance the treatment effects for two subpopulations) are strictly negative.}
\[
H_0:\ \mu_1\geq 0 \ \text{or} \ \mu_2\geq 0 
\quad \text{versus} \quad 
H_1:\ \mu_1<0,\ \mu_2<0.
\]
This hypothesis can be written in the form of \eqref{eqnARP} with
\begin{equation}
b=\begin{pmatrix}0\\0\\0\\1\\-1\end{pmatrix},
\quad 
A=\begin{pmatrix}
\mu_1 & \mu_2\\
0 & 1\\
1 & 0\\
1 & 1\\
-1 & -1
\end{pmatrix}.
\label{eqncomp1}
\end{equation}
When \(\mu_1=\mu_2=0\), the first, fourth, and fifth inequalities are always active, corresponding to the index set \(F=\{1,4,5\}\). In this case, the population matrix \(A_F\) has rank one. However, its sample analogue generically satisfies \(\mathrm{rank}(\hat{A}_F)=2\) with probability one for any finite \(N\) (this is true as \(Q(\hat{\mu}_1\neq \hat{\mu}_2)=1\)). Thus, the rank stability condition fails for this DGP, and it is ruled out by the assumptions needed to establish the validity of the procedure of \cite{CSS}. In contrast, our assumptions do not rule out this DGP. However, it can be shown that the condition \(\Lambda(A) < \infty\) in part (i) of Assumption~\ref{assAVEAM4} excludes neighborhoods of this DGP. Specifically, for \(r > 0\), define
\[
\mathcal{A}(r) := \left\{ A(\mu) \;\middle|\; \|\mu\| \leq r \text{ and } (\mu_1 \geq 0 \text{ or } \mu_2 \geq 0) \right\},
\]
where \(A(\mu)\) are the matrices appearing in \ref{eqncomp1}.
Then,\footnote{In fact, a stronger result can be established, namely that the Hoffman constant of matrices in \(\mathcal{A}(r)\) is unbounded.}
\[
\sup_{A \in \mathcal{A}(r)} \Lambda(A) = \infty \quad \text{for any } r > 0.
\]
Intuitively, when \(\mu\) is nonzero but close to zero, the first and fifth rows are nearly collinear, and $\Lambda(A(\mu))$ is large. Thus, although our method is guaranteed to be pointwise valid at the DGP considered above, our uniformity results rule out neighborhoods of such DGPs.



\section{Applications} \label{secApplication}

In this section we report simulations and empirical applications based on the two examples given in Section \ref{secApplications} in which the $A$ matrix contains estimated components.

\subsection{Application 1: the ATE with a quadratic marginal treatment effect} \label{secApplicationATE}

Following Example \ref{exMST}, we apply our method to a setting in which the parameter of interest is the average treatment effect (ATE) of a binary treatment, which is partially identified using instrumental variables (IV). The problem belongs to a class studied by \cite{MST} (MST) for identification in IV models with heterogeneous treatment effects. We build a simulation around an application studied in the working paper version of MST: \cite{MSTWP}. 

This application considers a field experiment from \cite{Dupas2014} that offered randomized prices $Z$ for an antimalarial bed net in Kenya. The treatment $D \in \{0,1\}$ indicates whether a given household purchased the bed net at the price they were offered, and the outcome $Y \in \{0,1\}$ whether the household reports actually using the bed net two months after the experiment. Given potential outcomes $Y(0),Y(1)$, we consider the average treatment effect $\text{ATE}=E[Y(1)-Y(0)]$ as the parameter of interest, which is partially identified in this setting given standard IV assumptions.\footnote{The ATE is not point identified under the standard assumptions of \cite{IA94} due to the presence of never-takers (who do not purchase the bed net even at the lowest price offered) and always-takers (who purchase the bed net even at the highest price offered). Since $Y \in \{0,1\}$ is bounded, the ATE remains partially identified.} A simplification afforded by the setting of \cite{Dupas2014} is that $Y(0)=0$ with probability one, because this particular bed net was not available at the time outside of the field experiment. Hence the ATE reduces to $\text{ATE}=E[Y(1)]$, the utilization rate that would occur if everyone purchased the bed net.

A strength of the MST approach to identification with instrumental variables is that additional assumptions about how selection is related to potential outcomes can be used to reduce the identified set for partially identified parameters like the ATE. Following MST, assume that purchases of the bed net are governed by $D=D(Z)$, where
\begin{equation} \label{eqnSimselection}
    D(z)=\mathbbm{1}(U \le p(z))
\end{equation}
where \( U \sim \mathrm{Unif}[0,1] \), and \( p(z) = \mathbb{E}[D \mid Z = z] \) is the propensity score function. Instrument validity implies that \( Z \perp (U, Y(1)) \). \cite{Vytlacil} establishes equivalence between this model and the LATE model of \cite{IA94}, which imposes monotonicity instead of Eq.~\eqref{eqnSimselection}.

Given the variable $U$ that governs selection into treatment (purchase of the net), MST propose leveraging assumptions on marginal treatment response (MTR) curves $m_0$ and $m_1$, where $m_d(u):=E[Y_i(d)|U=u]$ for $d \in \{0,1\}$ and $u \in [0,1]$. In our setting, $m_0(u)=E[Y(0)|U=u]=0$ for all $u$ so we focus on the treated MTR curve $m_1(u)$. A natural assumption is that $m_1(u)$ is weakly decreasing in $u$, indicating that individuals who would buy the bed net at a higher price are also more likely to use the net if they purchase it.

It is known that, even with a binary instrument $Z \in \{0,1\}$, the ATE is point identified if one imposes the assumption that the MTR curves $m_d(u)$ are linear in $u$ (\cite{BMW}). To allow for partial identification, we instead impose that $m_1(u)$ is quadratic: $m_1(u) = a+\beta \cdot u + \gamma \cdot u^2$ for some parameters $(a,\beta,\gamma)$. Given this assumption, the average treatment effect can be written as
\begin{equation} \label{eqnSimATE}
ATE=E[Y_i(1)] = \int_0^1 E[Y(1)|U=u] \cdot du = \int \{a+\beta \cdot u + \gamma \cdot u^2\}\cdot du = a+ \frac{\beta}{2}+\frac{\gamma}{3}
\end{equation}

For simplicity, we focus on a case with a binary instrument $Z \in\{0,1\}$ that reflects a lower versus higher price for the bed net, e.g. $Z=1$ corresponds to a price of $50$ Kenyan Shillings (Ksh) and $Z=0$ to $150$ Ksh. These prices are a subset of those randomized in the experiment of \cite{Dupas2014}, and correspond to the prices at which $p(0) \approx 1/3$ and $p(1) \approx 2/3$ of households buy the bed net. For either value of $z \in \{0,1\}$, we can write $E[Y \cdot D|Z=z]$ as:
\begin{equation} \label{eqnSimIVmoments}
E[Y_i(1) \mathbbm{1}(U \le p(z))] = \int_0^{p(z)} E[Y(1)|U=u] \cdot du  = p(z)\cdot a + \frac{p(z)^2}{2} \cdot \beta+\frac{p(z)^3}{3} \cdot \gamma
\end{equation}

\noindent Meanwhile, the researcher uses the estimands $E[Y \cdot D|Z=0]$ and $E[Y \cdot D|Z=1]$ to provide identifying information from the outcome data. They assume that $E[Y(1)|U=u] = a+\beta \cdot u + \gamma \cdot u^2$, that $E[Y(1)|U=u]$ is weakly decreasing in $u$, that $E[Y(1)|U=u] \in [0,1]$ for all $u \in [0,1]$ (as implied by $Y$ being binary), and that $E[Y(0)|U=u]=0$. Monotonicity implies that $\frac{d}{du}E[Y(1)|U=u]= \beta+2\gamma u$ is nonpositive on the unit interval, i.e. $\beta \le 0$ and $\beta+2\gamma \le 0$. Boundedness implies that $0 \le a \le 1$ and $0 \le a + \beta + \gamma \le 1$.\footnote{Note that $m_1(0) \ge 0$ is implied by $m_1(1) \ge 0$ and $m_1$ being non-increasing, and similarly $m_1(1) \le 1$ is implied by $m_1(0) \le 1$ and $m_1$ being non-increasing. Thus in the system of inequalities in Exmaple \ref{exMST} we have dropped two redundant inequality restrictions in defining $A$ and $b$.}

If $ATE=\theta$, then all together we obtain the system of linear equalities \eqref{eqMST} written in Example \ref{exMST}. We test the null that $A(Q) \lambda \ge b(Q,\theta)$ for a grid of $\theta$ values ranging from 0 to 1, using the procedure described in Section \ref{subsectionEAM}. The data is drawn from a DGP in which $P(Z=1) = 1/2$, $D$ is generated via Eq. \eqref{eqnSimselection} with $p(0)=1/3$, $p(z)=2/3$, and $U|Z \sim Unif[0,1]$, and $Y(1)=\mathbbm{1}(W \le a + \beta U + \gamma U^2),$ where $W|Z,U \sim Unif[0,1]$. In Appendix \ref{app:Suff}, we show that Assumption \ref{assAVEAM4} is satisfied in this setting, provided that for some $\delta>0$, we have a non-zero first stage with $p(1)-p(0) \ge \delta$, and meanwhile $p(0) \ge \delta$, $p(1) \le 1 - \delta$, uniformly over $Q \in \mathcal{Q}_0$.

\subsubsection{Simulations} \label{secSim1}
\begin{figure}[h!]
\begin{center}
    \includegraphics[width=6in]{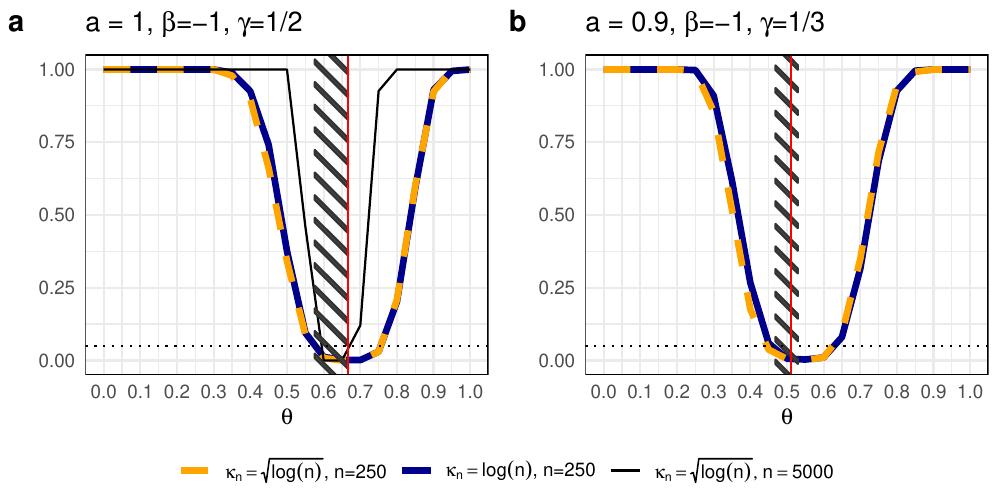}
\end{center}
\caption{Rejection probabilities for testing hypothetical values $\theta$ of the ATE.} \label{fig:MST2powercurve}
\end{figure}

Figure \ref{fig:MST2powercurve} plots rejection probabilities for 1000 Monte Carlo draws for a test with nominal size of 5\% and 250 bootstrap iterations, across two versions of this DGP. We begin by setting $n=250$.\footnote{This corresponds approximately to the sample size in \cite{Dupas2014} for our two instrument values: $Z=1$ (a price of $50$ Kenyan Shillings), and $Z=0$ (a price of $150$ Kenyan shillings), across all rural areas in the experiment.} In both versions, $p(0)=1/3$ and $p(1)=2/3$. In the left panel a), we set $a=1,\beta=-1,\gamma = 1/2$. The true value of the ATE in this $0.67$ (depicted by a vertical solid red line), and the identified set is $[0.58,0.67]$ (depicted by the hatched region). The horizontal dotted line depicts 5\%. In this version of the DGP, the slope of $m_1(u)$ is zero at $u=1$, and the true value is on the boundary of the identified set. The right panel b) of Figure \ref{fig:MST2powercurve} considers a second DGP in which $a = 0.9$, $\beta = -1$, and $\gamma = 1/3$. In this setting the true value of the ATE is $0.51$, which lies in the interior of the identified set $[0.47,0.53]$. In both panels, two choices of the tuning parameter $\kappa_n$ are compared, $\kappa_n = \sqrt{\log{n}}$ and $\kappa_n = \log{n}$. The results depicted in Figure \ref{fig:MST2powercurve} appear not to be sensitive to this choice, given the DGP and sample size.

In both DGPs depicted in Figure \ref{fig:MST2powercurve}, the test controls size well throughout the identified set for the ATE, and gains power relatively quickly outside of it. We note that the power gain is asymmetric. For example, in panel a), the rejection probability remains below $5\%$ until $\theta \approx 0.75$, well outside of the identified set. This is a small sample phenomenon only however: in panel a) we also show the power curve with a sample size of $n=5000$, which traces out the identified set exactly (up to our grid spacing for $\theta$:  multiples of $0.05$).

To demonstrate Remark \ref{remAVKAM4} regarding the second part of Theorem~\ref{thmAVKAM}, we consider an additional simulation in which point identification holds. We obtain this by setting $a=1, \beta=-1, \gamma=0$, so that the marginal treatment effect is linear in $u$. Given knowledge that $\gamma=0$, $\theta_0$ is point identified \citep{BMW}, and the condition in Remark \ref{remAVKAM4} holds, so the exact size of our test should be $95\%$. We evaluate this by considering a large sample size of $n=10,000$. As reported in Table \ref{tabMSTpointid}, our procedure covers the true value $\theta_0 = 0.5$ with close to probability $95\%$. We find that in this DGP the choice $\kappa_n = \sqrt{\log n}$ comes closer to the nominal size than $\kappa_n = \log n$ does. 

\begin{table}[h]
\centering{
\begin{tabular}{ccc}
& $\kappa_n = \sqrt{\log n}$ & $\kappa_n = \log n$\\
\midrule
Our procedure & 95.1\%& 96.3\%\\
\bottomrule
\end{tabular}}
\caption{Coverage of 95\% confidence intervals for the true value of the ATE, in a setting with point identification. $n=10,000$. \label{tabMSTpointid}}
\end{table}

\subsubsection{Empirical study} \label{secEmpirical}

Building off of the simulation results of the last section, we now apply our method to the data of \citet{Dupas2014}, calculating confidence intervals for the ATE under the assumption of quadratic MTR curves.

The experiment reported in \citet{Dupas2014} randomly assigned one of 17 bed net prices across households, with the price drawn from a smaller set of prices that varied based on which of six areas the household was in. To match the simulation of the last section, we first consider two prices: 50 Ksh ($Z=0$) and 150 Ksh ($Z=1$). We also consider a single of the eight rural areas from the experiment, since assignment probabilities for the price depends on area. These restrictions leave 116 observations, each one corresponding to a household with children. Of these, roughly one half purchase a bed net, and among the households that purchased the observed utilization rate was $0.72$. However this observed utilization rate $E[Y|D=1]$ does not provide a causal estimate of the proportion of all households that \textit{would} use an improved bed net were they to buy one, because the decision to buy at a given price may be correlated with usage behavior. That is, the MTR curve $E[Y(0)|U=u]$ may depend on $u$. Indeed, a point estimate for the identified set of the ATE given the quadratic MTR model is $[0.46, 0.52]$, which does not include the observed utilization rate $0.72$.

To construct a confidence interval for the ATE, we test whether $A(Q) \lambda \ge b(Q,\theta)$ for a grid of 101 values of $\theta$ from 0 to 1. Test inversion then yields a 90\% confidence interval for the ATE of $[0.35, 0.72]$, which just barely contains the naive estimate of 72\%. The results suggest that between roughly 1/3 and 2/3 of all households would use the improved bed net, were they to buy it.

These magnitudes are not directly comparable to results reported by \citet{Dupas2014}, who focus on the intent-to-treat effects of lowering price (rather than considering purchase as the treatment).\footnote{Nevertheless, our 95\% confidence interval notably does not contain the point estimate of $0.281$ reported as their main treatment effect estimate (Column 2 of Table II). Their estimate uses all price levels and geographical areas from the experiment, and estimates a linear probability model with the price instrument binarized into an indicator for a ``high subsidy'' (price of 50 Ksh and below).} Our CI for the ATE can be compared to those reported in \citet{MSTWP}, who consider our same parameter, but using the whole dataset and across choices of different identifying moments from the data. Our CI is similar to the one reported in Column (3) of MST's Table 3, which instead of assuming a quadratic MTR uses a Bernstein polynomial basis of order 10, and exploits the information from the traditional IV estimand for identification.

\subsection{Application 2: monotone instrumental variables} \label{secMonotoneIV}

In this section we provide simulation results and an empirical application for Example \ref{exMP}, which applies the monotone IV (MIV) assumption to bound the average treatment effect. Recall from Example \ref{exMP} that a \textit{monotone instrumental variable} $Z$ relaxes the stronger condition $E[Y(t)|Z=z'] = E[Y(t)|Z=z]$ for a valid instrument, replacing it with $E[Y(t)|Z=z'] \ge E[Y(t)|Z=z]$ when $z' \ge z$ \citep{MP}. Here $Y(t)$ denote potential outcomes for an ordered treatment $t$, and $Z$ an observed ordered scalar-valued variable. If for instance $Y$ is earnings, $Z$ a test score, and $T$ is a measure of education completed, then the MIV assumption is plausible because those with high aptitude as measured by the test may have abilities that would lead them to obtain higher wages even for a given fixed level of schooling. 

Suppose that the treatment $T$ and monotone IV $Z$ have finite supports $\mathcal{T}$ and $\mathcal{Z}$ with $|\mathcal{T}| = L$ and $|\mathcal{Z}|=J$. Introduce a $JL^2$ component vector $\eta$ with components $\eta_{tt'z}:=E[Y(t)|T=t',Z=z]$. For simplicity let $\mathcal{Z}$ be the integers from $1$ to $J$. Then we can write the MIV assumption as
$$\sum_{t'} P(T=t'|Z=j+1) \cdot \eta_{tt',j+1} \ge  \sum_{t'} P(T=t'|Z=j) \cdot \eta_{tt'j}$$ 
for each $j=1 \dots J-1$, using the law of iterated expectations. The identifying content of the data is that each $\eta_{ttz} = E[Y(t)|T=t,Z=z]$ is identified as $\eta_{ttz} = E[Y|T=t,Z=z]$.

A typical parameter of interest in the context of MIV is the average treatment effects between treatments $t_1$ and $t_2$: $E[Y(t_2)-Y(t_1)]$. By the law of iterated expectations $E[Y(t_2)-Y(t_1)] = \sum_{z, t'} P(T=t',Z=z) \cdot (\eta_{t_2,t'z}-\eta_{t_1,t'z})$. The hypothesis that $E[Y(t_2)-Y(t_1)]=\theta$ can be written as $c^T \eta = \theta$, where $c_{tt'z} = P(T=t',Z=z) \cdot \{1[t=t_2]-1[t=t_1]\}$. Finally, to achieve an informative identified set for $\theta$, one must also restrict the support of $Y$ to be bounded \citep{MP}. For example, if $Y$ has compact support $[Y_K,Y_U]$, then this implies the additional $2JL^2$ constraints $\eta_{tt'z} \ge Y_L$ and $-\eta_{tt'z} \ge -Y_U$ for each $t,z,t'$.

In the special case of $\mathcal{T}=\mathcal{Z}=\{0,1\}$ with the parameter of interest $\theta_0 = E[Y(1)-Y(0)]$, we obtain the system written in Example \ref{exMP}, along with the 16 additional inequalities $\eta_{tt'z} \ge Y_L$ and $-\eta_{tt'z} \ge -Y_U$ for each $t,z,t'$.

\subsubsection{Simulations}
We now investigate the performance of our inference procedure applied to the MIV setting described above. We calibrate our DGP to approximately match the empirical application in the next subsection, which is in turn motivated by the returns to schooling application of \citet{MP}. The empirical application of \citet{MP} assumes the special case of MIV in which $Z=T$, which they refer to as marginal treatment selection (MTS), coupled with an assumption known as marginal treatment response (MTR). A linear programming representation of the MTS and MTR assumptions is described in Appendix \ref{secMTSMTR}. We do not replicate the MTS-MTR approach here because, as \citet{MP} show, this assumption leads to simple analytic bounds on the identified set, without a need to solve the underlying LP. The bounds admit of straightforward asymptomatically normal estimators, so existing inference methods are well-suited to this case.

Instead, we consider the more general MIV assumption, for which closed form bounds on the ATE involve maximum and minimum operators over the support of $Z$ \citep{MP}.\footnote{For parameters of the form $\mathbbm{E}[Y(t)|Z=z]$, the sharp bounds arising from MIV take the form of intersection bounds, for which \citet{CLR} and \citet{FP2} consider inference. The ATE is a linear combination of such parameters.} Let $T \in \{0,1\}$ indicates completing college, $Y \in \{0,1\}$ indicates earning a ``high'' wage, and $Z \in \{0,1\}$ indicates receiving a ``high'' score on an aptitude test.

We consider two versions of the DGP. First, we calibrate the DGP to approximately match the data, letting
$$P(Y(0)=1|T=t,Z=z) = 0.2 + 0.1\cdot t+0.05\cdot z$$
and setting $E[Y(1)-Y(0)|T=t,Z=z] = 0.2$ for all $z$ and $t$, so that the average treatment effect is $\theta_0=0.2$. Meanwhile, $P(0,0)=0.3$, $P(0,1)=0.3$, $P(1,0)=0.1$, and $P(1,1)=0.3$. With this DGP (``DGP 1''), the identified set is $\theta_0 \in [-0.32,0.625]$. Meanwhile, the assumption-free bounds of \citet{M90} that impose only that $Y \in [0,1]$ are $\theta_0 \in [-0.32,0.68]$, so the MIV assumption does add identifying information to the upper bound.

We consider a second DGP designed to be more challenging for our procedure. In this DGP (``DGP 2'') we let $E[Y(1)-Y(0)|T=t,Z=z] = 0$ for all $z$ and $t$ (so that the average treatment effect is $\theta_0=0$), $E[Y(0)|T=t,Z=z] = t \cdot z$, and $P(0,0)=P(0,1)=P(1,0)=0.33$, while $P(1,1)=0.01$. In this setting, the sample estimators of $\mathbbm{E}[Y|T = 1, Z = 1]$ and $P(1,1)$ will have large asymptotic variances relative to the others, making $\bar{\sigma}>> \underline{\sigma}$ in the language of Assumptions \ref{assAVKAM2} and \ref{assAVEAM2}. 

\begin{figure}[h]
\centering{
\includegraphics[width=2.75in]{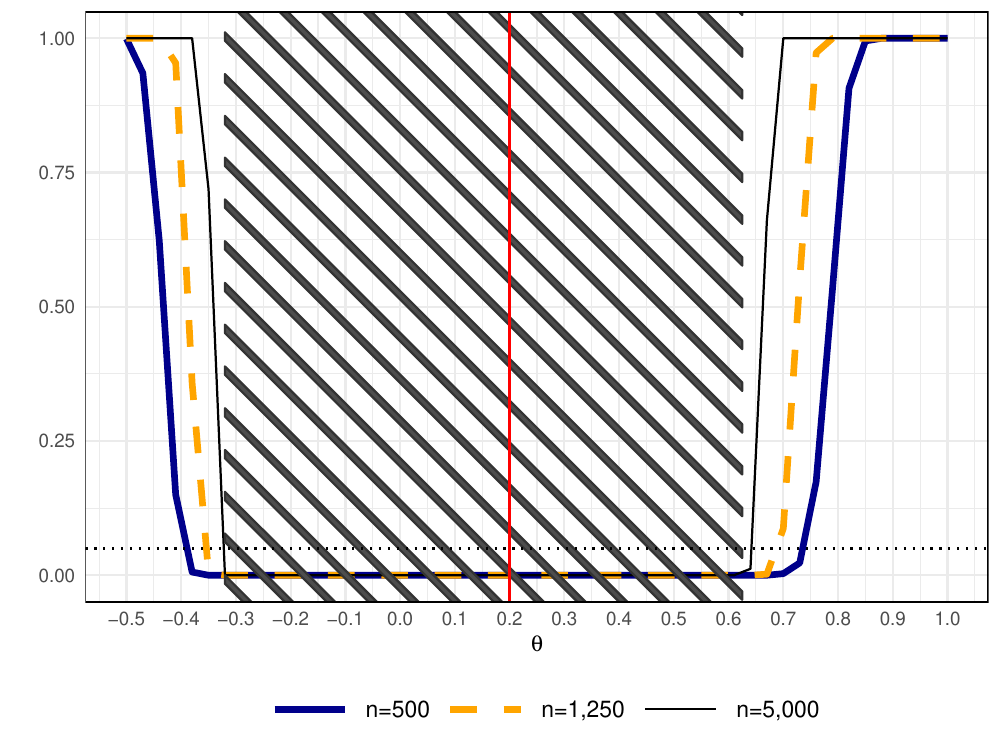}
\includegraphics[width=2.75in]{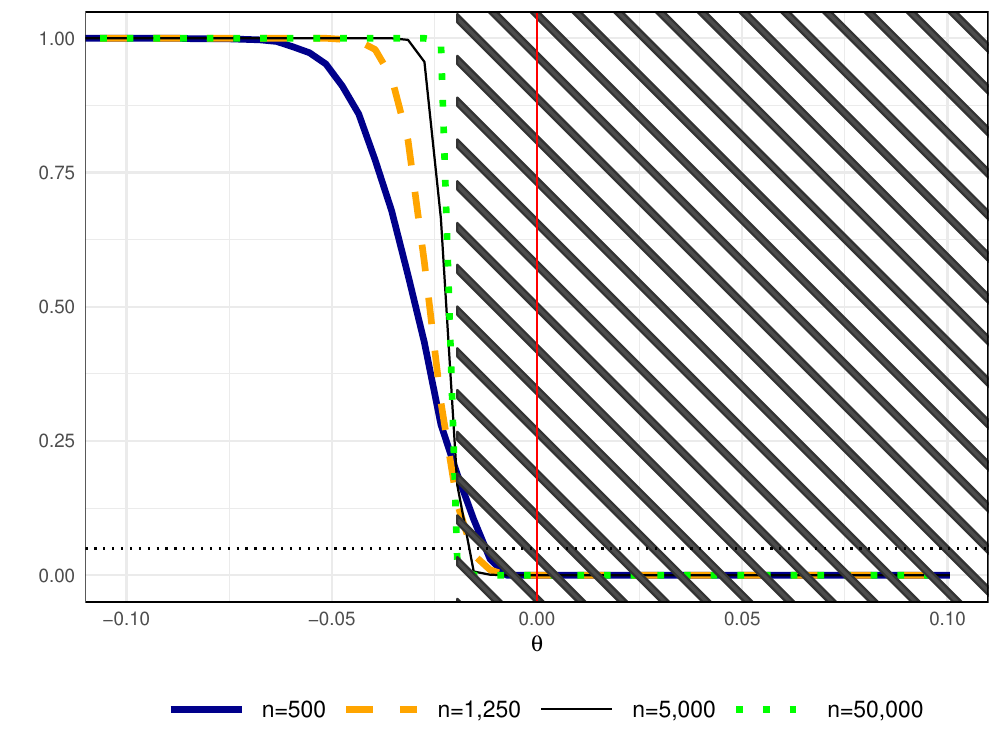}}
\caption{Rejection probabilities for the monotone IV example versus $\theta$, with DGP 1 (left) and DGP 2 (right). $\kappa_n=\sqrt{\log n}$. Shaded region depicts the identified set for $\theta_0$, while the vertical red line indicates its true value $\theta_0 = 0.2$ (in the case of DGP 1) and  $\theta_0 = 0$ (in the case of DGP 2).\label{figMIVSimulated}}
\end{figure}

Figure \ref{figMIVSimulated} reports rejection probabilities over $1000$ samples of our method for both DGPs, across values of $\theta$ with $k=\sqrt{\log n}$.\footnote{In the case of DGP 2, it occurs in a small number of samples (less than 1\%) that there are zero observations with $Z=T=1$, meaning that some of the quantities in the linear system (cf. Example \ref{exMP}) cannot be estimated. In these cases, we draw a new sample until we can proceed.} Table \ref{tabMIVSimulated} reports coverage probabilities and lengths of 95\% confidence intervals for the same $1000$ samples. The results demonstrate that our test exhibits good size control within the identified set, and that the power to detect violations of the null just outside the identified set increases with the sample size. For DGP 2 (right panel of Figure \ref{figMIVSimulated}), we focus on the lower boundary $\theta^L$ of the identified set, which occurs at $\theta^L \approx 0.019$. This is because the MIV inequality involving the poorly estimated quantities $\mathbbm{E}[Y|T = 1, Z = 1]$ and $P(1,1)$ binds at $\theta^L$. No such inequality binds at the upper bound $\theta^U$ of the identified set, and there performance for DGP 2 looks much the same as for DGP 1. Table \ref{tabMIVSimulated} shows that the rejection rate at $\theta^L$ is indeed still above the nominal level of $5\%$ with the moderate sample size $n=5,000$ (``Coverage at $\theta^L$'' is in the 80-90\% range, well below 95\%). Figure \ref{figMIVSimulated} reports a much higher sample size of $n=50,000$ (dashed green line) for reference. We found a rejection rate below $3\%$ at $\theta^L$ with this sample size.

\begin{table}[h]
\centering{
\resizebox{\columnwidth}{!}{\begin{tabular}{cccccccccccc}
& n=500&n=1,250&n=5,000& & n=500&n=1,250&n=5,000& & n=500&n=1,250&n=5,000\\
\midrule
&\multicolumn{3}{c}{CI length (normalized)} & & \multicolumn{3}{c}{Coverage at $\theta^U$}& & \multicolumn{3}{c}{Coverage at $\theta^L$}\\
\cmidrule{2-4} \cmidrule{6-8} \cmidrule{10-12} DGP 1 & 1.26 & 1.15 &	1.04 & & 100\% & 100\% & 100\% & & 100\% & 100\% & 99\%\\

\cmidrule{2-4} \cmidrule{6-8} \cmidrule{10-12} DGP 2 & 1.03 & 1.01 &	1.00 & & 95\% & 94\% & 95\% & & 81\% & 87\% & 83\%\\
\bottomrule
\end{tabular}}}
\caption{Length of 95\% confidence intervals for the ATE, expressed as a ratio of the length of the identified set, and coverage probabilities for the upper bound $\theta^U$ of the identified set. This represents a ``worst-case'' coverage probability when $\theta_0$ is on the boundary of the identified set-- coverage probabilities for the true $\theta_0=0.2$ are 100\% in all cases considered. \label{tabMIVSimulated}}
\end{table}

\subsubsection{Empirical study}
We next implement our procedure using the sample of white males working full time and reporting positive earnings used by \citet{MP}, drawn from the 1979 National Longitudinal Survey of Youth. We define ``college'' $T=1$ as at least 16 years of schooling, a ``high wage'' $Y=1$ in 1994 dollars as $\$15$ or greater, and a ``high score'' $Z=1$ as a score of 50 or greater on the Armed Forces Qualification Test (AFQT). With these definitions, $32\%$ of the sample complete college, $42.5\%$ earn a high wage, and $61\%$ recieve a high AFQT score. The sample size is $n=1,257$. We use test inversion on a grid of values $\theta$ with a spacing of $0.01$, with 250 bootstrap replications and $k=\sqrt{\log n}$.

\begin{table}[h]
\centering{
\begin{tabular}{c}
Our LP-based procedure \\
\midrule
$[-0.39, 0.73]$ \\
\bottomrule
\end{tabular}}
\caption{95\% confidence intervals for $E[Y(1)-Y(0)]$ using the data from \cite{MP} \label{tabMPempirical}}
\end{table}

\section{Conclusion}

We have developed a bootstrap-based procedure for testing whether a system of linear equality and inequality constraints admits a solution when the coefficients must be estimated. Our approach is uniformly valid over large classes of DGPs and accommodates semi–high-dimensional settings. Simulation evidence demonstrates strong finite-sample performance, and our empirical applications highlight the practical relevance of the method. One limitation of our procedure is its reliance on tuning parameters; developing theoretically justified, data-driven choices for these parameters is an important direction for future research. Another promising avenue is to explore alternative procedures that remain computationally feasible and are valid under a weakening of Assumption \ref{assAVEAM4}.

\begin{appendix}
\section{Proofs of Theorems \ref{thmAVKAM}, \ref{thmAVEAM}, and impossibility result}
In the proofs of Theorems \ref{thmAVKAM} and \ref{thmAVEAM}, we occasionally switch between norms from one line to the next. Such changes are justified under Assumptions \ref{assAVKAM0} and \ref{assAVEAM0}.
\begin{proof}{\textbf{(Proof of Theorem \ref{thmAVKAM})}}
\label{pthmAVKAM}
Let $L(\lambda, x, y)$ denote the Lagrangian associated with the LP in \eqref{eqnAVKAM0}, defined as
\[
L(\lambda, x, y) = b^\top \lambda + x(1 - \mathbbm{1}^\top D \lambda) - y^\top A^\top \lambda,
\]
where $x$ is a nonnegative scalar and $y \in \mathbb{R}^p$ are the Lagrange multipliers. Let $\Delta_0$ and $S_0$ denote, respectively, the sets of optimal solutions to the primal and dual problems associated with the LP~\eqref{eqnAVKAM0}.

To analyze the value function under perturbations, let $\mu$ denote the triple of inputs $\mu=(b,A,D)$, and for each such $\mu$, let $\phi(\mu)$ denote the value of the LP in \eqref{eqnAVKAM0} given these inputs. For perturbations $\xi = (\xi_b, \xi_A, \xi_D)$ and sample size $N$, define the perturbed inputs as $\mu_N(\xi) = (b_N(\xi), A_N(\xi), D_N(\xi)) = \left(b + \frac{\xi_b}{\sqrt{N}}, A + \frac{\xi_A}{\sqrt{N}}, D + \frac{\xi_D}{\sqrt{N}}\right)$.

Since the matrix $A$ is known, we will only consider perturbations with $\xi_A = 0$, and by abuse of notation, write $\xi = (\xi_b, \xi_D)$ and $\mu_N(\xi) = (b_N(\xi), D_N(\xi))$. Let $L_N(\lambda, x, y; \xi)$ denote the Lagrangian associated with the LP in \eqref{eqnAVKAM0} under the perturbed inputs $\mu_N(\xi)$, defined by
\[
L_N(\lambda, x, y; \xi) = b_N(\xi)^\top \lambda + x(1 - \mathbbm{1}^\top D_N(\xi) \lambda) - y^\top A^\top \lambda.
\]
Let $\Delta_{0,N}(\xi)$ and $S_{0,N}(\xi)$ denote, respectively, the sets of optimal solutions to the perturbed primal and dual problems. Note that the functions $L$ and $L_N$, as well as the sets $\Delta_0$, $\Delta_{0,N}$, and others, all depend on $\mu$ (or on $Q$); however, this dependence is omitted for notational simplicity.

Let $(\lambda_0, x_0, y_0) \in \Delta_0 \times S_0$ and $(\lambda_{\xi}, x_{\xi}, y_{\xi}) \in \Delta_{0,N}(\xi) \times S_{0,N}(\xi)$. By the saddle point property of the Lagrangian, for sufficiently small perturbations such that the LP remains feasible, we have
\begin{equation*}
\begin{aligned}
\sqrt{N} \bigl(\phi(\mu_N(\xi)) - \phi(\mu)\bigr)
&= \sqrt{N} \bigl(L_N(\lambda_{\xi}, x_{\xi}, y_{\xi}; \xi) - L(\lambda_0, x_0, y_0)\bigr) \\
&\leq \sqrt{N} \bigl(L_N(\lambda_{\xi}, x_0, y_0; \xi) - L(\lambda_{\xi}, x_0, y_0)\bigr) \\
&= \sqrt{N}(b_N(\xi) - b)^\top \lambda_{\xi} - x_0 \mathbbm{1}^\top \sqrt{N}(D_N(\xi) - D)\lambda_{\xi}.
\end{aligned}
\end{equation*}

Now, for $Q \in \mathcal{Q}_0$ we have $\phi(\mu) = 0$, and $(x_0, y_0) \in S_0$ implies that $x_0 = 0$ and $b - A y_0 \leq 0$. Therefore, the inequality simplifies to
\begin{equation}
\sqrt{N} \, \phi(\mu_N(\xi)) \leq \xi_b^\top \lambda_{\xi}.
\label{eqnPSSARP0}
\end{equation}

 Now, note that Assumptions \ref{assAVKAM1} and \ref{assAVKAM2}-i) imply that the family $\{\mathcal{N}(0,\Sigma(Q))\mid Q\in \mathcal{Q}\}$ is tight, and Assumption \ref{assAVKAM1} then implies that the family $\{\sqrt{N}(\hat{b}-b(Q))\mid Q\in {\cal Q}\}$ is asymptotically tight. Thus, Assumptions \ref{assAVKAM1}, \ref{assAVKAM2} and \ref{assAVKAM3} imply that there exists $\alpha_N\downarrow 0$, such that for any $M_N\uparrow \infty$, we have 
\[
\sup_{Q\in {\cal Q}_0}d_{\text{Pr}}\Bigl(\sqrt{N}(\hat{b}-b), \mathcal{N}(0,\Sigma(Q))\Bigr)\leq \alpha_N
\]
and for all $Q \in {\cal Q}_0$, the event $E_N(Q)$ defined by
\[
E_N(Q):=\Bigl\{\mathbb{W}\ \big| \  \|\hat{D}-D\|\leq \alpha_N,\  d_{\text{Pr}}\bigl(\mathcal{L}(\zeta^*\mid \mathbb{W}),\mathcal{N}(0,\Sigma(Q))\bigr)\leq \alpha_N, \sqrt{N}\|\hat{b}-b(Q)\|\leq M_N \Bigr\}
\]
 satisfies
\[
\underline{\lim}_{N\to \infty} \inf_{Q\in {\cal Q}_0} Q(E_N(Q))=1.
\]
Choose $M_N\uparrow \infty$ such that $M_N=o(\kappa_N)$ and $M_N \alpha_N=o(1)$. When $N$ is sufficiently large that $\alpha_N<\underline{\sigma}$, the indices of the diagonal entries of the matrices $\hat{D}$ and $D$ that are nonzero coincide, and correspond to the indices of the estimated components of $b$. In what follows, we let $\Pi^u$ denote the projection on the unknown components of b that must be estimated, and let $U\subseteq [p]$ denote the set of indices of the estimated components of \(b\).\footnote{\label{footPSSARP1}For example, if $b = (b_1, b_2, b_3, b_4)$, and $b_1$ and $b_3$ are known while $b_2$ and $b_4$ must be estimated, then $\Pi^u(b)$ denotes the vector $(0,b_2,0, b_4)$, and \(U=\{2,4\}\) Similarly, $\Pi^u(\lambda) = (0,\lambda_2, 0, \lambda_4)$.} Consider the set of pertubations $\mathcal{X}_N$ defined by
\[
\mathcal{X}_N := \left\{ \xi = (\xi_b, \xi_D) \;\middle|\; \|\xi_b\| \leq M_N, \; \|\xi_D\| \leq \sqrt{N} \alpha_N \right\},
\]
where, $\xi_D$ denotes diagonal matrices of the same dimensions as $D$, with diagonal entries corresponding to the deterministic components of $b$ equal to zero, and $\xi_b$ is a vector of the same dimension as $b$ with entries corresponding to the deterministic component of $b$ equal to zero. It is natural to consider such restrictive perturbations $\xi_D$, as they are similar to perturbations that arise from estimation ($\hat{D}-D$ has diagonal entries corresponding to known components of $b$ equal to zero).  Throughout the argument below, we use $C$ to represent a generic constant that does not depend on $Q \in \mathcal{Q}$ or on $N$, for all sufficiently large $N$. The value of $C$ may vary from line to line.

In the first three steps below, we establish inequality~\eqref{eqnAVKAM1}. The fourth step provides the proof of inequality~\eqref{eqnAVKAM2}.

\medskip
\underline{Step 1}
In this step, we show there exists $\beta_N=o(1)$ such that for all sufficiently large N's and for all $Q\in {\cal Q}_0$, we have on the event $E_N(Q)$:
\begin{equation}
\label{eqnPSSARP1}
\sup_{\xi \in {\cal X}_N}\sup_{\lambda \in \Delta_{0,N}(\xi)} \Bigl\{ \xi_b^\top \lambda-\sup_{\lambda \in \hat{\Delta}} \xi_b^\top \lambda \Bigr\}\leq \beta_N,
\end{equation}
where $\hat{\Delta}$ is as in equation \ref{eqnKAM4}. Let $\mathcal{D}$ and $\mathcal{D}_N(\xi)$ denote respectively the feasible regions of the unperturbed and perturbed LP. Note that since some diagonal entries of the matrices $D$ and $D_N(\xi)$ can be zero, it is possible for the sets ${\cal D}$ and ${\cal D}_N(\xi)$ to be unbounded. we first show that for all sufficiently large N, for all $Q\in {\cal Q}_0$, and $\xi \in {\cal X}_N$, we have that for each $\lambda \in {\cal D}_N(\xi)$ there exists $\Gamma(\lambda)\in \hat{{\cal D}}$ such that $\Gamma(\lambda)$ is a positive scalar multiple of $\lambda$ and 
\begin{equation}
\label{eqnPSSARP2}
\|\Pi^u(\lambda)-\Pi^u(\Gamma(\lambda))\|\leq C \alpha_N
\end{equation}
where $\Pi^u$ is as in Footnote \ref{footPSSARP1}. Note that for all large N such that $\alpha_N<\underline{\sigma}/2$ (with $\underline{\sigma}$ given in Assumption \ref{assAVKAM2}) and for $\xi\in {\cal X}_N$, we have 
\[
\|D_N(\xi)\lambda\|_1=\|\Pi^u(D_N(\xi)\lambda)\|_1\geq \underline{\sigma}/2 \|\Pi^u(\lambda)\|_1 \quad \quad \forall \lambda \in \mathbb{R}^p,
\]
which yields\footnote{The same is true on the event $E_N(Q)$, when ${\cal D}_N(\xi)$ is replaced by $\hat{\cal D}$.}
\begin{equation}
\label{eqnPSSARP4}
\sup_{ \lambda \in {\cal D}_N(\xi)} \|\Pi^u(\lambda)\|_1 \leq 2/\underline{\sigma}.
\end{equation}
Also, for $\xi \in {\cal X}_N$ and for all \(i\in U\), the ratios of the diagonal entries are bounded by
\[
1-\alpha_N/\underline{\sigma}\leq \frac{[D_N(\xi)]_{ii}}{D_{ii}}\leq 1+\alpha_N/\underline{\sigma}.
\]
As a consequence, we have $\forall \lambda \in \mathbb{R}^p$
\begin{equation}
\label{eqnPSSARP40}
(1-\alpha_N/\underline{\sigma})\|\Pi^u(D \lambda)\|_1\leq \|\Pi^u(D_N(\xi)\lambda)\|_1\leq (1+\alpha_N/\underline{\sigma})\|\Pi^u(D \lambda)\|_1.
\end{equation}
As a consequence, since the sets ${\cal D}$ and ${\cal D}_N(\xi)$ are star-shaped w.r.t. the origin,\footnote{We say that a set $\mathcal{C}$ is star-shaped (with respect to the origin) if $x \in \mathcal{C}$ implies that $tx\in \mathcal{C}$ for all $t\in[0,1]$.} for each $\lambda \in {\cal D}_N(\xi)$, we have $(1-\alpha_N/\underline{\sigma})\lambda \in {\cal D}$. Similarly, since $\|\hat{D}-D\|\leq \alpha_N$ on $E_N(Q)$, for all $\lambda \in \mathbb{R}^p$ and N such that $\alpha_N<\underline{\sigma}/2$, we have
\begin{equation}
\label{eqnPSSARP41}
(1-2\alpha_N/\underline{\sigma})\|\Pi^u(\hat{D} \lambda)\|_1\leq \|\Pi^u(D\lambda)\|_1\leq (1+2\alpha_N/\underline{\sigma})\|\Pi^u(\hat{D} \lambda)\|_1.
\end{equation}
Hence, for each $\lambda \in {\cal D}$, we have $(1-2\alpha_N/\underline{\sigma})\lambda \in \hat{{\cal D}}$. Combining the latter two observations implies that for all $\lambda \in {\cal D}_N(\xi)$, we can set $\Gamma(\lambda)=(1-\alpha_N/\underline{\sigma})(1-2\alpha_N/\underline{\sigma})\lambda \in \hat{{\cal D}}$, and we get
\[
\|\Pi^u(\Gamma(\lambda)) - \Pi^u(\lambda)\| \leq C \alpha_N \|\Pi^u(\lambda)\| \leq C \alpha_N,
\]
where we have used equation \ref{eqnPSSARP4}.
This proves equation \ref{eqnPSSARP2}.

Returning to the proof of equation \ref{eqnPSSARP1}, for $\xi\in {\cal X}_N$ and on the event $E_N(Q)$, we have
\begin{equation*}
\begin{aligned}
\sup_{\lambda \in \Delta_{0,N}(\xi)} \xi_b^\top \lambda &\leq \sup_{\lambda \in \Delta_{0,N}(\xi)} \xi_b^\top\bigl( \Pi^u(\lambda-\Gamma(\lambda))\bigr)+\sup_{\lambda \in \Delta_{0,N}(\xi)} \xi_b^\top\Gamma( \lambda)\\
&\leq CM_N\alpha_N+\sup_{\lambda \in \Delta_{0,N}(\xi)} \xi_b^\top\Gamma( \lambda),
\end{aligned}
\end{equation*}
where we have used equation \ref{eqnPSSARP2} and the fact that the part of $\xi_b$ that corresponds to the deterministic component of $b$ is equal to zero. Inequality \ref{eqnPSSARP1} then follows, with $\beta_N=CM_N\alpha_N$, if we show that for all $Q\in {\cal Q}_0$, $\xi \in {\cal X}_N$, and sufficiently large N, we have
\begin{equation}
\label{eqnPSSARP3}
\{\Gamma(\lambda)\mid \lambda \in \Delta_{0,N}(\xi)\}\subseteq \hat{\Delta}
\end{equation}
on the event $E_N(Q)$. We now proceed to prove \ref{eqnPSSARP3}. For $\lambda\in \hat{{\cal D}}$, for $Q\in {\cal Q}_0$ and on the event $E_N(Q)$, we have
\[
\langle \hat{b},\lambda\rangle=\langle \hat{b}-b,\Pi^u(\lambda)\rangle+\langle b,\lambda\rangle\leq CM_N/\sqrt{N},
\]
where we have used \ref{eqnPSSARP4} and the fact that for $Q\in {\cal Q}_0$, $b^\top \lambda\leq 0$ for all $\lambda \in {\cal D}$ (and thus in $\hat{{\cal D}}$). This yields
\begin{equation}
\label{eqnPSSARP04}
0\leq \hat{T}_N\leq CM_N/\sqrt{N}.
\end{equation}
Also, for $\xi \in {\cal X}_N$, $\lambda \in \Delta_{0,N}(\xi)$, and on the event $E_N(Q)$, we have
\begin{equation*}
\begin{aligned}
\langle \hat{b}, \Gamma(\lambda) \rangle&=\langle \hat{b}-b, \Pi^u(\Gamma(\lambda)) \rangle+\langle b-b_N(\xi), \Pi^u(\Gamma(\lambda))+\langle b_N(\xi), \Gamma(\lambda) \rangle \rangle\\
&\geq -CM_N/\sqrt{N}, 
\end{aligned}
\end{equation*}
where we have used equation \ref{eqnPSSARP4} and the fact that $b_N(\xi)^\top \Gamma(\lambda)\geq 0$ for all $\lambda\in \Delta_{0,N}(\xi)$, which holds since $b_N(\xi)^\top\lambda \geq 0$ for all $\lambda\in \Delta_{0,N}(\xi)$ (the origin is always feasible), and $\Gamma(\lambda)$ is a positive scalar multiple of $\lambda$. By our choice of $M_N$, we have $M_N=o(\kappa_N)$, and this combined with the preceding inequalities yields that for all sufficiently large N, for $\xi \in {\cal X}_N$ and $Q\in {\cal Q}_0$, we have
\[
\langle \hat{b}, \Gamma(\lambda) \rangle\geq \hat{T}_N-\kappa_N/\sqrt{N}
\]
for all $\lambda \in \Delta_{0,N}(\xi)$, which yields equation \ref{eqnPSSARP3}, and concludes the proof of equation \ref{eqnPSSARP1}.

\medskip

\underline{Step 2} \quad In this step, we use equation~\eqref{eqnPSSARP1} and a coupling argument to upper bound the test statistic by another statistic whose distribution can be uniformly estimated for all $Q \in \mathcal{Q}_0$. Let $\hat{\xi}^*=(\sqrt{N}(\hat{b}^*-b(Q)),\sqrt{N}(\hat{D}^*-D(Q)))$ be an independent and identically distributed version of the root $\hat{\xi}(Q)=(\sqrt{N}(\hat{b}-b(Q)),\sqrt{N}(\hat{D}-D(Q)))$, computed from a sample $\mathbb{W}_N^*$, such that $\mathbb{W}_N^*$ and $\mathbb{W}_N$ are independent and identically distributed. For $Q\in {\cal Q}$, let the event $F^*_N=F^*_N(Q)$ be defined by 
\[
F_N^*(Q) := \left\{ \mathbb{W}^* \;\middle|\; \|\hat{\xi}^*_b\| > M_N \;\text{or}\; \|\hat{\xi}^*_D\| > \sqrt{N} \alpha_N \right\}.
\]
By our choice of $M_N$ and $\alpha_N$, we have 
\[
\delta_N:=\sup_{Q\in {\cal Q}_0} Q\bigl(F^*_N(Q)\bigr) \quad \text{satisfies}\quad \lim_{N\to \infty} \delta_N=0
\]
Equations \ref{eqnPSSARP0} and \ref{eqnPSSARP1} then imply that on the event $E_N(Q)$, $Q\in {\cal Q}_0$, we have

\begin{equation}
\label{eqnPSSARP5}
\sqrt{N}\phi(\mu_N(\hat{\xi}^*))\leq \sup_{\lambda \in \hat{\Delta}} \langle \hat{\xi}^*_b, \lambda \rangle +\beta_N+\infty\cdot \mathbbm{1}_{\{F_N^*\}}.
\end{equation}
Let \( J_N(\cdot; Q) \) denote the CDF of \( \sqrt{N} \hat{T}_N \). Define
\[
\hat{G}_N(\cdot; Q) := \mathcal{L} \left( \sup_{\lambda \in \hat{\Delta}} \langle \hat{\xi}^*_b, \lambda \rangle + \beta_N + \infty \cdot \mathbbm{1}_{\{F_N^*\}} \,\middle|\, \mathbb{W}_N \right)
\]
as the conditional (on \( \mathbb{W}_N \)) CDF of the right-hand side of inequality \eqref{eqnPSSARP5}, viewed as a random element taking values in \( \mathbb{R} \cup \{\infty\} \).  
Similarly, define
\[
\hat{G}_\infty(\cdot; Q) := \mathcal{L} \left( \sup_{\lambda \in \hat{\Delta}} \langle \zeta, \lambda \rangle \,\middle|\, \mathbb{W}_N \right),
\]
where \( \zeta \sim N(0, \Sigma(Q)) \). And let $\hat{H}_N(;Q)$ be defined as in Theorem \ref{thmAVKAM}
\[
\hat{H}_N(\cdot; Q) := \mathcal{L} \left( \sup_{\lambda \in \hat{\Delta}} \langle \zeta^*_N, \lambda \rangle \,\middle|\, \mathbb{W}_N \right),
\]
where $\zeta^*$ is the bootstrap estimate of of $\zeta$ that is given in Assumption \ref{assAVKAM3}.

By Strassen's theorem and our choice of $\alpha_N$, we have on the event $E_N(Q)$
\begin{equation}
\label{eqnPSSARP6}
\begin{aligned}
d_{\text{Pr}}\Big( \hat{G}_N, \hat{G}_{\infty} \Big) 
&\leq d_{\text{Pr}}\Bigg( \hat{G}_N, \mathcal{L}\left( \sup_{\lambda \in \hat{\Delta}} \langle \hat{\xi}_b^*, \lambda \rangle \,\middle|\, \mathbb{W}_N \right) \Bigg) \\
&\quad + d_{\text{Pr}}\Bigg( \hat{G}_{\infty}, \mathcal{L}\left( \sup_{\lambda \in \hat{\Delta}} \langle \hat{\xi}_b^*, \lambda \rangle \,\middle|\, \mathbb{W}_N \right) \Bigg)\\
&\leq \beta_N+Q(F_N^*)+C \alpha_N.
\end{aligned}
\end{equation}
Here, we have in part used the fact that equation \ref{eqnPSSARP4}, the inequality \( d_{\text{Pr}}(\hat{\xi}_b^*, \zeta) \leq \alpha_N \) on $E_N(Q)$, and Strassen's theorem, imply that\footnote{Strassen's theorem implies that if $d_{Pr}(X,Y)\leq \epsilon$ and $f$ is a Lipschitz function that satisfies $\|f\|_{\text{Lip}}\leq C$, then $d_{Pr}(f(X),f(Y))\leq (C\vee 1) \epsilon$. Here, Lipschitz boundedness is a consequence of equation \ref{eqnPSSARP4}.}
\begin{equation}
\resizebox{\hsize}{!}{$ d_{\text{Pr}}\Bigg( \hat{G}_{\infty}, \mathcal{L}\left( \sup_{\lambda \in \hat{\Delta}} \langle \hat{\xi}_b^*, \lambda \rangle \,\middle|\, \mathbb{W}_N \right) \Bigg)=d_{\text{Pr}}\left( \mathcal{L}\left( \sup_{\lambda \in \hat{\Delta}} \langle \zeta, \Pi^u(\lambda) \rangle \,\middle|\, \mathbb{W}_N \right), \mathcal{L}\left( \sup_{\lambda \in \hat{\Delta}} \langle \hat{\xi}_b^*, \Pi^u(\lambda) \rangle \,\middle|\, \mathbb{W}_N \right) \right)\\
$}\end{equation}
is less than or equal to $C \alpha_N$. Similarly, as
\[
d_{\text{Pr}}\bigl( \mathcal{L}(\zeta^* \mid \mathbb{W}), \mathcal{N}(0, \Sigma(Q)) \bigr) \leq \alpha_N
\]
holds on the event \( E_N(Q) \), it follows from Strassen's theorem that
\[
d_{\text{Pr}}\left( \hat{G}_{\infty}, \hat{H}_N \right) \leq C \alpha_N
\]
on \( E_N(Q) \), for all sufficiently large \( N \).

\medskip
In conclusion, there exists $\gamma_N=o(1)$ such that for all sufficiently large N and all $Q \in {\cal Q}_0$, on the event $E_N(Q)$ we have
\begin{equation}
\label{eqnPSSARP7}
d_{\text{Pr}}\bigl(\hat{G}_N,\hat{G}_{\infty}\bigr)\vee d_{\text{Pr}}\bigl(\hat{H}_N,\hat{G}_{\infty}\bigr)\leq \gamma_N.
\end{equation}
Let the significance level \( \alpha \) and the critical value \( \hat{c}_N(1-\alpha) \) be as in Theorem~\ref{thmAVKAM}.
For now, assume that there exists a constant \( C \) such that for all sufficiently large \( N \), for all \( Q \in \mathcal{Q}_0 \), and for all sample realizations \( \mathbb{W}_N \) on the event \( E_N(Q) \cap \{ \sqrt{N} \hat{T}_N > \hat{c}_N(1-\alpha) \} \), we have the anti-concentration condition
\begin{equation}
\label{eqnPSSARP8}
\hat{G}_{\infty}\big( \hat{c}_N(1-\alpha) + \gamma_N \big) - \hat{G}_{\infty}\big( \hat{c}_N(1-\alpha) - \gamma_N \big) \leq C \gamma_N.
\end{equation}
We prove in the next step that inequality~\eqref{eqnPSSARP8} indeed holds. For \( \mathbb{W}_N \in E_N(Q) \cap \{ \sqrt{N} \hat{T}_N > \hat{c}_N(1-\alpha) \} \), and $Q\in {\cal Q}_0$, we then have\footnote{This argument is similar to the proof of Theorem 2.1 in \cite{JRAS}}
\begin{equation}
\label{eqnPSSARP9}
\begin{aligned}
J_N\big( \hat{c}_N(1-\alpha) \big) 
&\geq \hat{G}_N\big( \hat{c}_N(1-\alpha) \big) \\
&\geq \hat{G}_{\infty}\big( \hat{c}_N(1-\alpha) - \gamma_N \big) - \gamma_N\\
&\geq \hat{G}_{\infty}\big( \hat{c}_N(1-\alpha) + \gamma_N \big) - C\gamma_N\\
&\geq \hat{H}_{N}\big( \hat{c}_N(1-\alpha)\big) - C\gamma_N\\
&\geq 1-\alpha-C\gamma_N
\end{aligned}
\end{equation}
where we have used ~\eqref{eqnPSSARP8} to derive the third inequality, and where the other inequalities follow from ~\eqref{eqnPSSARP7} and the fact that $\hat{H}_N(\hat{c}_N(1-\alpha))\geq 1-\alpha$ by the definition of $\hat{c}_N(1-\alpha)$. 
 Let $q_{N}(\cdot)$ denote the quantiles of $J_N(\cdot)$. Inequality \ref{eqnPSSARP9} then implies that on the event \(E_N(Q) \cap \{ \sqrt{N} \hat{T}_N > \hat{c}_N(1-\alpha) \} \), with $Q\in {\cal Q}_0$, we have
\begin{equation}
\label{eqnPSSARP10}
q_N(1-\alpha-C\gamma_N)\leq \hat{c}_N(1-\alpha).
\end{equation}
 We have
\begin{equation}
\label{eqnPSSARP11}
\begin{aligned}
\sup_{Q\in {\cal Q}_0}Q\Bigl(\{ \sqrt{N} \hat{T}_N > \hat{c}_N(1-\alpha) \} \Bigr)&\leq \sup_{Q\in {\cal Q}_0}Q\Bigl(\{ \sqrt{N} \hat{T}_N > \hat{c}_N(1-\alpha) \} \cap E_N(Q)\Bigr)+\delta_N\\
&\leq \sup_{Q\in {\cal Q}_0}Q\Bigl(\{ \sqrt{N} \hat{T}_N > q_N(1-\alpha-C\gamma_N) \}\Bigr)+\delta_N\\
&\leq \alpha+C\gamma_N+\delta_N.
\end{aligned}
\end{equation}
which yields equation \ref{eqnAVKAM1} from Theorem \ref{thmAVKAM}, as $\gamma_N=o(1)$ and $\delta_N=o(1)$.

\medskip 

\underline{Step 3} In this step, we establish the anti-concentration condition \ref{eqnPSSARP8}. We first recall that as $0\in \hat{{\cal D}}$ and $\lambda\geq 0$ for all $\lambda \in \hat{{\cal D}}$, $0$ is an extreme point of $\hat{{\cal D}}$, and $\operatorname{extr}(\hat{{\cal D}})\neq \emptyset$. It then follows that any face of $\hat{{\cal D}}$ contains at least one extreme point of $\hat{{\cal D}}$ (see Section 8.5 of \cite{ALS}). On the event of a rejection, we necessarily have $\hat{v}(=\hat{T}_N)>0$, in which case $\hat{\Delta}$ contains an element of $\operatorname{extr}(\hat{{\cal D}})\backslash \{0\}$. Indeed, as the set $\hat{\Delta}_0=\arg\max\{\hat{b}^\top \lambda \mid \lambda \in \hat{{\cal D}}\}=\{\lambda \in \hat{{\cal D}}\mid \hat{b}^\top \lambda =\hat{v}\}$ is a face of $\hat{{\cal D}}$, it must contain an element $\lambda$ of $\operatorname{extr}(\hat{{\cal D}})$, and it must be the case that $\lambda\neq 0$, as $\hat{v}>0$. Since $\hat{\Delta}_0\subseteq \hat{\Delta}$, we conclude that $\hat{\Delta}\cap [\operatorname{extr}(\hat{{\cal D}})\backslash \{0\}]\neq \emptyset$. 

\medskip

We first consider the case where $Q\in {\cal Q}_0$ is such that $\Delta_0(Q)=\{0\}$. Suppose that the following event occurs
\[
 E_N(Q) \cap \{ \sqrt{N} \hat{T}_N > \hat{c}_N(1-\alpha) \}
\]
and let $\hat{\lambda}\in \hat{\Delta}\cap [\operatorname{extr}(\hat{{\cal D}})\backslash \{0\}]$. Then Lemma \ref{lemEPC} implies that there exists $\lambda \in \operatorname{extr}({\cal D}(Q))\backslash \{0\} $ such that 
\begin{equation}
\label{eqnPSSARP12}
\|\Pi^u(\hat{\lambda})-\Pi^u(\lambda)\|\leq C\|\Pi^u(\hat{\lambda})\| \|\hat{D}-D\|\leq C\alpha_N
\end{equation}
where we have used equation \ref{eqnPSSARP4} and the fact that $\|\hat{D}-D\|\leq \alpha_N$ on $E_N(Q)$. The latter inequality implies that 
\begin{equation}
\label{eqnPSSARP013}
\begin{aligned}
\hat{\lambda}^\top \Sigma(Q)\hat{\lambda}&=\Pi^u(\hat{\lambda})^\top \Sigma(Q)\Pi^u(\hat{\lambda})=\|\Sigma(Q)^{1/2}\Pi^u(\hat{\lambda})\|^2\\
&\geq (1/2)\|\Sigma(Q)^{1/2}\Pi^u(\lambda)\|^2-\|\Sigma(Q)^{1/2}(\Pi^u(\hat{\lambda})-\Pi^u(\lambda))\|^2\\
&=(1/2)\lambda^\top \Sigma(Q)\lambda-\|\Sigma(Q)^{1/2}(\Pi^u(\hat{\lambda})-\Pi^u(\lambda))\|^2\\
&\geq (1/2)\lambda^\top \Sigma(Q)\lambda-C\alpha_N^2
\end{aligned}
\end{equation}
where we have used equation \ref{eqnPSSARP12} and part $ii)$ of Assumption \ref{assAVKAM4}. Hence, using part $ii)$ of Assumption \ref{assAVKAM4}, for all sufficiently large $N$ and for $Q\in {\cal Q}_0$ such that $\Delta_0(Q)=\{0\}$, there exists $\hat{\lambda}\in \hat{\Delta}$ such that 
\begin{equation}
\label{eqnPSSARP13}
\hat{\lambda}^\top \Sigma(Q)\hat{\lambda}\geq \rho/4
\end{equation}
on the event $ E_N(Q) \cap \{ \sqrt{N} \hat{T}_N > \hat{c}_N(1-\alpha) \}$.

\medskip 

We now establish the analogue of equation \ref{eqnPSSARP13} for $Q\in {\cal Q}_0$ such that $\Delta_0(Q)\neq \{0\}$. By part i) of Assumption \ref{assAVKAM4}, there exists $\lambda_Q\in \Delta_0(Q)$ such that 
\begin{equation}
\label{eqnPSSARP14}
\lambda_Q^\top \Sigma(Q)\lambda_Q\geq \rho.
\end{equation}
Equation \ref{eqnPSSARP2} (with $\xi=0$) implies that for all sufficiently large N, and on the event $E_N(Q)$, there exist $\hat{\lambda}\in \hat{{\cal D}}(Q)$ such that
\begin{equation}
\label{eqnPSSARP15}
\|\Pi^u(\hat{\lambda})-\Pi^u(\lambda_Q)\|\leq C\|\Pi^u(\lambda_Q)\| \|\hat{D}-D\|\leq C\alpha_N.
\end{equation}
Moreover, since equation \ref{eqnPSSARP2} gives $\hat{\lambda}$ as a positive scalar multiple of $\lambda_Q$ and $\lambda_Q\in \Delta_0(Q)$, for sufficiently large $N$ and on the event $E_N(Q)$, we have 
\[
|\hat{b}^\top \hat{\lambda}|=|(\hat{b}-b)^\top \hat{\lambda}+b^\top \hat{\lambda}|=|(\hat{b}-b)^\top \Pi^u(\hat{\lambda})|\leq CM_N/\sqrt{N}
\]
where we have used homogeneity and the fact that $\lambda_Q\in \Delta_0(Q)$ implies that $b^\top \lambda_Q=0$, as well as equation \ref{eqnPSSARP4}. The latter inequality, equation \ref{eqnPSSARP04}, and the fact that $M_N=o(\kappa_N)$, then imply that for sufficiently large $N$ and on the event $E_N(Q)$, we have 
\begin{equation}
\label{eqnPSSARP16}
\Bigl[\hat{b}^\top \hat{\lambda}\geq \hat{v}-\kappa_N/\sqrt{N}\Bigr] \quad \quad \text{which implies that} \quad \quad \Bigl[\hat{\lambda}\in \hat{\Delta}\Bigr].
\end{equation}
Reasoning as in equation \ref{eqnPSSARP013}, equation \ref{eqnPSSARP15} and the part $i)$ of Assumption \ref{assAVKAM4}, then yield
\[
\hat{\lambda}^\top \Sigma(Q)\hat{\lambda}\geq (1/2)\lambda_Q^\top \Sigma(Q)\lambda_Q-C\alpha_N^2.
\]
For all sufficiently large $N$, the latter inequality and equation \ref{eqnPSSARP14} then give
\[
\hat{\lambda}^\top \Sigma(Q)\hat{\lambda}\geq \rho/4.
\]
In conclusion, for all sufficiently large $N$ and for $Q\in {\cal Q}_0$ such that $\Delta_0(Q)\neq \{0\}$, on the event $E_N(Q)$, there exists a $\hat{\lambda}\in  \hat{\Delta}$ such that
\begin{equation}
\label{eqnPSSARP17}
\hat{\lambda}^\top \Sigma(Q)\hat{\lambda}\geq \rho/4.
\end{equation}

\medskip
Combining equations \ref{eqnPSSARP13} and \ref{eqnPSSARP17}, we conclude that for all sufficiently large $N$ and for all $Q\in {\cal Q}_0$, on the event $ E_N(Q) \cap \{ \sqrt{N} \hat{T}_N > \hat{c}_N(1-\alpha) \}$, there exists a $\lambda \in \hat{\Delta}$ such that
\[
\Pi^u(\hat{\lambda})^\top \Sigma(Q)\Pi^u(\hat{\lambda})=\hat{\lambda}^\top \Sigma(Q)\hat{\lambda}\geq \rho/4,
\]
and the anti-concentration condition \ref{eqnPSSARP8} then follows from Proposition \ref{prop1}, using the fact that the sets $\Pi^u(\hat{\Delta})$ are eventually uniformly bounded on the events $E_N(Q)$, and
\[
\sup_{\lambda \in \hat{\Delta}} \langle \zeta, \lambda \rangle=\sup_{\lambda \in \hat{\Delta}} \langle \Pi^u(\zeta), \Pi^u(\lambda) \rangle.
\]

\underline{Step 4} We now prove inequality \ref{eqnAVKAM2}.
 Fix $Q \in \mathcal{Q}_0$. Under Assumptions~\ref{assAVKAM0}, \ref{assAVKAM1} and \ref{assAVKAM2}, Lemma \ref{lemAD} implies that the asymptotic distribution of $\sqrt{N} \hat{T}_N$ is given by
\begin{equation}
\label{eqnPSSARP18}
\max \left\{ \zeta^\top \lambda \;\middle|\; \lambda \in \Delta_0(Q) \right\}=\max \left\{ \zeta^\top \Pi^u(\lambda) \;\middle|\; \lambda \in \Delta_0(Q) \right\},
\end{equation}
where $\Delta_0(Q) := \arg\max \{ b(Q)^\top \lambda \mid \lambda \in \mathbb{R}^p_+, A^\top \lambda=0, \ \|D(Q)\lambda\|_1 \leq 1 \}$ and $\zeta \sim \mathcal{N}(0, \Sigma(Q))$. Let $G_\infty(\cdot)$ denote the CDF of the value of the program in equation \ref{eqnPSSARP18}. 
Then $d_{\mathrm{Pr}}(\mathcal{L}(\sqrt{N} \hat{T}_N), G_\infty) = o(1)$. To prove~\eqref{eqnAVKAM2}, it suffices to show
\begin{equation}
\label{eqnPSSARP19}
d_{\mathrm{Pr}}(\hat{H}_N, G_\infty) = o_p(1),
\end{equation}
where $\hat{H}_N$ is the conditional (given $\mathbb{W}_N$) CDF in~\eqref{eqnKAM3}. We show below that 
\begin{equation}
\label{eqnPSSARP20}
d_H(\Pi^u(\Delta_0(Q)), \Pi^u(\hat{\Delta})) = o_p(1),
\end{equation}
from which~\eqref{eqnPSSARP19} follows from Assumption \ref{assAVKAM3}, via an almost sure representation argument and Lemma~\ref{lem2}.

To establish equation \ref{eqnPSSARP20}, we first show that
\begin{equation}
\label{eqnPSSARP21}
\overset{\rightarrow}{d}_H\bigl(\Pi^u(\Delta_0),\Pi^u(\hat{\Delta})\bigr)=o_p(1),
\end{equation}
and then establish the converse in equation \ref{eqnPSSARP24} below. Arguing as in the proof of equation \ref{eqnPSSARP41}, for all $N$ sufficiently large $N$ that $\alpha_N<\underline{\sigma}/2$, and on the event $E_N(Q)$, we have that
\begin{equation}
\label{eqnPSSARP22}
\forall \lambda \in {\cal D}\quad \hat{\Gamma}(\lambda):=(1-(2/\underline{\sigma})\alpha_N)\lambda\in \hat{{\cal D}} \quad \quad \text{and}\quad \quad \sup_{\lambda \in {\cal D}}\|\Pi^u(\lambda)-\Pi^u(\hat{\Gamma}(\lambda))\|\leq C\alpha_N.
\end{equation}
We now show that, on $E_N(Q)$, $\hat{\Gamma}(\Delta_0)$ is included in $\hat{\Delta}$ for all sufficiently large N. Equation \ref{eqnPSSARP22} would then imply that for all sufficiently large N and on the event $E_N(Q)$, we have
\begin{align*}
\overset{\rightarrow}{d}_H\bigl(\Pi^u(\Delta_0),\Pi^u(\hat{\Delta})\bigr)&\leq \overset{\rightarrow}{d}_H\bigl(\Pi^u(\Delta_0),\Pi^u(\hat{\Gamma}(\Delta_0)))\bigr)+\overset{\rightarrow}{d}_H\bigl(\hat{\Gamma}(\Delta_0),\hat{\Delta})\bigr)\\
&=\overset{\rightarrow}{d}_H\bigl(\Pi^u(\Delta_0),\Pi^u(\hat{\Gamma}(\Delta_0)))\bigr)\leq C\alpha_N,
\end{align*}
which would yield equation \ref{eqnPSSARP21}. Indeed, on $E_N(Q)$ and for all sufficiently large N, we have
\begin{equation}
\label{eqnPSSARP23}
\begin{aligned}
\inf_{\lambda \in \Delta_0} \langle \hat{b}, \hat{\Gamma}(\lambda) \rangle &= \langle \hat{b}-b, \Pi^u(\hat{\Gamma}(\lambda)) \rangle+\langle b, \hat{\Gamma}(\lambda) \rangle\\
&\geq -\frac{M_N}{\sqrt{N}} \sup_{\lambda \in \hat{{\cal D}}}\|\Pi^u(\lambda)\|\\
&\geq -C\frac{M_N}{\sqrt{N}},
\end{aligned}
\end{equation}
where we have used the fact that $\langle b, \hat{\Gamma}(\lambda)\rangle=0$ since $Q\in {\cal Q}_0$, $\lambda\in \Delta_0$, and $\hat{\Gamma}(\lambda)$ is a positive scalar multiple of $\lambda$; we have also used the analogue of equation \ref{eqnPSSARP4}, for $\hat{{\cal D}}$ on the event $E_N(Q)$. Since by equation \ref{eqnPSSARP04} we have $0\leq \hat{v}\leq C M_N/\sqrt{N}$, equation \ref{eqnPSSARP23} and the fact that $M_N=o(\kappa_N)$ imply that on the event $E_N(Q)$, and for all sufficiently large N, we have
\[
\hat{\Gamma}(\Delta_0)\subseteq \{\lambda \in \hat{{\cal D}}\mid \langle \hat{b},\lambda\rangle\geq \hat{v}-\kappa_N/\sqrt{N} \}.
\]
Hence, $\hat{\Gamma}(\Delta_0)$ is eventually a subset of $\hat{\Delta}$ on the event $E_N(Q)$, and this completes the proof of equation \ref{eqnPSSARP21}.

We now establish the converse of equation \ref{eqnPSSARP21}, and show that 
\begin{equation}
\label{eqnPSSARP24}
\overset{\rightarrow}{d}_H\bigl(\Pi^u(\hat{\Delta}),\Pi^u(\Delta_0)\bigr)=o_p(1).
\end{equation}
As in equation \ref{eqnPSSARP22}, it can be shown that on the event $E_N(Q)$ and for all sufficiently large $N$ such that $\alpha_N<\underline{\sigma}$, we have
\begin{equation}
\label{eqnPSSARP25}
\forall \lambda \in \hat{{\cal D}}\quad \Gamma(\lambda):=(1-(1/\underline{\sigma})\alpha_N)\lambda \in {\cal D} \quad \quad \text{and}\quad \quad \sup_{\lambda \in \hat{{\cal D}}}\|\Pi^u(\lambda)-\Pi^u(\Gamma(\lambda))\|\leq C\alpha_N.
\end{equation}
Moreover, on $E_N(Q)$ and for all sufficiently large $N$, we have
\begin{equation}
\label{eqnPSSARP26}
\begin{aligned}
0\geq \sup_{\lambda \in \hat{\Delta}} \langle b,\Gamma(\lambda)\rangle&\geq \inf_{\lambda \in \hat{\Delta}} \langle b,\Gamma(\lambda)\rangle\\
&=\inf_{\lambda \in \hat{\Delta}} \langle b-\hat{b},\Pi^u(\Gamma(\lambda))\rangle+ \langle \hat{b},\Gamma(\lambda)\rangle\\
&\geq \inf_{\lambda \in \hat{\Delta}} \langle b-\hat{b},\Pi^u(\Gamma(\lambda))\rangle-\kappa_N/\sqrt{N}\\
&\geq \inf_{\lambda \in {\cal D}} \langle b-\hat{b},\Pi^u(\lambda)\rangle-\kappa_N/\sqrt{N}\\
&\geq -\frac{M_N}{\sqrt{N}}\sup_{\lambda \in {\cal D}} \|\Pi^u(\lambda)\|-\kappa_N/\sqrt{N}\\
&\geq -C\kappa_N/\sqrt{N}
\end{aligned}
\end{equation}
where the first inequality follows since $v=0$ (for $Q\in {\cal Q}_0$); the third inequality follows from the fact that $\hat{v}\geq 0$ (we always have $0\in \hat{{\cal D}}$), $\langle \hat{b}, \lambda\rangle \geq \hat{v}-\kappa_N/\sqrt{N}\geq -\kappa_N/\sqrt{N}$ for all $\lambda \in \hat{\Delta}$, and that $\Gamma(\lambda)$ is a scalar multiple of $\lambda$ with scalar factor in $(0,1)$; the fourth inequality follows from the fact that $\Gamma(\hat{\Delta})\subseteq {\cal D}$; and the last inequality follows from the fact that as in equation \ref{eqnPSSARP4} we can show that $\sup_{\lambda \in {\cal D}} \|\Pi^u(\lambda)\|<1/\underline{\sigma}$, and that $M_N=o(\kappa_N)$.
As $\kappa_N=o(\sqrt{N})$, equation \ref{eqnPSSARP26} implies that for any $\epsilon>0$, and for all sufficiently large $N$, we have 
\[
\Gamma(\hat{\Delta})\subseteq \{\lambda \in {\cal D} \mid \langle b,\lambda \rangle \geq -\epsilon\}.
\]
As a consequence, Lemma \ref{lemOSC} then implies that on $E_N(Q)$ we have
\begin{equation}
\label{eqnPSSARP27}
\overset{\rightarrow}{d}_H(\Gamma({\hat{\Delta}}), \Delta_0)=o(1).
\end{equation}
Combining equations \ref{eqnPSSARP25} and \ref{eqnPSSARP27}, we conclude that, on the event $E_N(Q)$ and for all sufficiently large $N$, we have
\begin{equation*}
\begin{aligned}
\overset{\rightarrow}{d}_H(\Pi^u(\hat{\Delta}), \Pi^u(\Delta_0))&\leq  \overset{\rightarrow}{d}_H(\Pi^u(\hat{\Delta}), \Pi^u(\Gamma(\hat{\Delta})))+\overset{\rightarrow}{d}_H(\Gamma(\hat{\Delta}), \Delta_0)\\
&\le C\alpha_N+o(1)
\end{aligned}
\end{equation*}
which establishes equation \ref{eqnPSSARP24}, and completes the proof of inequality \ref{eqnAVKAM2}.
\end{proof}

\begin{proof}{\textbf{(Proof of Theorem \ref{thmAVEAM})}}
\label{pthmAVEAM}
In this section, we provide proofs of equations~\eqref{eqnAVEAM1} and~\eqref{eqnAVEAM2}. First note that the dual and the associated Lagrangian of the LP \ref{eqnAVEAM0} are given by
\begin{equation}
\label{eqnPSSLSUC01}
\min\{t\mid t\Omega(Q) \mathbbm{1}+A(Q)\eta\geq b(Q),\ t\geq 0\}
\end{equation}
and
\[
L(\lambda,\eta,t)=b^\top \lambda-\eta^\top A^\top \lambda+t\bigl(1-\mathbbm{1}^\top \Omega\lambda\bigr),
\]
where $t\geq 0$ and $\eta$ are the Lagrange multipliers.

Let $\xi=(\xi_A,\xi_b,\xi_{\Omega})$ denote generic pertubations of the input parameters $\mu=(A,b,\Omega)$ of our LPs, where $\xi_A\in \mathbb{R}^{p\times d}$, $\xi_b\in \mathbb{R}^p$, and $\xi_{\Omega}\in \mathbb{R}^{p\times p}$ is a diagonal matrix. Given $\xi$, let the resulting perturbed inputs $\mu_N(\xi)=(A_N(\xi),b_N(\xi),\Omega_N(\xi))$ be defined by
\[
\mu_N(\xi)=\Bigl(A+\frac{\xi_A}{\sqrt{N}}, b+\frac{\xi_b}{\sqrt{N}}, \Omega +\frac{\xi_{\Omega}}{\sqrt{N}}\Bigr).
\]
We assume here that the pertubation matrices $\xi_{\Omega}$, like the matrices $\Omega$, have all their $(i,i)$-th diagonal entries that correspond to indices $i\in K$ equal to zero.\footnote{See the paragraph that precedes Assumption \ref{assAVEAM0}, for a definition of the set $K$, $U$, and for a definition of the projection operator $\Pi^u$.} Let $\phi_N(\mu)$ denote the value of the LP \ref{eqnAVEAM0}, for arbitrary inputs $\mu=(A,b,\Omega)$. Let $L_N(\lambda,\eta,t; \xi)$ denote the Lagrangian for the LP \ref{eqnAVEAM0} with inputs $\mu_N(\xi)$, given by
\[
L_N(\lambda,\eta,t;\xi)=b_N(\xi)^\top \lambda-\eta^\top A_N(\xi)^\top \lambda+t\bigl(1-\mathbbm{1}^\top \Omega_N(\xi)\lambda\bigr).
\]
Let $\Delta_N(0,N)(\xi)$ and $S_{0,N}(\xi)$ denote respectively the set of optimal solutions for the LP \ref{eqnAVEAM0} and its dual, when the inputs are given by $\mu_N(\xi)$. Let $\Delta_{0}$ and $S_{0}$ denote the analoguous objects when $\xi=0$. 

For $(\lambda_{\xi}, \eta_{\xi}, t_{\xi}) \in \Delta_{0,N}(\xi) \times S_{0,N}(\xi)$ and $(\lambda_0, \eta_0, t_0) \in \Delta_{0} \times S_{0}$, the saddle point property of the Lagrangian implies that, for all sufficiently small perturbations $\xi$ such that $\phi_N(\mu_N(\xi))$ is finite, we have
\begin{equation*}
\begin{aligned}
\sqrt{N} \bigl(\phi_N(\mu_N(\xi)) - \phi(\mu)\bigr)
&= \sqrt{N} \bigl(L_N(\lambda_{\xi}, \eta_{\xi}, t_{\xi}; \xi) - L(\lambda_0, \eta_0, t_0)\bigr) \\
&\leq \sqrt{N} \bigl(L_N(\lambda_{\xi}, \eta_0, t_0; \xi) - L(\lambda_{\xi}, \eta_0, t_0)\bigr) \\
&= \xi_b^\top \lambda_{\xi} - \eta_0^\top \xi_A^\top \lambda_{\xi} - t_0 \mathbbm{1}^\top \xi_{\Omega} \lambda_{\xi}.
\end{aligned}
\end{equation*}
Now, for any $Q \in \mathcal{Q}_0$, $\phi(\mu)=0$ and the condition $(\eta_0, t_0) \in S_{0}$ implies $t_0 = 0$. In fact, $S_0=\{(t,\eta)\mid t=0, \ b(Q)-A(Q)\eta\leq 0\}$. Therefore, the inequality above yields, for all sufficiently large $N$, for all sufficiently small perturbations $\xi$\footnote{For sufficiently small perturbations $\xi$ such that $\Omega_{\xi}$ has strictly positive diagonal entries for all $i \in U$, it follows from Farkas' lemma that $\phi(\mu_N(\xi))$ remains finite for all $Q \in \mathcal{Q}_0$.}, and for $Q \in \mathcal{Q}_0$:
\begin{equation}
\label{eqnPSSLSUC02}
\sqrt{N} \phi_N(\mu_N(\xi)) \leq \xi_b^\top \lambda_{\xi} - \eta_0^\top \xi_A^\top \lambda_{\xi}=\langle \xi_b-\xi_A \eta_0, \lambda_{\xi}\rangle.
\end{equation}

Let $d=vec(\begin{bmatrix}A& b\end{bmatrix})$ and $\hat{d}=vec(\begin{bmatrix}\hat{A}& \hat{b}\end{bmatrix})$. Assumptions \ref{assAVEAM0} through \ref{assAVEAM3} imply the existence of $\alpha_N\downarrow 0$, such that
\[
\sup_{Q\in {\cal Q}}d_{\text{Pr}}\Bigl(\sqrt{N}(\hat{d}-d(Q)), \mathcal{N}(0,\Sigma(Q))\Bigr)\leq \alpha_N,
\]
and such that for any $M_N\uparrow \infty$, and for all $Q \in {\cal Q}_0$, the event $E_N(Q)$ defined by\footnote{Since assumption \ref{assAVEAM0} implies that the number of estimated components of $b$ and $A$ are uniformly bounded w.r.t. $N$, the uniform consistency of the estimator $\hat{D}$ of the standard deviation matrices implies the uniform consistency of the normalizing matrices $\hat{\Omega}$.}
\[
E_N(Q) := \left\{ \mathbb{W} \;\middle|\; 
\begin{array}{l}
\|\hat{\Omega} - \Omega(Q)\| \leq \alpha_N,\quad
d_{\mathrm{Pr}}\bigl(\mathcal{L}(\zeta^* \mid \mathbb{W}), \mathcal{N}(0, \Sigma(Q))\bigr) \leq \alpha_N,\\[0.5em]
\sqrt{N} \|\hat{b} - b(Q)\| \leq M_N,\quad
\sqrt{N} \|\hat{A} - A(Q)\| \leq M_N
\end{array}
\right\}.
\]
 satisfies
\[
\underline{\lim}_{N\to \infty} \inf_{Q\in {\cal Q}} Q(E_N(Q))=1.
\]
We choose $M_N\uparrow \infty$ such that $M_N=o(\kappa_N)$, $M_N \alpha_N=o(1)$, and $M_N^2 \Bigl(\kappa_N/\sqrt{N}\Bigr)=o(1)$.
Such a choice is possible since $\kappa_N=o(\sqrt{N})$, where $\kappa_N$, here and in what follows, denotes either $\kappa_{1N}$, $\kappa_{2N}$, or $\kappa_{3N}$.

Let ${\cal X}_N$ denote the set of pertubation $\xi$ defined by
\[
{\cal X}_N=\Bigl\{\xi \mid \|\xi_b\|\leq M_N, \ \|\xi_A\|\leq M_N, \ \|\xi_{\Omega}\|\leq \sqrt{N}\alpha_N\Bigr\}.
\]
We now proceed to derive inequalities~\eqref{eqnAVEAM1} and~\eqref{eqnAVEAM2} from inequality~\eqref{eqnPSSLSUC02}. The proof is structured in four steps, which mirror the steps used in the proof of Theorem~\ref{thmAVKAM} in Section \ref{pthmAVKAM}. In particular, inequality~\eqref{eqnAVEAM1} is established in Steps~1 through~3, while Step~4 contains the proof of inequality~\eqref{eqnAVEAM2}. Throughout the argument, we let \( C \) denote a generic constant that is uniform over all data-generating processes \( Q \in \mathcal{Q}_0 \), for sufficiently large \( N \). The constant \( C \) may vary from one equation to another.

\underline{Step 1} \quad In this step, we show that there exists a sequence $\beta_N = o(1)$ such that, for all sufficiently large $N$ and for all $Q \in \mathcal{Q}_0$, the following inequality holds on the event $E_N(Q)$:
\begin{equation}
\label{eqnPSSLSUC03}
\sup_{\xi \in \mathcal{X}_N} \sup_{\lambda \in \Delta_{0,N}(\xi)} 
\left\{ 
\langle \xi_b - \xi_A \eta_0, \lambda \rangle 
- \sup_{\lambda \in \hat{\Delta}} \langle \xi_b - \xi_A \hat{\eta}, \lambda \rangle 
\right\}
\leq \beta_N,
\end{equation}
where $\hat{\Delta}$ is defined in equation~\ref{eqnEAM5}, and $\hat{\eta}$ is as given in equation~\ref{eqnEAM6}.

Let ${\cal D}(Q)$ and $\hat{{\cal D}}$ denote respectively the feasible region of the LPs \ref{eqnAVEAM0} and \ref{eqnEAM2}. Note that on the event $E_N(Q)$, the condition $\|\hat{D} - D\| \leq \alpha_N$ implies $\|\hat{\Omega} - \Omega\| \leq \alpha_N$. Consequently, for all sufficiently large $N$ such that $\alpha_N \leq \underline{\sigma}/2$, we have $[\hat{\Omega}]_{ii} > \underline{\sigma}/2$ for all $i \in U$. The same reasoning shows that for all such $N$ and for all $\xi \in \mathcal{X}_N$, we also have $[\Omega_N(\xi)]_{ii} \geq \underline{\sigma}/2$ for all $i \in U$. As a consequence, since $\|\Omega_N(\xi) \lambda\|_1 \leq 1$ for all $\lambda \in \Delta_{0,N}(\xi)$, and $\|\hat{\Omega}\lambda\|_1\leq 1$ for all $\lambda\in \hat{\cal D}$, it follows that for all N such that $\alpha_N\leq \underline{\sigma}/2$, we have
\begin{equation}
\label{eqnPSSLSUC04}
\sup_{\xi \in \mathcal{X}_N} \sup_{\lambda \in \Delta_{0,N}(\xi)} \|\Pi^u(\lambda)\|_1 \leq \frac{2}{\underline{\sigma}}, \quad \quad \text{and}\quad \quad \sup_{\lambda \in \hat{{\cal D}}} \|\Pi^u(\lambda)\|_1\leq 2/\underline{\sigma}\quad \text{on the event } E_N(Q).
\end{equation}
In the derivations that follow, we assume that N is sufficiently large that $\alpha_N\leq \underline{\sigma}/2$.

As $\eta_0$ in equation \ref{eqnPSSLSUC02} is any solution to $b(Q)-A(Q)\eta\leq 0$, we can assume that $\eta_0=\eta_Q$, where $\eta_Q$ is as in Assumption \ref{assAVEAM4}. We first show that on the events $E_N(Q)$, for all sufficiently large N and for $Q\in {\cal Q}_0$, we can replace $\eta_Q$ by $\hat{\eta}$, and we have
\begin{equation}
\label{eqnPSSLSUC05}
\sup_{\xi \in \mathcal{X}_N} \sup_{\lambda \in \Delta_{0,N}(\xi)} 
\left\{ 
\langle \xi_b - \xi_A \eta_Q, \lambda \rangle 
- \langle \xi_b - \xi_A \hat{\eta}, \lambda \rangle 
\right\}
\leq C M_N\Bigl(\kappa_N/\sqrt{N}\Bigr)^{1/2},
\end{equation}
We first show that on the event $E_N(Q)$, for all sufficiently large N and for $Q\in {\cal Q}_0$, $\eta(Q)$ is in the feasible region of the quadratic program \ref{eqnEAM6}, and satisfies
\begin{equation}
\label{eqnPSSLSUC06}
\hat{b}-\hat{A}\eta(Q)\leq \frac{\kappa_N}{\sqrt{N}} \hat{\Omega}\mathbbm{1}.
\end{equation}
 
Indeed, we have
\begin{equation*}
\begin{aligned}
\hat{b}-\hat{A}\eta_Q&\leq \hat{b}-\hat{A}\eta_Q+A\eta(Q)-b\\
&\leq \Bigl(\|\hat{b}-b\|+\|\hat{A}-A\|\|\eta_Q\|\Bigr)\Pi^u(\mathbbm{1}).
\end{aligned}
\end{equation*}
Since Assumption \ref{assAVEAM4} implies that $\eta_Q$ is uniformly (in $Q\in {\cal Q}_0$) bounded, we conclude that on the event $E_N(Q)$, we have
\[
\hat{b} - \hat{A} \eta(Q) \leq \frac{C M_N}{\sqrt{N}}\Pi^u(\mathbbm{1}),
\]
which yields equation \ref{eqnPSSLSUC06} for all sufficiently large $N$ as $M_N=o(\kappa_N)$, and $[\hat{\Omega}]_{ii}\geq \underline{\sigma}/2$ for all $i\in U$. In turn, this implies that for all sufficiently large $N$ and all $Q \in \mathcal{Q}_0$, we have
\begin{equation}
\label{eqnPSSLSUC07}
\|\hat{\eta}\| \leq \|\eta_Q\|\leq M
\end{equation}
where $M>0$ is such that $\sup_{Q\in {\cal Q}_0} \|\eta_Q\|\leq M$.

By Lemma~\ref{lemhoff}, if ${\cal C}_Q:=\{\eta\mid b(Q)-A(Q)\eta\leq 0\}$, then inequality \ref{eqnPSSLSUC07}  and Assumption \ref{assAVEAM4} imply that
\begin{equation}
\label{eqnPSSLSUC08}
d\bigl(\hat{\eta},{\cal C}_Q\bigr)\leq C\|[b(Q)-A(Q)\hat{\eta}]_+\|\leq C\kappa_N/\sqrt{N}.
\end{equation}
Indeed, we have
\begin{equation*}
\begin{aligned}
[b-A\hat{\eta}]_+&\leq \Bigl[b-A\hat{\eta}+\hat{A}\hat{\eta}-\hat{b}+\bigl(\hat{T}_N+\frac{\kappa_N}{\sqrt{N}}\bigr)\hat{\Omega}\mathbbm{1}\Bigr]_+\\
&\leq [b-\hat{b}]_++[(\hat{A}-A)\hat{\eta}]_++C\frac{\kappa_N}{\sqrt{N}}\hat{\Omega}\mathbbm{1}\\
&\leq C\frac{\kappa_N}{\sqrt{N}}\Pi^u(\mathbbm{1}),
\end{aligned}
\end{equation*}
where we have used the fact that $\hat{\eta}$ is in the feasible region of the quadratic program \ref{eqnEAM6}, equation \ref{eqnPSSLSUC07}, the fact that $M_N=o(\kappa_N)$, and the additional fact, which we derive further below, that $\hat{T}_N=o(\kappa_N/\sqrt{N})$.

Inequalities \ref{eqnPSSLSUC07} and \ref{eqnPSSLSUC08} imply that
\begin{equation}
\label{eqnPSSLSUC09}
\|\hat{\eta}-\eta(Q)\|\leq C \Bigl(\kappa_N/\sqrt{N}\Bigr)^{1/2}.
\end{equation}
 Indeed, if $\Pi_{{\cal C}}(\cdot)$ denotes the (Euclidean) projection operator on ${\cal C}(Q)$, then equations \ref{eqnPSSLSUC07} and \ref{eqnPSSLSUC08} imply that
\[
\langle \eta(Q),\eta\rangle \geq \|\eta(Q)\|^2 \quad \forall \eta\in {\cal C}(Q) \quad \quad{and}\quad \quad \|\Pi_{{\cal C}}(\hat{\eta})\|\leq \|\eta(Q)\|+C\frac{\kappa_N}{\sqrt{N}}.
\]
This gives 
\[
\|\eta(Q)-\Pi_{{\cal C}}(\hat{\eta})\|^2=\|\eta(Q)\|^2+\|\Pi_{{\cal C}}(\hat{\eta})\|^2-2\langle\eta(Q),  \Pi_{{\cal C}}(\hat{\eta})\rangle\leq  C \frac{\kappa_N}{\sqrt{N}},
\]
and equation \ref{eqnPSSLSUC09} then follows from the latter inequality, and equation \ref{eqnPSSLSUC08}, noting that $\|\hat{\eta}-\eta(Q)\|\leq\|\hat{\eta}-\Pi_{{\cal C}}(\hat{\eta})\|+\|\Pi_{{\cal C}}(\hat{\eta})-\eta(Q)\|$.

Equation \ref{eqnPSSLSUC04} implies that $\|\xi_A^\top \lambda \|=\|\xi_A^\top \Pi^u(\lambda)\|\leq C M_N$ for $\xi \in {\cal X}_N$ and $\lambda \in \Delta_{0,N}(\xi)$, and equation \ref{eqnPSSLSUC09} then implies that on the events $E_N(Q)$, for all sufficiently large N and for $Q\in {\cal Q}_0$, we have
\begin{equation*}
\sup_{\xi \in \mathcal{X}_N} \sup_{\lambda \in \Delta_{0,N}(\xi)} 
\big|\left\{ 
\langle \xi_b - \xi_A \eta_0, \lambda \rangle 
- \langle \xi_b - \xi_A \hat{\eta}, \lambda \rangle 
\right\}\big|
\leq C M_N\Bigl(\kappa_N/\sqrt{N}\Bigr)^{1/2}.
\end{equation*}
Note also that, since $M_N$ is chosen such that $M_N^2 (\kappa_N/\sqrt{N}) = o(1)$, the right-hand side of the latter inequality is asymptotically negligible. This yields equation \ref{eqnPSSLSUC05}.

We now show that for all sufficiently large $N$, and for all $Q \in \mathcal{Q}_0$, the following holds on the event $E_N(Q)$:
\begin{equation}
\label{eqnPSSLSUC010}
\sup_{\xi \in {\cal X}_N}\overset{\rightarrow}{d}_H\bigl(\Pi^u(\Delta_{0,N}(\xi)), \Pi^u(\hat{\Delta})\bigr) \leq C \alpha_N.
\end{equation}
Assuming that inequality~\ref{eqnPSSLSUC010} holds, and invoking equation~\ref{eqnPSSLSUC04}, it follows that on the event $E_N(Q)$ we have
\begin{equation}
\label{eqnPSSLSUC011}
\sup_{\xi \in \mathcal{X}_N} \sup_{\lambda \in \Delta_{0,N}(\xi)} 
\left\{ 
\langle \xi_b - \xi_A \hat{\eta}, \lambda \rangle 
- \sup_{\lambda \in \hat{\Delta}} \langle \xi_b - \xi_A \hat{\eta}, \lambda \rangle 
\right\}
\leq C \alpha_N M_N.
\end{equation}
Combining inequalities~\ref{eqnPSSLSUC05} and~\ref{eqnPSSLSUC011} yields inequality~\ref{eqnPSSLSUC03} with
\[
\beta_N = C M_N\left( \frac{\kappa_N}{\sqrt{N}} \right)^{1/2} + C \alpha_N M_N = o(1).
\]
It therefore remains to establish inequality~\ref{eqnPSSLSUC010} in order to complete Step~1.

We first show that on the event $E_N(Q)$, for all sufficiently large $N$ and for $Q\in {\cal Q}_0$, we have
\begin{equation}
\label{eqnPSSLSUC012}
\forall \xi \in {\cal X}_N,\quad \Delta_{0,N}(\xi)\subseteq \Bigl\{\lambda\in \mathbb{R}^p_+\mid -\frac{\kappa_{N}}{\sqrt{N}}\mathbbm{1}\leq \hat{A}^\top \lambda\leq \frac{\kappa_{N}}{\sqrt{N}}\mathbbm{1}\Bigr\}.
\end{equation}
Indeed, on $E_N(Q)$ we have
\[
\sup_{\lambda\in \Delta_{0,N}(\xi)}\|\hat{A}^\top\lambda\|_{\infty}=\sup_{\lambda\in \Delta_{0,N}(\xi)}\|(\hat{A}-A_N(\xi))^\top \Pi^u(\lambda)\|_{\infty}\leq CM_N/\sqrt{N},
\]
where we have use equation \ref{eqnPSSLSUC04}. Equation \ref{eqnPSSLSUC012} then follows since $M_N=o(\kappa_N)$. Arguing as in \ref{eqnPSSARP41}, we easily show that on $E_N(Q)$ and for all $N$ sufficiently large that $\alpha_N< \underline{\sigma}/4$, we have
\[
(1-4\alpha_N/\underline{\sigma})\|\hat{\Omega}\lambda\|_1\leq \|\Omega_N(\xi)\lambda\|_1\leq (1+4\alpha_N/\underline{\sigma})\|\hat{\Omega}\lambda\|_1,
\]
for all $\lambda\in \mathbb{R}^p$. Hence if $\hat{\Gamma}(\lambda):=(1-4\alpha_N/\underline{\sigma})\lambda$, then the latter inequality and equation \ref{eqnPSSLSUC012} imply that on the event $E_N(Q)$, for all sufficiently large $N$ and for $Q\in {\cal Q}_0$, we have
\[
\forall \xi \in {\cal X}_N,\quad \hat{\Gamma}(\Delta_{0,N}(\xi))\subseteq  \Bigl\{\lambda\in \mathbb{R}^p_+\mid -\frac{\kappa_{N}}{\sqrt{N}}\mathbbm{1}\leq \hat{A}^\top \lambda\leq \frac{\kappa_{N}}{\sqrt{N}}\mathbbm{1}, \ \|\hat{\Omega}(\lambda)\|_1\leq 1\Bigr\}.
\]
Moreover, equation \ref{eqnPSSLSUC04} implies that 
\[
\forall \xi \in {\cal X}_N, \quad \sup_{\lambda\in \Delta_{0,N}(\xi)}\|\Pi^u(\lambda-\hat{\Gamma}(\lambda))\|\leq C\alpha_N.
\]

We now show that for all sufficiently large $N$, we have
\[
\inf_{\lambda\in \Delta_{0,N}(\xi)}\hat{b}^\top \hat{\Gamma}(\lambda)\geq \hat{T}_N-\kappa_N/\sqrt{N},
\]
which when combined with the preceding two inequalities yields \ref{eqnPSSLSUC010}, and completes the proof of Step 1.
We first show that $\hat{T}_N=o(\kappa_N/\sqrt{N})$. Indeed, first note that Assumption \ref{assAVEAM4} implies that there exists a constant $M>0$ such that for all $Q\in {\cal Q}_0$, we have 
\begin{equation}
\label{eqnPSSLSUC013}
b^\top \lambda\leq \langle \lambda, A\eta_Q\rangle=\langle A^\top \lambda, \eta_Q\rangle\leq M\|A^\top \lambda\|, \ \quad \forall \lambda\geq0.
\end{equation}
On the event $E_N(Q)$, for all sufficiently large $N$ and for $Q\in {\cal Q}_0$, we have
\begin{equation}
\label{eqnPSSLSUC014}
\begin{aligned}
0\leq \hat{T}_N=\sup_{\lambda\in \hat{{\cal D}}}\langle \hat{b},\lambda\rangle&\leq \sup_{\lambda\in \hat{{\cal D}}}\langle \hat{b}-b,\lambda\rangle+\sup_{\lambda\in \hat{{\cal D}}}\langle b,\lambda\rangle\\
&\leq \sup_{\lambda\in \hat{{\cal D}}}\langle \hat{b}-b,\Pi^u(\lambda)\rangle+\sup_{\lambda\in \hat{{\cal D}}} M\|A^\top \lambda\|\\
&=\sup_{\lambda\in \hat{{\cal D}}}\langle \hat{b}-b,\Pi^u(\lambda)\rangle+\sup_{\lambda\in \hat{{\cal D}}} M\|(A-\hat{A})^\top \Pi^u(\lambda)\|\\
&\leq CM_N/\sqrt{N}=o\bigl(\kappa_N/\sqrt{N}\bigr),
\end{aligned}
\end{equation}
where we have used equations \ref{eqnPSSLSUC04} and \ref{eqnPSSLSUC013}, and the fact that $M_N=o(\kappa_N)$.

Now on $E_N(Q)$ and for $\xi \in {\cal X}_N$, we have
\begin{equation}
\label{eqnPSSLSUC015}
\begin{aligned}
\inf_{\lambda\in \Delta_{0,N}(\xi)}\hat{b}^\top \hat{\Gamma}(\lambda)&\geq \inf_{\lambda\in \Delta_{0,N}(\xi)} \langle \hat{b}-b_N(\xi), \Pi^u(\hat{\Gamma}(\lambda))\rangle+ \inf_{\lambda\in \Delta_{0,N}(\xi)} \langle b_N(\xi), \hat{\Gamma}(\lambda)\rangle\\
&\geq \inf_{\lambda\in \Delta_{0,N}(\xi)} \langle \hat{b}-b_N(\xi), \Pi^u(\hat{\Gamma}(\lambda))\rangle\\
&\geq -CM_N/\sqrt{N}
\end{aligned}
\end{equation}
where we have used the fact that $\langle b_N(\xi), \hat{\Gamma}(\lambda)\rangle\geq 0$ for all $\lambda \in \Delta_{0,N}(\xi)$, as $\hat{\Gamma}(\lambda)$ is a positive scalar multiple of $\lambda$ and $\inf_{\lambda\in \Delta_{0,N}(\xi)}\langle b_N(\xi),\lambda\rangle\geq 0$ (since 0 is always feasible). Combining equations \ref{eqnPSSLSUC014} and \ref{eqnPSSLSUC015}, and the fact that $M_N=o(\kappa_N)$, implies that on $E_N(Q)$ and for all sufficiently large N we have
\[
\inf_{\lambda\in \Delta_{0,N}(\xi)}\hat{b}^\top \hat{\Gamma}(\lambda)\geq \hat{T}_N-\kappa_N/\sqrt{N},
\]
thus completing the proof of Step 1.

\underline{Step 2} \quad In this step, we use equation~\eqref{eqnPSSLSUC03} and a coupling argument to upper bound the test statistic by another statistic whose distribution can be uniformly estimated for all $Q \in \mathcal{Q}_0$. The proof closely parallels Step 2 in the proof of Theorem~\ref{thmAVKAM}. Let $\hat{\xi}^*=(\sqrt{N}(\hat{A}^*-A(Q)),\sqrt{N}(\hat{b}^*-b(Q)),\sqrt{N}(\hat{\Omega}^*-\Omega(Q)))$ be an independent and identically distributed version of the root $\hat{\xi}=(\sqrt{N}(\hat{A}-A(Q)),\sqrt{N}(\hat{b}-b(Q)),\sqrt{N}(\hat{\Omega}-\Omega(Q)))$, computed from a sample $\mathbb{W}_N^*$, such that $\mathbb{W}_N^*$ and $\mathbb{W}_N$ are independent and identically distributed. For $Q\in {\cal Q}$, let the event $F^*_N=F^*_N(Q)$ be defined by 
\[
F_N^*(Q) := \left\{ \mathbb{W}^* \;\middle|\; \|\hat{\xi}^*_A\|>M_N \;\text{or}\; \|\hat{\xi}^*_b\| > M_N \;\text{or}\; \|\hat{\xi}^*_{\Omega}\| > \sqrt{N} \alpha_N \right\}.
\]
By our choice of $M_N$ and $\alpha_N$, we have 
\[
\delta_{1N}:=\sup_{Q\in {\cal Q}} Q\bigl(F^*_N(Q)\bigr) \quad \text{satisfies}\quad \lim_{N\to \infty} \delta_{1N}=0
\]
Equations \ref{eqnPSSLSUC02} and \ref{eqnPSSLSUC03} then imply that for all sufficiently large $N$ and on the event $E_N(Q)$, for $Q\in {\cal Q}_0$, we have
\begin{equation}
\label{eqnPSSLSUC016}
\sqrt{N}\phi(\mu_N(\hat{\xi}^*))\leq \sup_{\lambda \in \hat{\Delta}}\langle \hat{\xi}^*_b - \hat{\xi}^*_A \hat{\eta}, \lambda \rangle+\beta_N+\infty\cdot \mathbbm{1}_{\{F_N^*\}}.
\end{equation}
Let \( J_N(\cdot; Q) \) denote the CDF of \( \sqrt{N} \hat{T}_N \). Define
\[
\hat{G}_N(\cdot; Q) := \mathcal{L} \left( \sup_{\lambda \in \hat{\Delta}} \langle \hat{\xi}^*_b - \hat{\xi}^*_A \hat{\eta}, \lambda \rangle + \beta_N + \infty \cdot \mathbbm{1}_{\{F_N^*\}} \,\middle|\, \mathbb{W}_N \right)
\]
as the conditional (on \( \mathbb{W}_N \)) CDF of the right-hand side of inequality \eqref{eqnPSSLSUC016}, viewed as a random element taking values in \( \mathbb{R} \cup \{\infty\} \).  
Similarly, define
\[
\hat{G}_\infty(\cdot; Q) := \mathcal{L} \left( \sup_{\lambda \in \hat{\Delta}} \langle \zeta_b-\zeta_A\eta_Q, \lambda \rangle \,\middle|\, \mathbb{W}_N \right),
\]
where \( \zeta \sim N(0, \Sigma(Q)) \) (as in Assumption \ref{assAVEAM1}).\footnote{Here, $\zeta_A$ denotes the matrix formed by extracting the components corresponding to $A$ from the random vector $\zeta$, arranged in a manner compatible with the dimensions of $A$. $\zeta_b$ is defined analogously.} And let $\hat{H}_N(;Q)$ be defined as in Theorem \ref{thmAVEAM}
\[
\hat{H}_N(\cdot; Q) := \mathcal{L} \left( \sup_{\lambda \in \hat{\Delta}} \langle \zeta^*_b - \zeta^*_A \hat{\eta}, \lambda \rangle \,\middle|\, \mathbb{W}_N \right),
\]
where $\zeta^*$ is the bootstrap estimate of of $\zeta$ that is given in Assumption \ref{assAVEAM3}.
Note that Assumptions \ref{assAVEAM0} and \ref{assAVEAM2} (part~(i)) imply that the random variables $\bigl\{\|\zeta\|\mid \zeta\sim \mathcal{N}(0,\Sigma(Q)),\ Q\in {\cal Q}_0\bigr\}$ are tight. As a consequence, Strassen's coupling, and equations \ref{eqnPSSLSUC04} and \ref{eqnPSSLSUC09}, imply that the sequence
\[
\delta_{2N} = \sup_{Q \in \mathcal{Q}_0} \sup_{\mathbb{W}_N \in E_N(Q)} 
d_{\text{Pr}}\left(
\mathcal{L} \left( \sup_{\lambda \in \hat{\Delta}} \langle \zeta_b - \zeta_A \eta_Q, \lambda \rangle \,\middle|\, \mathbb{W}_N \right),
\mathcal{L} \left( \sup_{\lambda \in \hat{\Delta}} \langle \hat{\xi}^*_b - \hat{\xi}^*_A \hat{\eta}, \lambda \rangle \,\middle|\, \mathbb{W}_N \right)
\right)
\]
satisfies
\[
\lim_{N\to \infty} \delta_{2N}=0.
\]
The preceding equation, together with Strassen's theorem, implies that on the event \( E_N(Q) \), we have
\begin{equation*}
\begin{aligned}
d_{\text{Pr}}\Big( \hat{G}_N, \hat{G}_{\infty} \Big) 
&\leq d_{\text{Pr}}\Bigg( \hat{G}_N, \mathcal{L}\left( \sup_{\lambda \in \hat{\Delta}} \langle \hat{\xi}^*_b - \hat{\xi}^*_A \hat{\eta}, \lambda \rangle \,\middle|\, \mathbb{W}_N \right) \Bigg) \\
&\quad + d_{\text{Pr}}\Bigg( \hat{G}_{\infty}, \mathcal{L}\left( \sup_{\lambda \in \hat{\Delta}} \langle \hat{\xi}^*_b - \hat{\xi}^*_A \hat{\eta}, \lambda \rangle \,\middle|\, \mathbb{W}_N \right) \Bigg)\\
&\leq \beta_N+Q(F_N^*)+\delta_{2N}.
\end{aligned}
\end{equation*}
Similarly, by Strassen's theorem and equations~\ref{eqnPSSLSUC04} and~\ref{eqnPSSLSUC09}, it follows that on the event \( E_N(Q) \), we have
\[
\lim_{N\to \infty} \delta_{3N}=0,
\]
where the sequence $\delta_{3N}$ is defined by 
\[
\delta_{3N} = \sup_{Q \in \mathcal{Q}_0} \sup_{\mathbb{W}_N \in E_N(Q)} 
d_{\text{Pr}}\left(
\mathcal{L} \left( \sup_{\lambda \in \hat{\Delta}} \langle \zeta_b - \zeta_A \eta_Q, \lambda \rangle \,\middle|\, \mathbb{W}_N \right),
\mathcal{L} \left( \sup_{\lambda \in \hat{\Delta}} \langle \zeta^*_b - \zeta^*_A \hat{\eta}, \lambda \rangle \,\middle|\, \mathbb{W}_N \right)
\right).
\]
We have
\[
d_{\text{Pr}}\left( \hat{G}_{\infty}, \hat{H}_N \right) \leq \delta_{3N}
\]
on $E_N(Q)$.

In conclusion, there exists $\gamma_N=o(1)$ such that for all sufficiently large N and all $Q \in {\cal Q}_0$, on the event $E_N(Q)$ we have
\begin{equation}
d_{\text{Pr}}\bigl(\hat{G}_N,\hat{G}_{\infty}\bigr)\vee d_{\text{Pr}}\bigl(\hat{H}_N,\hat{G}_{\infty}\bigr)\leq \gamma_N.
\end{equation}
Let the significance level \( \alpha \) and the critical value \( \hat{c}_N(1-\alpha) \) be as in Theorem~\ref{thmAVEAM}.
For now, assume that there exists a constant \( C \) such that for all sufficiently large \( N \), for all \( Q \in \mathcal{Q}_0 \), and for all sample realizations \( \mathbb{W}_N \) on the event \( E_N(Q) \cap \{ \sqrt{N} \hat{T}_N > \hat{c}_N(1-\alpha) \} \), we have the anti-concentration condition
\begin{equation}
\label{eqnPSSLSUC017}
\hat{G}_{\infty}\big( \hat{c}_N(1-\alpha) + \gamma_N \big) - \hat{G}_{\infty}\big( \hat{c}_N(1-\alpha) - \gamma_N \big) \leq C \gamma_N.
\end{equation}
We prove in the next step that inequality~\eqref{eqnPSSLSUC017} indeed holds. For \( \mathbb{W}_N \in E_N(Q) \cap \{ \sqrt{N} \hat{T}_N > \hat{c}_N(1-\alpha) \} \), and $Q\in {\cal Q}_0$, arguing as in \ref{eqnPSSARP9}, we have
\begin{equation}
\label{eqnPSSLSUC018}
J_N\big( \hat{c}_N(1-\alpha) \big) \geq 1-\alpha-C\gamma_N.
\end{equation}
 Let $q_{N}(\cdot)$ denote the quantiles of $J_N(\cdot)$. Inequality \ref{eqnPSSLSUC018} then implies that on the event \(E_N(Q) \cap \{ \sqrt{N} \hat{T}_N > \hat{c}_N(1-\alpha) \} \), with $Q\in {\cal Q}_0$, we have
\begin{equation}
\label{eqnPSSLSUC019}
q_N(1-\alpha-C\gamma_N)\leq \hat{c}_N(1-\alpha).
\end{equation}
 Arguing as in equation \ref{eqnPSSARP11}, we have
\begin{equation}
\label{eqnPSSLSUC020}
\sup_{Q\in {\cal Q}_0}Q\Bigl(\{ \sqrt{N} \hat{T}_N > \hat{c}_N(1-\alpha) \} \Bigr)\leq \alpha+C\gamma_N+\delta_{1N}.
\end{equation}
which yields equation \ref{eqnAVEAM2}, as $\gamma_N=o(1)$ and $\delta_{1N}=o(1)$.

\medskip

\underline{Step 3} \quad In this step, we establish inequality~\ref{eqnPSSLSUC017}, which is required in the derivation of equation~\ref{eqnPSSLSUC020}. We begin by noting that for all sufficiently large \( N \) and for all \( Q \in \mathcal{Q}_0 \), equation~\ref{eqnPSSLSUC014} implies that \( 0 \in \hat{\Delta} \), as $\hat{T}_N=o(\sqrt{N}/N)$. Hence for all such sufficiently large $N$, the distributions $\hat{H}(\cdot)$ are supported on $\mathbb{R}_+$, and rejection occurs, i.e. $\{\hat{T}_N>\hat{c}_N(1-\alpha)\}$, only if $\hat{T}_N>0$.

We first consider the case where \( \Delta_0(Q) \neq \{0\} \), with
\[
\Delta_0(Q) := \arg\max \left\{ b(Q)^\top \lambda \ \middle| \ \lambda \geq 0,\ A(Q)^\top \lambda = 0,\ \|\Omega(Q)\lambda\|_1 \leq 1 \right\}.
\]
Equation \ref{eqnPSSLSUC010}, with $\xi=0$, then implies that on the event $E_N(Q)$, we have
\[
\overset{\rightarrow}{d}_H\bigl(\Pi^u(\Delta_{0}(Q)), \Pi^u(\hat{\Delta})\bigr) \leq C \alpha_N.
\]
By Assumption~\ref{assAVEAM5}, there exists \( \lambda_Q \in \Delta_0(Q) \) such that
\[
\lambda_Q^\top V(Q) \lambda_Q \geq \rho.
\]
This, combined with inequality the preceding inequality, implies that there exists \( \hat{\lambda} \in \hat{\Delta} \) such that
\[
\left\| \Pi^u(\hat{\lambda}) - \Pi^u(\lambda_Q) \right\| \leq C \alpha_N.
\]
Hence, for all sufficiently large \( N \), we have
\begin{equation}
\label{eqnPSSLSUC021}
\sup_{\lambda \in \hat{\Delta}} \lambda^\top V(Q) \lambda=\sup_{\lambda \in \hat{\Delta}} \Pi^u(\lambda)^\top V(Q)\Pi^u( \lambda) \geq  \Pi^u(\hat{\lambda})^\top V(Q)\Pi^u(\hat{ \lambda})\geq \rho / 2.
\end{equation}

We now consider the case where $\Delta_0(Q)= \{0\}$. Since for all sufficiently large N, rejection implies that $\hat{T}_N>0$, with $\hat{T}_N$ finite, there must exist $\hat{\lambda} \in \operatorname{extr}(\hat{D})\setminus\{0\}$ such that $\|\hat{\Omega}\hat{\lambda}\|_1=1$, and  
\[
\hat{\lambda}\in \arg\max \left\{ \hat{b}^\top \lambda \ \middle| \ \lambda \geq 0, \ \hat{A}^\top \lambda=0, \  \|\hat{\Omega}\lambda\|_1 \leq 1 \right\}.
\]
Note that we have $\hat{T}_N= \hat{b}^\top \hat{\lambda}$, and we clearly have that $\hat{\lambda}\in \hat{\Delta}$. By Assumption \ref{assAVEAM5}, for all $\lambda\geq 0$ such that $\|\Pi^u(\lambda)\|_1\geq 1$ we have
\begin{equation}
\label{eqnPSSLSUC022}
\lambda^\top V(Q) \lambda \geq \rho.
\end{equation}

For all sufficiently large \( N \), we eventually have \( \alpha_N \leq \overline{\sigma} \). Therefore, on the event \( E_N(Q) \), it follows that \( \|\hat{\Omega}\| \leq 2\overline{\sigma} \), which implies
\[
1 = \|\hat{\Omega} \hat{\lambda}\|_1 \leq 2\overline{\sigma} \|\Pi^u(\hat{\lambda})\|_1.
\]
Using equation~\ref{eqnPSSLSUC022}, we then obtain
\begin{equation}
\label{eqnPSSLSUC023}
\hat{\lambda}^\top V(Q) \hat{\lambda} \geq \rho \left( \frac{1}{2\overline{\sigma}} \right)^2.
\end{equation}

In conclusion, equations~\ref{eqnPSSLSUC021} and~\ref{eqnPSSLSUC023} together imply that for all sufficiently large \( N \), and for all \( Q \in \mathcal{Q}_0 \), whenever the event \( E_N(Q) \cap \{\hat{T}_N > \hat{c}_N(1-\alpha)\} \) occurs, there exists some \( \lambda \in \hat{\Delta} \) such that
\[
\lambda^\top V(Q) \lambda \geq \rho \left( \frac{1}{2\overline{\sigma}} \right)^2 \wedge \frac{\rho}{2}.
\]
Inequality~\ref{eqnPSSLSUC017} then follows from Proposition~\ref{prop1}, by applying the above inequality together with the identity
\[
\langle \zeta_b - \zeta_A \eta_Q, \lambda \rangle = \langle \Pi^u(\zeta_b - \zeta_A \eta_Q), \Pi^u(\lambda) \rangle,
\]
and equation~\ref{eqnPSSLSUC04}.

\medskip

\underline{Step 4} \quad In this step, we establish equation ~\eqref{eqnAVEAM2}. Fix $Q\in {\cal Q}_0$, and suppose that $\eta_Q$ is the unique solution to the inequality system 
\[
b(Q)-A(Q)\eta\leq 0.
\]
Lemma~\ref{lem1AD} implies that
\[
\sqrt{N}\hat{T}_N\overset{d}{\longrightarrow} \max_{\lambda\in \Delta_0} \langle \zeta_b-\zeta_A\eta_Q,\Pi^u(\lambda) \rangle,
\]
where $\Delta_0$ is the set of optimal solutions to the LP given by
\[
\Delta_0:=\bigl\{\lambda\in \mathbb{R}^p \mid \lambda \geq 0, \ \langle b, \lambda \rangle \geq 0, \  A^\top \lambda=0, \mathbbm{1}\Omega\lambda\leq 1 \bigr\}.
\]
Assumption \ref{assAVEAM3} and equation \ref{eqnPSSLSUC09} imply that
\[
d_{Pr}\Bigl(\mathcal{L}\Bigl(\zeta^*_b-\zeta^*_A\hat{\eta}\mid \mathbb{W}_N\Bigr), \mathcal{L}(\zeta_b-\zeta_A\eta_Q)\Bigr)=o_p(1).
\]
Reasoning using an almost sure representation and Lemma~\ref{lem2}, equation \ref{eqnAVEAM2} then follows if we show that 
\[
d_{H}(\Pi^u(\hat{\Delta}),\Pi^u(\Delta_0))=o_p(1).
\] 
Equation \ref{eqnPSSLSUC010}, with $\xi=0$, implies that 
\[
\overset{\rightarrow}{d}_{H}(\Pi^u(\Delta_0),\Pi^u(\hat{\Delta}))=o_p(1).
\]
It thus remains to show that 
\begin{equation}
\label{eqnPSSLSUC024}
\overset{\rightarrow}{d}_{H}(\Pi^u(\hat{\Delta}),\Pi^u(\Delta_0))=o_p(1).
\end{equation}

Using the definition of $\hat{\Delta}$ and equations \ref{eqnPSSLSUC04} and \ref{eqnPSSLSUC014}, on the event $E_N(Q)$, we have
\begin{equation}
\label{eqnPSSLSUC025}
\begin{aligned}
\sup_{\lambda\in \hat{\Delta}}\|A^\top \lambda\| &\leq \sup_{\lambda\in \hat{\Delta}}\|(A-\hat{A})^\top \Pi^u(\lambda)\|+(\kappa_N/\sqrt{N})\|\mathbbm{1}\|\\
&\leq C\kappa_N/\sqrt{N}=o(1),
\end{aligned}
\end{equation}
and 
\begin{equation}
\label{eqnPSSLSUC026}
\begin{aligned}
\sup_{\lambda\in \hat{\Delta}} (\mathbbm{1}^\top \Omega \lambda-1)_+&\leq \sup_{\lambda\in \hat{\Delta}} (\mathbbm{1}^\top (\Omega-\hat{\Omega}) \lambda)_+\\
&\leq C\alpha_N=o(1),
\end{aligned}
\end{equation}
and 
\begin{equation}
\label{eqnPSSLSUC027}
\begin{aligned}
\sup_{\lambda \in \hat{\Delta}} (b^\top \lambda)_- 
&\leq \sup_{\lambda \in \hat{\Delta}} \left[ (b - \hat{b})^\top \Pi^u(\lambda) + \hat{T}_N - \frac{\kappa_N}{\sqrt{N}} \right]_- \\
&\leq \sup_{\lambda \in \hat{\Delta}} \|b - \hat{b}\| \cdot \|\Pi^u(\lambda)\| + |\hat{T}_N| + \frac{\kappa_N}{\sqrt{N}} \\
&\leq C \frac{\kappa_N}{\sqrt{N}} = o(1).
\end{aligned}
\end{equation}
Equations \ref{eqnPSSLSUC025}, \ref{eqnPSSLSUC025}, and \ref{eqnPSSLSUC025}, in combination with Lemma \ref{lemhoff}, then imply that we have 
\[
\overset{\rightarrow}{d}_{H}(\hat{\Delta},\Delta_0)=o_p(1),
\]
which in turn implies equation \ref{eqnPSSLSUC024}. This completes the proof.
\end{proof}
\subsection{Impossibility result}
\label{subsectionImp}

Consider the setting of Section \ref{subsectionAVEAM}, i.e. in which the null hypothesis follows Eq. \eqref{eqnARP} with
\[\label{eq:nullARP}\mathcal{Q}_0 = \left\{ Q \in \mathcal{Q} \mid \ \text{there exists} \ \eta \in \mathbb{R}^d \ \text{such that} \ b(Q) \leq A(Q) \eta \right\}.\]
Our impossibility result exhibits a simple parametric class ${\cal Q}$ in which all the assumptions of Section~\ref{subsectionAVEAM} are satisfied---except for Assumption~\ref{assAVEAM4} ---and where every asymptotically uniformly valid inference procedure is trivial (i.e., has power no greater than its size).

Let $\operatorname{vec}\!\bigl([A(Q)\;\; b(Q)]\bigr) = \mathbb{E}_{Q}[W]$.  
Our data consist of an i.i.d.\ sample $\{W_i\}_{i=1}^N$,  
where $W_i \sim Q_\mu$ and $Q_\mu$ denotes the distribution  
$\mathcal{N}(\mu, I_m)$ with $m = p(d+1)$.  
Here $\mu$ is unknown, while the common covariance matrix $I_m$ is assumed to be known.

The parametric class we consider is
\begin{equation}
\label{exNormalARP}
\mathcal{Q} = \bigl\{\, Q_\mu : \|\mu\| \le C \,\bigr\},
\end{equation}
for some fixed constant $C > 0$. Let \(\tilde{{\cal Q}} \subseteq {\cal Q} \setminus {\cal Q}_0\) be defined by
\[
\tilde{{\cal Q}} = \{ P \in {\cal Q} \setminus {\cal Q}_0 \mid \{\eta \neq 0 \mid A(P)\eta \ge 0\} \neq \emptyset \}.
\]

\begin{proposition} \label{propImp}
    For the class \ref{exNormalARP}, let \(\psi_N : (\mathbb{R}^{m})^{\otimes N} \to [0,1]\) denote a sequence of tests for the null 
\(H_0 : Q \in {\cal Q}_0\) that are asymptotically uniformly valid at level \(\alpha\) (\(\alpha \in (0,1)\)), i.e.,
\begin{equation}
\label{eqnIR3}
\overline{\lim}_{N \to \infty} \sup_{Q \in {\cal Q}_0} E_{Q^{\otimes N}}[\psi_N] \le \alpha.
\end{equation}
Then we have
\begin{equation}
\label{eqnIR4}
\overline{\lim}_{N \to \infty} \sup_{P \in \tilde{{\cal Q}}} E_{P^{\otimes N}}[\psi_N] \le \alpha.
\end{equation}
In particular, if $d>p$, then it follows that any uniformly asymptotically valid test over $Q_0$ has trivial power everywhere in the alternative, as the sets $\{\eta \mid \eta\neq 0, \ A(Q)\eta\geq 0\}$ are always non-empty, and we have $\tilde{{\cal Q}}={\cal Q}\setminus {\cal Q}_0$. \footnote{If \({\cal Q}_0\) is replaced by \(\{Q \in {\cal Q} \mid \text{there exists } \eta \geq 0, \text{ such that } A(Q)\eta = b(Q)\}\), then a similar argument shows that the analogous set of alternatives $\tilde{\cal Q}$ where uniform tests have trivial power consists of those \(P\) such that \(\{\eta \ge 0 \mid \eta\neq 0, \ A(P)\eta = 0\} \neq \emptyset\). Note however, that in this case, for any values of the dimensions p and d, we have now  $\tilde{{\cal Q}}\subsetneq {\cal Q}\setminus {\cal Q}_0$. } 
\end{proposition}

\noindent
It is straightforward to verify that Assumptions~\ref{assAVEAM1} and \ref{assAVEAM3} are satisfied with $\hat c$ taken as the sample average of the $W_i$’s and $\zeta_N^*$ obtained via either the parametric or nonparametric bootstrap. Assumption~\ref{assAVEAM2} is trivially satisfied because the covariance matrix is known to equal the identity. Assumption \ref{assAVEAM0} concerns the dimensions of the problem, and is satisfied if p and d do not grow with N. Assumption \ref{assAVEAM5} can be dispensed with by modifying the rejection rule. Assumption \ref{assAVEAM4} is violated here since the DGPs that we consider allow for the set of matrices \{$A(Q)\mid Q\in {\cal Q}_0\}$ to contain a neighborhood of the zero matrix, and $\Lambda(A)$ is unbounded on such neighborhoods.

Proposition \ref{propImp} implies that for any test to control size uniformly but have non-trivial power even in the simple case of the class \ref{exNormalARP}, it must restrict $\mathcal{Q}$ in some further way, e.g. by restricting the allowable $A(Q)$ or the dimensions $d,p$ of the problem. The introduction of Assumption \ref{assAVEAM4} is one way to accomplish this with a high degree of generality, though alternative restrictions may be natural in particular settings.

Note that Proposition \ref{propImp} admits a sort of converse, as the set $\tilde{\cal Q}$ exactly consists of all alternatives that cannot be separated from the null. That is, if \(Q_{\mu} \in {\cal Q} \setminus {\cal Q}_0\) is such that there exists a sequence 
\(\mu_N \to \mu\) with $Q_{\mu_N}\in {\cal Q}_0$, then $Q_{\mu}\in \tilde{\cal Q}$. To see this, consider a sequence \(\{\eta_N\}\) that satisfies \(b(Q_{\mu_N}) \le A(Q_{\mu_N}) \eta_N\). We claim that the sequence \(\{\eta_N\}\) must be unbounded.  
Otherwise, after passing to a convergent subsequence \(\eta_{N_k} \to \eta\) and taking limits, we would obtain
\(b(Q_{\mu}) \le A(Q_{\mu}) \eta\), contradicting the assumption that 
\(Q_{\mu} \in {\cal Q} \setminus {\cal Q}_0\). Since \(\{\eta_N\}\) is unbounded, divide the inequalities
\(b(Q_{\mu_N}) \le A(Q_{\mu_N}) \eta_N\) by \(\|\eta_N\|\) and consider a subsequence of
\(\{\eta_N / \|\eta_N\|\}\) converging to some \(\eta\) with \(\|\eta\| = 1\).  
Passing to the limit yields \(A(Q_{\mu}) \eta \ge 0\), establishing that $Q_{\mu}\in \tilde{{\cal Q}}$ as desired.

The proof of Proposition \ref{propImp} is based on a general argument of \citealt{BS56}. We show that for any $P\in \tilde{{\cal Q}}$ and $\epsilon>0$, there exists a distribution \(Q_{\epsilon} \in {\cal Q}_0\) such that \(Q_{\epsilon}^{\otimes N}\) is \(\epsilon\)-close to \(P^{\otimes N}\) in total variation. It follows that if a level-\(\alpha\) test achieves asymptotic uniform size control over \({\cal Q}_0\), then asymptotically, its power over $\tilde{{\cal Q}}$ must be uniformly bounded by $\alpha$. When \(d > p\), $\tilde{{\cal Q}}$ coincides with the set of all alternatives $\tilde{Q}\setminus\tilde{Q}_0$, so any test with uniform size control has asymptotically trivial power and performs no better than a test that rejects with probability \(\alpha\) without using the data.

\begin{proof}
Let \(\{P_N \mid N \ge 1\}\) denote a sequence of DGPs in $\tilde{{\cal Q}}$, and let \(\{\epsilon_N\}_{N \ge 1}\) be an arbitrary sequence of positive numbers converging to zero.

We first show that for each \(N \ge 1\), there exists \(Q_N \in {\cal Q}_0\) such that 
\begin{equation}
\label{eqnIR1}
d_{TV}(P_N^{\otimes N}, Q_N^{\otimes N}) \le \epsilon_N.\footnote{Here $d_{TV}(P,Q)=\sup\{|P(A)-Q(A)|\mid \text{over all measurable sets A}\}$.}
\end{equation}
Toward that end, for $P\in \tilde{{\cal Q}}$ and $\delta>0$, we first show that there exists a perturbation $A_{\delta}$ of the matrix $A$ such that \footnote{Here $\|A\|_F$ denotes the Frobenius norm of A.}
\begin{equation}
\label{eqnIR2}
\|A(P) - A_{\delta}\|_F \le \delta \quad \text{and} \quad \{\eta \mid \ b(P) \le A_{\delta} \eta\} \neq \emptyset.
\end{equation}
Hence, if \(\mu_{\delta} = \mathrm{vec}([A_{\delta} \ b(P)])\), then \(Q_{\mu_{\delta}} \in {\cal Q}_0\).  
By definition of $\tilde{{\cal Q}}$, there exists $\eta_P\neq 0$ such that $ A(P)\eta_P\geq 0$. We can assume without loss of generality that the first component of \(\eta_P\), denoted \(\eta_{P,1}\), is nonzero. Let \(B\) be a matrix of the same dimension as \(A(P)\), with its first column equal to \(\mathrm{sgn}(\eta_{P,1}) \, b(P)\) and all other columns equal to zero. Let \(t^* > 0\) be a small constant such that \(\|t^* B\|_F \le \delta\), and set \(A_{\delta} = A(P) + t^* B\). Let \(\tilde{\eta} = \eta_P / (t^* |\eta_{P,1}|)\).  
Then one can easily check that \(b(P) \le A_{\delta} \tilde{\eta}\) and \(\|A_{\delta} - A(P)\|_F \le \delta\), proving \eqref{eqnIR2}. 

Now, for any \(Q_{\mu}\) and \(Q_{\mu'}\) in \({\cal Q}\), and any \(N \ge 1\), we have (see p.~25 of \citealt{cam2000asymptotics}) \footnote{Here 
\(d_H^2(P,Q) := \tfrac{1}{2} \int (\sqrt{p} - \sqrt{q})^{2} \, d\mu\),
where \(p\) and \(q\) are the respective densities of \(P\) and \(Q\) with respect to a dominating measure \(\mu\).}

\[
d_{TV}^2(Q_{\mu}^{\otimes N}, Q_{\mu'}^{\otimes N}) \le 2 \Bigl( 1 - (1 - d_H^2(Q_{\mu}, Q_{\mu'}))^N \Bigr),
\]
and (see p.~51 of \citealt{pardollorente})
\[
d_H^2(Q_{\mu}, Q_{\mu'}) = 1 - \exp\Bigl(-\frac{1}{8} \|\mu - \mu'\|^2\Bigr),
\]
where we have used the fact that the covariance matrix for any $Q_{\mu}\in {\cal Q}$ is the identity.
Therefore, for any $N\geq 1$, $P_N$, and $\epsilon_N>0$, if $Q_N$ is constructed as in equation \ref{eqnIR2}, with $\delta_N$ sufficiently small that 
\[
2(1-e^{-\frac{N}{8}\delta_N^2})\leq \epsilon_N^2,
\]
then $d_{TV}(P_N^{\otimes N},Q_N^{\otimes N})\leq \epsilon_N$. This establishes equation \ref{eqnIR1}.

To derive Equation \ref{eqnIR4}, note that if \(\{P_N\}\) is any sequence of DGPs in \(\tilde{{\cal Q}}\), then \eqref{eqnIR1} with \(\epsilon_N = 1/N\) implies that there exists a sequence \(\{Q_N\} \subset {\cal Q}_0\) such that \(d_{TV}(P_N^{\otimes N}, Q_N^{\otimes N}) \to 0\). Since \(\psi_N\) are bounded in absolute value by 1, it follows that
\[
\lim_{N \to \infty} \big| E_{P_N^{\otimes N}}[\psi_N] - E_{Q_N^{\otimes N}}[\psi_N] \big| = 0,
\]
which when combined with Equation \ref{eqnIR3} establishes \eqref{eqnIR4}.
\end{proof}

\section{Supporting Propositions and Lemmas}

The following proposition builds on Theorems 1 and 2 of \cite{VST} to derive an anti-concentration result, which is useful for establishing our uniformity results and may be of independent interest.
\begin{proposition}[Anti-concentration]
\label{prop1}
Let $T$ be an arbitrary index set. Let $\{K_t \mid t \in T\}$ be a family of compact subsets of $\mathbb{R}^p$ $(K_t \subseteq \mathbb{R}^p)$, and let $\{Z_t \mid t \in T\}$ be a family of mean-zero Gaussian vectors in $\mathbb{R}^p$, with $Z_t \sim \mathcal{N}(0, \Sigma_t)$. Suppose there exist strictly positive constants $M$ and $\sigma$ such that
\begin{equation}
\label{eqnprop1}
\sup_{t \in T} \|\Sigma_t\| \leq M \quad \text{and} \quad \sup_{t \in T} \max_{\lambda \in K_t} \|\lambda\| \leq M,
\end{equation}
and that for all $t \in T$, there exists $\lambda_t \in K_t$ such that
\begin{equation}
\label{eqnprop2}
\lambda_t^\top \Sigma_t \lambda_t > \sigma^2.
\end{equation}
Define random variables $\{Y_t\}_{t \in T}$ by
\[
Y_t = \max_{\lambda \in K_t} Z_t^\top \lambda.
\]
Then, for any $\delta > 0$, the distributions $\{F_t\}_{t \in T}$ of the random variables $\{Y_t\}_{t \in T}$ are absolutely continuous on the interval $[\delta, +\infty)$ and possess densities $\{f_t\}_{t \in T}$ that are uniformly bounded on $[\delta, +\infty)$; that is,
\begin{equation}
\label{eqnprop3}
\sup_{t \in T} \sup_{x \in [\delta, +\infty)} f_t(x) < \infty.
\end{equation}
As a consequence, if $q_t(\cdot)$ denotes the quantile function of $F_t$, then for any $\overline{\alpha} \in (0, 1/2)$, there exists $\underline{x} > 0$ such that
\begin{equation}
\label{eqnprop4}
\inf_{\alpha \in (0, \overline{\alpha})} \inf_{t \in T} q_t(1-\alpha) > 2\underline{x}, \quad \text{and}\quad \sup_{t\in T}\sup_{x\in [\underline{x},+\infty)}f_t(x)<\infty.
\end{equation}

\end{proposition}

\begin{proof}[\bf{Proof of Proposition \ref{prop1}}]
We proceed in two steps. In the first step, we establish inequality \eqref{eqnprop3} using \eqref{eqnprop1} and Theorem 2 of \cite{VST}. In the second step, we use \eqref{eqnprop2} and \eqref{eqnprop3} to establish inequality \eqref{eqnprop4}.

\medskip
\underline{Step 1: Proof of \eqref{eqnprop3}.}  
Given $\delta > 0$ as in the statement of the proposition, let $C_{\delta}>0$ be a constant satisfying $M C_{\delta} < \delta/2$.
We first show that inequality \eqref{eqnprop2} implies
\begin{equation}
\label{eqnpprop1}
\inf_{t \in T} P(\|Z_t\| < C_{\delta}) > \eta
\end{equation}
for some $\eta > 0$. Indeed, since $Z_t \sim \Sigma_t^{1/2} Z$, where $Z \sim \mathcal{N}(0, \mathbb{I}_p)$ is a standard normal vector in $\mathbb{R}^p$, we have
\[
\inf_{t \in T} P(\|Z_t\| < C_{\delta}) \geq P(\|Z\| < C_{\delta}/\sqrt{M}),
\]
where $M$ is as in \eqref{eqnprop1}, and we have used the fact that $\|\Sigma_t^{1/2}\| = \|\Sigma_t\|^{1/2} \leq M^{1/2}$, as $\Sigma_t$ is symmetric and positive semidefinite. We can thus set $\eta = P(\|Z\| < C_{\delta}/\sqrt{M})$ in \eqref{eqnpprop1}.

Now, by inequality \eqref{eqnpprop1} and the definition of $Y_t$, it follows that
\begin{equation}
\label{eqnpprop2}
F_t(\delta/2) = P(Y_t \leq \delta/2) \geq P(|Y_t|\leq \delta/2)\geq P(\|Z_t\| \leq C_{\delta}) \geq \eta.
\end{equation}
Let $\Phi(\cdot)$ denote the CDF of a standard normal variable, and for each $t \in T$, define $\tau_t$ by
\[
\tau_t = \Phi^{-1}(F_t(\delta/2)).
\]
Assuming without loss of generality that $\eta < 1/2$, inequality \eqref{eqnpprop2} yields
\begin{equation}
\label{eqnpprop3}
(\tau_t)_{-} := \max\{-\tau_t, 0\} \leq -\Phi^{-1}(\eta).
\end{equation}

Applying Theorem 2 of \cite{VST}, for each $t \in T$ and for $x \in [\delta, +\infty)$, we obtain
\[
f_t(x) \leq \bigl( (\tau_t)_- + 2 \bigr)^2 \frac{x}{(x-\delta/2)^2} + \bigl( (\tau_t)_- + 2 \bigr) \frac{1}{x-\delta/2}.
\]
Combining this with \eqref{eqnpprop3}, we deduce that
\[
\sup_{t \in T} \sup_{x \in [\delta, +\infty)} f_t(x) \leq \left( -\Phi^{-1}(\eta) + 2 \right)^2 \frac{4}{\delta} + \left( -\Phi^{-1}(\eta) + 2 \right) \frac{2}{\delta} < \infty,
\]
which establishes \eqref{eqnprop3}.

\medskip
\underline{Step 2: Proof of \eqref{eqnprop4}.}  
We now prove inequality \eqref{eqnprop4}. Observe that by the definition of $Y_t$, it stochastically dominates $\lambda_t^\top Z_t$, where $\lambda_t \in K_t$ is as in the statement of the proposition. Hence, the $1-\alpha$ quantiles of $Y_t$ are at least as large as those of $\lambda_t^\top Z_t$.

Moreover, for each $\alpha \in (0,1/2)$, inequality \eqref{eqnprop2} implies that the $1-\alpha$ quantiles of $\lambda_t^\top Z_t$ are bounded below by those of a $\mathcal{N}(0, \sigma^2)$ distribution, with $\sigma$ as in \eqref{eqnprop2}. Therefore,
\[
\inf_{\alpha \in (0, \overline{\alpha})} \inf_{t \in T} q_t(1-\alpha) \geq \inf_{t \in T} q_t(1-\overline{\alpha})\geq  \sigma \Phi^{-1}(1-\overline{\alpha}) := 2\underline{x}.
\]
The second part of \eqref{eqnprop4} then follows from inequality \eqref{eqnprop3}, with $\delta = \underline{x}$.
\end{proof}
The following two lemmas are useful in deriving the (pointwise) asymptotic distributions of our test statistics. The first lemma gives the asymptotic distribution of the test statistics in Theorem \ref{thmAVKAM}, while the second lemma is used to derive the asymptotic distribution of the test statistic in \ref{thmAVEAM}. Both are variations on the results of \cite{AS}.
\begin{lemma}
\label{lemAD}
Consider the linear program $v = \max\{ b^\top \lambda \mid A^\top \lambda \leq 0,\ \|D\lambda\|_1 \leq 1 \}$. Let $\hat{b}$ and $\hat{D}$ denote estimators of $b$ and $D$, respectively, such that $\sqrt{N}(\hat{b} - b) \Rightarrow \zeta^b$ and $\hat{D} - D = o_p(1)$. Let \( K \) and \( U \) form a partition of \([p]\), where \( p \) denotes the dimension of \( b \). We allow \( K \) to be empty, but require \( U \) to be nonempty. The set \( K \) corresponds to the indices of the deterministic or known components of \( b \), and we assume that
\begin{equation}
\label{eqnlemAD0}
\forall i \in K, \quad \hat{b}_i = b_i \quad \text{and} \quad \hat{D}_{ii} = D_{ii} = 0, 
\qquad \text{and} \qquad \exists\, \underline{\sigma} > 0 \;\; \text{s.t.} \;\; D_{ii} > \underline{\sigma} \quad \forall i \in U.
\end{equation}

 Define $\Delta_0 := \arg\max\{ b^\top \lambda \mid A^\top \lambda \leq 0,\ \|D\lambda\|_1 \leq 1 \}$. If $v = 0$, then
\begin{equation}
\label{eqnlemAD1}
\sqrt{N}(\hat{v} - v)=\sqrt{N}\hat{v} \overset{d}{\longrightarrow} \max_{\lambda \in \Delta_0} \langle \zeta^b, \lambda \rangle.
\end{equation}
Here $\hat{v} = \max\{ \hat{b}^\top \lambda \mid A^\top \lambda \leq 0,\ \|\hat{D}\lambda\|_1 \leq 1 \}$.
\end{lemma}
\begin{proof}[\bf{Proof of Lemma \ref{lemAD}}] 
Let $\hat{\zeta}=(\sqrt{N}(\hat{b}-b),\hat{D}-D)$. The hypothesis of the theorem implies that $\hat{\zeta}\overset{d}{\rightarrow} \zeta:=(\zeta^b,0)$. By the Skorokhod representation theorem, there exists a sufficiently rich probability space, and versions $\xi_N \overset{d}{=}\hat{\zeta}$ and $\xi:=(\xi^b,0)\overset{d}{=}\zeta$ such that $a.s.-\lim_{N\to \infty} \xi_N=\xi$. Using the notation of Section~\ref{pthmAVKAM}, let $\mu=(b,D)$ and $\mu_N(\xi_N)=(b_N(\xi_N),D_N(\xi_N))=(b+\frac{\xi_N^b}{\sqrt{N}}, D+\xi_N^D)$. We have $v=\phi(\mu)$ and $\hat{v}\overset{d}{=}\phi(\mu_N(\xi_N))$. For $\lambda\in \mathbb{R}^p$, let $\Pi^u:\mathbb{R}^p\rightarrow \mathbb{R}^p$ be the projection defined by
\[
[\Pi^u(\lambda)]_i=\lambda_i \quad \forall i\in U \quad \quad \text{and} \quad \quad [\Pi^u(\lambda)]_i=0 \quad \forall i \in K,
\]
where $K$ and $U$ are as in equation \ref{eqnlemAD0}. That is, $\Pi^u(\cdot)$ is the projection on the subspace spanned by the cannonical basis vectors with indices in $U$. Condition \ref{eqnlemAD0} then implies that we necessarily have (with prob. 1)
\[
[\xi^b]_i=[\xi_N^b]_i=0 \quad \forall i \in K \quad \quad \text{and}\quad \quad [\xi_N^D]_{ii}=0\quad \forall i\in K.
\]

We show below, using an argument similar to that of Theorem 17 in \cite{YG}, that 
\[
a.s.-\lim_{N\to\infty}\sqrt{N}\phi(\mu_N(\xi_N))=\max_{\lambda\in\Delta_0}\langle \lambda,\xi^b\rangle
\]
from which the conclusion of the lemma follows.

Note that the value of our LP has the following equivalent representations
\begin{equation}
\label{eqnPlemAD0}
\begin{aligned}
v&=\max\{b^\top\lambda \mid A^\top \lambda\leq 0,\ D\lambda =u-v, \ \mathbbm{1}^\top(u+v)\leq 1, \ u,v\geq 0\}\\
&=\min\{x\mid x\geq 0, \ z\geq 0, \ Az-b=Dy, \ \|y\|_{\infty}\leq x\}.
\end{aligned}
\end{equation}
Let $L(\lambda,u,v,x,y,z)$ denote the Lagrangian associated with the unperturbed LP, and given by
\[
L(\lambda,u,v,x,y,z)=\langle \lambda,b\rangle+y^\top(D\lambda-u+v)+x(1-\mathbbm{1}^\top (u+v))-z^\top A^\top \lambda.
\]
Similarly, let $L_N(\lambda,u,v,x,y,z;\xi_N)$ denote the Lagrangian associated with the perturbed LP, and defined by
\[
L_N(\lambda,u,v,x,y,z;\xi_N)=\langle \lambda,b_N(\xi_N)\rangle+y^\top(D_N(\xi_N)\lambda-u+v)+x(1-\mathbbm{1}^\top (u+v))-z^\top A^\top \lambda.
\]
Let $(\lambda_0, u_0, v_0, x_0, y_0, z_0)$ (resp. $(\lambda(\xi_N), u(\xi_N), v(\xi_N), x(\xi_N), y(\xi_N), z(\xi_N))$) denote the optimal primal and dual solutions to the unperturbed (resp. perturbed) linear programs. Then, by the saddle point property of the Lagrangian, and by noting that the condition $v=0$ impilies that $x_0=0$ and $y_0=0$, we have
\begin{equation}
\label{eqnPlemAD1}
\begin{aligned}
\sqrt{N}\phi(\mu_N(\xi_N))&=\sqrt{N}\Bigl(L(\lambda(\xi_N), u(\xi_N), v(\xi_N), x(\xi_N), y(\xi_N), z(\xi_N);\xi_N)-L(\lambda_0, u_0, v_0, x_0, y_0, z_0)\Bigr)\\
&\leq \sqrt{N}\Bigl(L(\lambda(\xi_N), u(\xi_N), v(\xi_N), x_0, y_0, z_0;\xi_N))-L(\lambda(\xi_N), u(\xi_N), v(\xi_N), x_0, y_0, z_0)\Bigr)\\
&=\langle \xi_N^b,\lambda(\xi_N)\rangle=\langle \xi_N^b,\Pi^u(\lambda(\xi_N))\rangle .
\end{aligned}
\end{equation}

We now work toward establishing equation \ref{eqnPlemAD3}. To that end, we want to show that
\begin{equation}
\label{eqnPlemAD20}
\overset{\rightarrow}{d}_H\bigl(\Pi^u(\mathcal{\Delta}_{0,N}(\xi_N)),\Pi^u(\Delta_0)\bigr)\overset{a.s.}{\rightarrow}0,
\end{equation}
where 
\[
\mathcal{\Delta}_{0,N}(\xi_N):=\arg\max\{ b_N(\xi_N)^\top \lambda \mid A^\top \lambda \leq 0,\ \|D_N(\xi_N)\lambda\|_1 \leq 1 \}.
\]
Note that, almost surely, for all sufficiently large $N$ such that $\|\xi_N^D\|\leq \underline{\sigma}/2$, we have 
\[
\|D_N(\xi_N)\lambda\|_1=\|\Pi^u(D_N(\xi_N)\lambda)\|_1\geq (\underline{\sigma}/2)\|\Pi^u(\lambda)\|_1.
\]
with
\[
\mathcal{D}_N(\xi_N):=\bigl\{\lambda \in \mathbb{R}^p \mid A^\top \lambda\leq 0, \ \|D_N(\xi_N)\lambda\|_1\leq 1\bigr\}.
\]
As a consequence, we have
\begin{equation}
\label{eqnPlemAD02}
a.s.-\overline{\lim}_{N\to \infty}  \sup_{\lambda \in {\cal D}_N(\xi_N)} \|\Pi^u(\lambda)\|_1\leq 2/\underline{\sigma}.
\end{equation}
A similar argument gives
\begin{equation}
\label{eqnPlemAD002}
 \sup_{\lambda \in {\cal D}} \|\Pi^u(\lambda)\|_1\leq 1/\underline{\sigma}.
\end{equation}

Since, the smallest nonzero diagonal element of $D$ is bounded away from zero, for all N sufficiently large that $\|\xi_N^D\|< \underline{\sigma}$, we have
\begin{equation}
\label{eqnPlemAD21}
\bigl(1-\|\xi_N^D\|/\underline{\sigma}\bigr)\|D \lambda\|_1\leq \|D_N(\xi_N)\lambda\|_1\leq (1+\|\xi_N^D\|/\underline{\sigma})\|D \lambda\|_1.
\end{equation}
Therefore, as the feasible regions are star-shaped w.r.t the origin, if $\lambda \in \Delta_{0,N}(\xi_N)$, then 
\[
\Gamma_N(\lambda):=(1-\|\xi_N^D\|/\underline{\sigma})\lambda\in {\cal D} \quad \quad \text{and}\quad \quad  \|\Pi^u(\lambda)-\Pi^u(\Gamma_N(\lambda))\|\leq  \|\Pi^u(\lambda)\|\times \|\xi_N^D\|/\underline{\sigma}.
\]
The latter inequality and equation \ref{eqnPlemAD02} then yields
\begin{equation}
\label{eqnPlemAD0021}
\overset{\rightarrow}{d}_H(\Pi^u(\Delta_{0,N}(\xi_N)), \Pi^u(\Gamma_N(\Delta_{0,N}(\xi_N))))\overset{a.s.}{\longrightarrow} 0.
\end{equation}

We have
\begin{equation}
\label{eqnPlemAD22}
\begin{aligned}
0\geq \sup_{\lambda \in \Delta_{0,N}(\xi_N)}\langle b, \Gamma_N(\lambda)\rangle&=\inf_{\lambda \in \Delta_{0,N}(\xi_N)}\langle b, \Gamma_N(\lambda)\rangle\\
&=\inf_{\lambda \in \Delta_{0,N}(\xi_N)}\langle b-b_N(\xi_N), \Gamma_N(\lambda)\rangle+\langle b_N(\xi_N), \Gamma_N(\lambda)\rangle\\
&\geq \inf_{\lambda \in \Delta_{0,N}(\xi_N)}\langle b-b_N(\xi_N), \Gamma_N(\lambda)\rangle=\inf_{\lambda \in \Delta_{0,N}(\xi_N)}\langle \xi_N^b/\sqrt{N}, \Pi^u(\Gamma_N(\lambda))\rangle\\
&\geq \inf_{\lambda \in {\cal D}}\langle \xi_N^b/\sqrt{N}, \Pi^u(\lambda)\rangle\\
&\geq -\sup_{\lambda \in {\cal D}} \|\Pi^u(\lambda)\| \frac{\|\xi_N^b\|}{\sqrt{N}}\overset{a.s.}{\longrightarrow}0
\end{aligned}
\end{equation}
where the first inequality follows from the condition \( v = 0 \); the second inequality follows from the fact that \( \langle b_N(\xi_N), \lambda \rangle \geq 0 \) for all \( \lambda \in \Delta_{0,N}(\xi_N) \), since \( 0 \in \mathcal{D}_N(\xi_N) \) and \( \Gamma_N(\lambda) \) is a positive scalar multiple of \( \lambda \); the third inequality follows from the inclusion \( \Gamma_N(\Delta_{0,N}(\xi_N)) \subseteq \mathcal{D} \); and the almost sure convergence follows from equation~\ref{eqnPlemAD002} and the fact that \( \xi_N^b \overset{a.s.}{\rightarrow} 0 \). By Lemma \ref{lemOSC}, Inequality \ref{eqnPlemAD22} and the fact that $v=0$ then imply
\[
\overset{\rightarrow}{d}_H(\Gamma_N(\Delta_{0,N}(\xi_N)), \Delta_0)\overset{a.s.}{\longrightarrow} 0, 
\]
and we thus have
\begin{equation}
\label{eqnPlemAD0022}
\overset{\rightarrow}{d}_H(\Pi^u(\Gamma_N(\Delta_{0,N}(\xi_N))), \Pi^u(\Delta_0))\overset{a.s.}{\longrightarrow} 0.
\end{equation}
Equation \ref{eqnPlemAD20} then follows from equations \ref{eqnPlemAD0021} and \ref{eqnPlemAD0022}.
 
We can now take the limit of both sides of equation \ref{eqnPlemAD1}, and using equation \ref{eqnPlemAD20} gives
\begin{equation}
\label{eqnPlemAD3}
\begin{aligned}
a.s.-\overline{\lim}_{N\to \infty}\sqrt{N}\phi(\mu_N(\xi_N))&\leq a.s.-\overline{\lim}_{N\to \infty} \langle \xi_N^b,\Pi^u(\lambda(\xi_N))\rangle \\
&\leq a.s.-\overline{\lim}_{N\to \infty} \sup_{\lambda\in \Delta_0}\langle \xi_N^b,\Pi^u(\lambda)\rangle +\|\xi_N^b\|\overset{\rightarrow}{d}_H(\Pi^u(\Delta_{0,N}(\xi_N)), \Pi^u(\Delta_0))\\
&=\sup_{\lambda\in \Delta_0}\langle \xi^b,\lambda\rangle.
\end{aligned}
\end{equation}
To complete the argument, it remains to establish the converse of inequality \ref{eqnPlemAD3}. To that end, using the saddle point property and adopting the same notation as in the derivation of inequality \ref{eqnPlemAD1}, we have
\begin{equation}
\label{eqnPlemAD4}
\begin{aligned}
\sqrt{N}\phi(\mu_N(\xi_N))&=\sqrt{N}\Bigl(L(\lambda(\xi_N), u(\xi_N), v(\xi_N), x(\xi_N), y(\xi_N), z(\xi_N);\xi_N)-L(\lambda_0, u_0, v_0, x_0, y_0, z_0)\Bigr)\\
&\geq \sqrt{N}\Bigl(L(\lambda_0, u_0, v_0, x(\xi_N), y(\xi_N), z(\xi_N);\xi_N)-L(\lambda_0, u_0, v_0, x(\xi_N), y(\xi_N), z(\xi_N))\Bigr)\\
&=\langle \xi_N^b,\lambda_0\rangle +\sqrt{N}\langle y(\xi_N), \xi_N^D\lambda_0\rangle. 
\end{aligned}
\end{equation}
The dual representation of our LP (equation \ref{eqnPlemAD0}) and inequality \ref{eqnPlemAD3} imply that 
\[
a.s.-\overline{\lim}_{N\to \infty} \sqrt{N}\ \|y(\xi_N)\|_{\infty}= a.s.-\overline{\lim}_{N\to \infty} \sqrt{N}\ x(\xi_N)=a.s.-\overline{\lim}_{N\to \infty}\sqrt{N}\phi(\mu_N(\xi_N))<\infty,
\]
and since $\xi_N^D\overset{a.s.}{\rightarrow}0$, we conclude that 
\[
a.s.-\lim_{N\to \infty}\sqrt{N}\langle y(\xi_N), \xi_N^D\lambda_0\rangle=0.
\]
Equation \ref{eqnPlemAD4} thus gives
\begin{equation}
a.s.-\underline{\lim}\sqrt{N}\phi(\mu_N(\xi_N))\geq \max_{\lambda \in \Delta_0}\langle \xi^b,\lambda \rangle,
\end{equation}
which establishes the converse to \ref{eqnPlemAD3}, and proves the claim.
\end{proof}

\begin{lemma}
\label{lem1AD}
Consider the linear program $v = \max\{ b^\top \lambda \mid\lambda\geq 0,\ A^\top \lambda = 0,\ \|\Omega\lambda\|_1 \leq 1 \}$ in \ref{eqnAVEAM0}. And let $\hat{d}$ be an estimator of $d=vec([A\ \ b])$ such that 
\begin{equation}
\sqrt{N}(\hat{d}-d)\overset{d}{\longrightarrow}\mathcal{N}(0,\Sigma) \quad \quad \text{and}\quad \quad \hat{\Omega}-\Omega \overset{p}{\longrightarrow}0
\end{equation}
where $\Omega$ and $\hat{\Omega}$ are as defined in Section \ref{subsectionEAM}. Let $U\subset [p]$ denote the set of indices of positive diagonal elements of $\Omega$. Assume that the system
\[
b-A\eta\leq 0
\]
has a unique solution $\eta^*$. For $\hat{T}_N$ as defined in equation \ref{eqnEAM2}, we have
\begin{equation}
\label{eqnlem1AD1}
\sqrt{N}\hat{T}_N\overset{d}{\longrightarrow} \max_{\lambda \in \Delta_0} \langle \zeta_b-\zeta_A\eta^*, \lambda \rangle,
\end{equation}
where $\zeta=vec([\zeta_A \ \ \zeta_b])\sim \mathcal{N}(0,\Sigma)$, and $\Delta_0=\arg\max\{ b^\top \lambda \mid\lambda\geq 0,\ A^\top \lambda = 0,\ \|\Omega\lambda\|_1 \leq 1 \}$.
\end{lemma}

\begin{proof}[\bf{Proof of Lemma \ref{lem1AD}}]
We proceed using an almost sure representation argument as in the preceding lemma. Let $\hat{\zeta}=(\sqrt{N}(\hat{A}-A),\sqrt{N}(\hat{b}-b),\ \hat{\Omega}-\Omega)$. The hypothesis of the theorem implies that $\hat{\zeta}=(\hat{\zeta}_A,\hat{\zeta}_b,\hat{\zeta}_{\Omega})\overset{d}{\rightarrow} \zeta:=(\zeta_A,\zeta_b,0)$ where $vec([\zeta_A\ \ \zeta_b])\sim \mathcal{N}(0,\Sigma)$. By the Skorokhod representation theorem, there exists a sufficiently rich probability space, and versions $\xi_N \overset{d}{=}\hat{\zeta}$ and $\xi:=(\xi_A,\xi_b,0)\overset{d}{=}\zeta$ such that $a.s.-\lim_{N\to \infty} \xi_N=\xi$. Using the notation used in Section~\ref{pthmAVEAM}, let $\mu=(A,b,\Omega)$ and $\mu_N(\xi_N)=(A_N(\xi_N),b_N(\xi_N),\Omega_N(\xi_N))=\bigl(A+\frac{\xi_{AN}}{\sqrt{N}}, b+\frac{\xi_{bN}}{\sqrt{N}}, \Omega+\xi_{\Omega N}\bigr)$. We have $v=\phi(\mu)$ and $\hat{v}\overset{d}{=}\phi(\mu_N(\xi_N))$. Let the Lagrangians $L_N(\lambda,\eta,t;\xi)$ and $L(\lambda,\eta,t)$ be as defined in Section \ref{pthmAVEAM}.

Reasoning as in the derivation of equation \ref{eqnPSSLSUC02}, we have
\begin{equation}
\label{eqnPlem1AD1}
\sqrt{N} \phi_N(\mu_N(\xi_N)) \leq \langle \xi_{bN}-\xi_{AN} \eta^*, \lambda(\xi_N)\rangle
\end{equation}
where $\lambda(\xi_N)$ is any element in $\Delta_{0,N}(\xi_N)=\arg\max\{b_N(\xi_N)^\top \lambda \mid \lambda\geq 0, \ A_N(\xi_N)^\top \lambda=0, \ \mathbbm{1}^\top \Omega_N(\xi_N)\lambda\leq 1\}$, and $\eta^*$ is as in the statement of the lemma.

We now show that
\begin{equation}
\label{eqnPlem1AD2}
a.s.-\lim_{N\to \infty} \overset{\rightarrow}{d}_H(\Delta_{0,N}(\xi_N),\Delta_0)=0.
\end{equation}
We have
\[
\sup_{\lambda\in \Delta_{0,N}(\xi_N)}\|A^\top\lambda\|=\sup_{\lambda\in \Delta_{0,N}(\xi_N)}\|(A-A_N(\xi_N))^\top\Pi^u(\lambda)\| \leq \|A-A_N(\xi_N)\|\sup_{\lambda\in \Delta_{0,N}(\xi_N)}\|\Pi^u(\lambda)\|\overset{a.s.}{\longrightarrow} 0,
\]
and
\begin{equation*}
\begin{aligned}
\sup_{\lambda\in \Delta_{0,N}(\xi_N)} (\mathbbm{1}^\top \Omega \lambda-1)_+ &\leq \sup_{\lambda\in \Delta_{0,N}(\xi_N)} (\mathbbm{1}^\top \Omega \lambda-1+1-\mathbbm{1}^\top \Omega_N(\xi_N) \lambda)_+\\
&\leq \|\Pi^u(\mathbbm{1})\|\|\Omega-\Omega_N(\xi_N)\|\sup_{\lambda\in \Delta_{0,N}(\xi_N)}\|\Pi^u(\lambda)\|\overset{a.s.}{\longrightarrow} 0,
\end{aligned}
\end{equation*}
where we have used equation \ref{eqnPSSLSUC04}. The latter two inequalities and Lemma \ref{lemhoff} yield equation \ref{eqnPlem1AD2}.

Equations \ref{eqnPlem1AD1} and \ref{eqnPlem1AD2} imply that
\begin{equation}
\label{eqnPlem1AD3}
\begin{aligned}
a.s.-\overline{\lim}_{N\to \infty} \sqrt{N} \phi_N(\mu_N(\xi_N)) &\leq a.s.-\overline{\lim}_{N\to \infty}  \sup_{\lambda\in \Delta_0} \langle \xi_{bN}-\xi_{AN} \eta^*, \lambda\rangle+\\
&\overset{\rightarrow}{d}_H(\Delta_{0,N}(\xi_N),\Delta_0)\|\xi_{bN}-\xi_{AN} \eta^*\| \\
&=\sup_{\lambda\in \Delta_0} \langle \xi_{b}-\xi_{A} \eta^*, \lambda\rangle.
\end{aligned}
\end{equation}

We now derive the converse to equation \ref{eqnPlem1AD3}. Reasoning as in the derivation of equation \ref{eqnPlemAD4}, we have
\begin{equation}
\label{eqnPlem1AD4}
\begin{aligned}
\sqrt{N} \, \phi(\mu_N(\xi_N)) 
&\geq \sqrt{N} \bigl( L_N(\lambda_0, \eta(\xi_N), t(\xi_N); \xi_N) - L(\lambda_0, \eta(\xi_N), t(\xi_N)) \bigr) \\
&= \xi_{bN}^\top \lambda_0 - \eta(\xi_N)^\top \xi_{AN}^\top \lambda_0 - t(\xi_N) \mathbbm{1}^\top \xi_{\Omega N} \lambda_0 ,
\end{aligned}
\end{equation}
where $\lambda_0 \in \Delta_0$ and $(\eta(\xi_N), t(\xi_N)) \in S_{0,N}(\xi_N)$, the set of optimal solutions to the (perturbed) dual LP \ref{eqnPSSLSUC01} with input $\mu_N(\xi_N)$ (see Proof of Theorem \ref{thmAVEAM}). Note that all elements of the set $S_{0,N}(\xi_N)$ have the same second component, given by the value of the perturbed LP: i.e., $t(\xi_N)=\phi(\mu_N(\xi_N))$. We first show that 
\begin{equation}
\label{eqnPlem1AD5}
a.s.-\overline{\lim}_{N\to \infty} t(\xi_N))=0.
\end{equation}
Indeed, we have
\[
\phi(\mu_N(\xi_N))=\min\{t\mid t\Omega_N(\xi_N) \mathbbm{1}+A_N(\xi_N)\eta\geq b_{N}(\xi_N),\ t\geq 0\}.
\]
Since $b-A\eta^*\leq 0$, we have
\begin{equation*}
\begin{aligned}
b_N(\xi_N)-A_N(\xi_N)\eta^*&\leq \bigl(b_N(\xi_N)-b\bigr)-\bigl(A_N(\xi_N)-A\bigr)\eta^*\\
&\leq \Bigl(\|b-b_N(\xi_N)\|+\|\eta^*\| \|A-A_N(\xi_N)\|\Bigr)\Pi^u(\mathbbm{1}).
\end{aligned}
\end{equation*}
Equation \ref{eqnPlem1AD5} then follows since $\|b-b_N(\xi_N)\|+\|\eta^*\| \|A-A_N(\xi_N)\|\overset{a.s.}{\longrightarrow}0$ and $[\Omega]_{ii}>0$ for all $ i\in U$. 

We now show that 
\begin{equation}
\label{eqnPlem1AD6}
a.s.-\lim \overset{\rightarrow}{d}_H(\Pi_1(S_{0,N}(\xi_N)),\{\eta^*\})=0
\end{equation}
where $\Pi_1(S_{0,N(\xi_N)}):=\bigl\{\eta\mid (\eta,t(\xi_N))\in S_{0,N}(\xi_N)\bigr\}$ is the projection of $S_{0,N}(\xi_N)$ on its first component. Indeed, for every $\eta\in \Pi_1(S_{0,N}(\xi_N))$, we have
\[
b_{N}(\xi_N)-A_N(\xi_N)\eta\leq t(\xi_N)\Omega_N(\xi_N)\mathbbm{1}.
\]
Note that if $(A_N,b_N,t_N)$ is any sequence such that $A_N\rightarrow A$ and $b_N\rightarrow b$, and $t_N\rightarrow 0$, then the set $\mathcal{C}_N=\{\eta \mid b_N-A_N\eta\leq t_N\}$ satisfies
\[
\overline{lim}_{N\to \infty} \sup_{\eta \in {\cal C}_N}\|\eta\|<\infty.
\]
If not, then, after passing to a subsequence, we can assume that there exist $\eta_N\in {\cal C}_N$ such that $\|\eta_N\|\to \infty$.
But then we have
\[
b_N/\|\eta_N\|-A_N(\eta_N/\|\eta_N\|)\leq t_N/\|\eta_N\|.
\]
After passing to a further subsequence, we can assume that $\eta_N/\|\eta_N\|\to \eta_0\neq 0$, and we have
\[
-A\eta_0\leq 0,
\]
which contradicts the assumption of the lemma that the inequality system $b-A\eta\leq 0$ has a unique solution $\eta^*$. Thus, we have
\[
a.s.-\overline{\lim}_{N\to \infty} \sup_{\eta \in \Pi_1(S_{0,N}(\xi_N))}\|\eta\| <\infty,
\]
and
\begin{equation*}
\begin{aligned}
a.s.-\overline{lim} \sup_{\eta\in \Pi_1(S_{0,N}(\xi_N))} \|[b-A\eta]_+\|&\leq a.s.-\overline{lim} \Bigl\{ \|b-b_N(\xi_N)\|+\|A-A_N(\xi_N)\| \sup_{\eta\in \Pi_1(S_{0,N}(\xi_N))}\|\eta\|+\\
&t(\xi_N)\|\Omega_N(\xi_N)\| \|\mathbbm{1}\|\Bigr\}\\
&=0.
\end{aligned}
\end{equation*}
The latter inequality and Lemma \ref{lemhoff} then imply equation \ref{eqnPlem1AD6}.

Combining equations~\ref{eqnPlem1AD4}, ~\ref{eqnPlem1AD5}, and ~\ref{eqnPlem1AD6}, we get 
\begin{equation}
\label{eqnPlem1AD7}
\begin{aligned}
a.s.-\underline{\lim}_{N\to \infty} \sqrt{N}\phi(\mu_N(\xi_N))&\geq a.s.-\underline{\lim}_{N\to \infty}\sup_{\eta\in \Pi_1(S_{0,N}(\xi_N))} \sup_{\lambda \in \Delta_0} \Bigl\{\xi_{bN}^\top \lambda - \eta^\top \xi_{AN}^\top \lambda - t(\xi_N) \mathbbm{1}^\top \xi_{\Omega N} \lambda\Bigr\}\\
&=\sup_{\lambda\in \Delta_0}\langle \xi_b-\xi_A\eta^*,\lambda \rangle.
\end{aligned}
\end{equation}
Combining equations \ref{eqnPlem1AD3} and \ref{eqnPlem1AD7} yield equation \ref{eqnlem1AD1}, and the proof is complete.

\end{proof}

The following lemma establishes the set of maximizers of a LP is well separated in the sense that values of the objective function at feasible points that are at some fixed distance from the set of optimizers are strictly below the optimal value, even if the feasible region is not compact.

\begin{lemma}
\label{lemOSC}
Consider the LP $v=\max\{b^\top \lambda \mid A \lambda \leq d\}$, where $v$ is finite, and the feasible region ${\cal D}:=\{\lambda\in \mathbb{R}^p\mid A \lambda \leq d \}$ is not necessarily bounded. Let $\Delta_0=\arg\max\{b^\top \lambda \mid A \lambda \leq d\}$ denote the set of optimal solutions, and given $\epsilon>0$, let $\Delta_{\epsilon}$ denote the set of $\epsilon-$optimal solutions defined by
$\Delta_{\epsilon}:=\{\lambda \in {\cal D}\mid b^\top \lambda \geq v-\epsilon\}$.
Then there exists a constant $C$ that depends on the inputs $(A,b,d)$, such that
\begin{equation}
\label{eqnlemOSC1}
\overset{\rightarrow}{d}_H(\Delta_{\epsilon},\Delta_0)\leq C\epsilon.
\end{equation}
\end{lemma}
\begin{proof}[\bf{Proof of Lemma \ref{lemOSC}}]
The dual LP is given by 
\[
\min\{d^\top z\mid z\geq 0,\ A^\top z=b\}.
\]
Let $z^*$ denote an optimal dual solution, and let $J\subseteq [m]$ denote the set of nonzero indices of $z^*$, where $m$ denotes the number of rows of the matrix $A$. If $J=\emptyset$, then $z^*=0$, and since $A^\top z^*=b$, we have $b=0$, from which we deduce that  $\Delta_0={\cal D}$, and the claim trivially holds. So for the remainder of the proof, let us assume that $J\neq \emptyset$.

Let the face $F$ of ${\cal D}$ be given by 
\[
F:=\{\lambda \in \mathbb{R}^p\mid A_J\lambda=d_J \quad \text{and}\quad A_{J^c}\lambda\leq d_{J^c}\}
\]
where $J^c=[m]\backslash J$, and $A_J$ is the matrix obtained from A by deleting the rows not in $J$ (with a similar definition for $d_J$, $A_{J^c}$ and $d_{J^c}$).
We assert that \( F \) is equal to the set of optimal solutions, i.e., \( F = \Delta_0 \). Indeed, if $\lambda\in F$ then 
\[
b^\top \lambda=\langle A^\top z^*, \lambda \rangle=\langle  A^\top z^*, \lambda \rangle=\langle   z_{J}^*, A_J\lambda \rangle=\langle  z_{J}^*, d_J \rangle=\langle  z^*, d \rangle=v,
\]
and we have $F\subseteq \Delta_0$. For the reverse inclusion, note that if $\lambda^*\in \Delta_0$, then 
\[
\langle z^*,d \rangle=v=\langle \lambda^*, b\rangle= \langle \lambda^*, A^\top z^*\rangle \quad \Longrightarrow \quad \langle z^*,A\lambda^*-d\rangle=0\Longleftrightarrow \langle z_J^*,A_J\lambda^*-d_J\rangle=0.
\]
The last equation then implies that $A_J\lambda^*-d_J=0$, since we necessarily have $A_J\lambda^*-d_J\leq 0$, and all entries of $z_J*$ are strictly positive.\footnote{This corresponds to the complementary slackness condition in linear programming.}. As $\lambda^*\in {\cal D}$, we clearly have $A_{J^c}\lambda^*\leq d_{J^c}$. Therefore, $\lambda^*\in \Delta_0$ implies that $A_J\lambda^*-d_J=0$ and $A_{J^c}\lambda^*\leq d_{J^c}$, which gives the second inclusion $\Delta_0\subseteq F$, and we thus conclude that $F=\Delta_0$.

Now for $\lambda\in \Delta_{\epsilon}$ we have 
\[
\big[0\leq v-b^\top \lambda\leq \epsilon\big] \Longleftrightarrow  \big[0\leq \langle z^*,d\rangle- \langle A^\top z^*,\lambda \rangle \leq \epsilon\big] \Longleftrightarrow  \big[0\leq \langle z^*,d-A\lambda \rangle \leq \epsilon\big]
\Longleftrightarrow  \big[0\leq \langle z_J^*,d_J-A_J\lambda \rangle \leq \epsilon\big].
\]
If $\underline{z}:=\min\{z^*_j\mid  j \in J\}$ denotes the smallest value of $z^*$ on its support, then the last inequality implies that
for all $\lambda \in \Delta_{\epsilon}$, we have 
\begin{equation}
\label{eqnPlemOSC1}
\|d_J-A_J\lambda\|_1\leq \epsilon/\underline{z}.
\end{equation}
We recall that Hoffman's Theorem (Lemma \ref{lemhoff}) implies that there exists a constant $C>0$ such that for all $\lambda \in \mathbb{R}^p$ we have
\[
d(\lambda,F)\leq C\Bigl\{\|d_J-A_J\lambda\|+\|(d_{J^c}-A_{J^c}\lambda)_-\| \Bigr\}.
\] 
For $\lambda \in \Delta_{\epsilon}$, we necessarily have $(d_{J^c}-A_{J^c}\lambda)_-=0$, and Hoffman's Theorem and equation \ref{eqnPlemOSC1} then implies that 
\[
\sup_{\lambda \in \Delta_{\epsilon}}d(\lambda,F)\leq C\Bigl\{\|d_J-A_J\lambda\|\Bigr\} \leq (C/\underline{z})\epsilon
\] 
which completes the proof of inequality \ref{eqnlemOSC1}.
\end{proof}

The following lemma establishes the Hausdorff continuity of the set of extreme points of our feasible regions under perturbations of the normalizing matrix $D$. It extends Lemma 4 in \cite{ARP} to settings in which the points $\lambda$ in the feasible region are not restricted to lie in an orthant of $\mathbb{R}^p$, and where the constraint used for the normalization may correspond to intersecting the feasible region with more than one halfspace. \footnote{When, as in \cite{ARP}, $\lambda \in \mathbb{R}^p_+$ for all $\lambda$ in the feasible region, the normalizing constraint $\|D\lambda\|_1\leq 1$ is equivalent to the inequality $\mathbbm{1}^\top D\lambda\leq 1$. When $\lambda$ is not restricted to lie in a specific orthant, the constraint $\|D\lambda\|_1\leq 1$ is no longer equivalent to one inequality, and may be equivalent with up to $2^p$ inequalities.}

\begin{lemma}
\label{lemEPC}
Given a diagonal matrix $D \in \mathbb{R}^{n \times n}$ and a matrix $B \in \mathbb{R}^{m_B \times n}$, let ${\cal D}(B,D)$ denote a polyhedral set defined by
\[
\mathcal{D}(B, D) := \left\{ \lambda \in \mathbb{R}^n \,\middle|\, B\lambda \leq 0,\ \|D\lambda\|_1 \leq 1 \right\}.
\]
For $D$ a diagonal matrix, let $\operatorname{supp}(D)$ denote the support of $D$, defined as the set of indices corresponding to nonzero diagonal entries of $D$:
\[
\operatorname{supp}(D) := \left\{ i \in [n] \,\middle|\, D_{ii} \neq 0 \right\}.
\]

Let $\operatorname{extr}(\mathcal{D})$ denote the (possibly empty) set of extreme points of $\mathcal{D}$. Let $\underline{\sigma}$ and $\overline{\sigma}$ be two fixed positive constants such that $\underline{\sigma}<\overline{\sigma}$ . Then for all matrices $B$ and diagonal matrices $D$ and $D'$ satisfying
\[
\operatorname{supp}(D) = \operatorname{supp}(D'), \quad 
\|D - D'\| \leq \underline{\sigma}/2, \quad 
\text{and} \quad 
\underline{\sigma} \leq D_{ii} \leq \overline{\sigma}, \quad \forall i \in \operatorname{supp}(D)
\]

 the following hold:

\begin{enumerate}[i)]
\item For all $\lambda \in \operatorname{extr}(\mathcal{D}(B, D))\backslash \{0\}$, we have $\|D'\lambda\|_1>0$, and if we set $\lambda' := \lambda/\|D'\lambda\|_1$, then  $\lambda' \in \operatorname{extr}(\mathcal{D}(B, D'))\backslash \{0\}$, and we have
\begin{equation}
\label{eqnlemEPC1}
   \lambda - \lambda'=\Bigl(1-\frac{1}{\|D'\lambda\|_1}\Bigr)\lambda   \quad \quad \text{and}\quad \quad \bigl|1-\frac{1}{\|D'\lambda\|_1}\bigr| \leq  \frac{2}{\underline{\sigma}} \|D' - D\|.
\end{equation}

\item For all $\lambda' \in \operatorname{extr}(\mathcal{D}(B, D'))\backslash \{0\}$, we have $\|D\lambda'\|_1>0$, and if we set $\lambda := \lambda'/\|D\lambda'\|_1$, then  $\lambda \in \operatorname{extr}(\mathcal{D}(B, D))\backslash \{0\}$, and we have
\begin{equation}
\label{eqnlemEPC2}
   \lambda' - \lambda=\Bigl(1-\frac{1}{\|D\lambda'\|_1}\Bigr)\lambda'   \quad \quad \text{and}\quad \quad \bigl|1-\frac{1}{\|D\lambda'\|_1}\bigr| \leq  \frac{1}{\underline{\sigma}} \|D' - D\|.
\end{equation}

\end{enumerate}
\end{lemma}

\begin{proof}[\bf{Proof of Lemma \ref{lemEPC}}]
We establish \ref{eqnlemEPC1}. The proof of \ref{eqnlemEPC2} is similar. Note that as the sets ${\cal D}(B,D)$ and ${\cal D}(B,D')$ are such that we have $\lambda\in {\cal D}(B,D)$ iff $\exists t>0$ such that $t\lambda \in {\cal D}(B,D')$. Hence $0 \in \operatorname{extr}(\mathcal{D}(B, D))$ iff $0 \in \operatorname{extr}(\mathcal{D}(B, D'))$, and it suffices to consider the nonzero elements of $\operatorname{extr}(\mathcal{D}(B, D))$. To that end, we first note that the inequality $\|\lambda\|_1\leq 1$ can be expressed as a set of $2^n$ inequality constraints, given by $H \lambda\leq 1$, where the rows of $H$ are given by the elements of $\{-1,1\}^n$. Thus the set ${\cal D}(B,D)$ has the equivalent representation
\[
{\cal D}(B,D)=\bigl\{\lambda \in \mathbb{R}^n \mid B\lambda \leq 0, \ HD\lambda\leq \mathbbm{1}_{2^n}\bigr\}
\]
where $\mathbbm{1}_{2^n}$ represent the vector of all ones of dimension $2^n$. For $\lambda \in \mathbb{R}^n$, define the support of $\lambda$ as $\operatorname{supp}(\lambda) = \{i \in [n] \mid \lambda_i \neq 0\}$.
Let $e_{n,j}$, for $j \in [n]$, denote the canonical basis vectors in $\mathbb{R}^n$.

By the characterization of extreme points of a polyhedron, we have $\lambda \in \mathcal{D}(B, D) \setminus \{0\}$ if and only if $\lambda \in \mathcal{D}(B, D)$ and there exist subsets $J_1 \subseteq [m_B]$ and $J_2 \subseteq [2^n]$ such that $J_2\neq \emptyset$ and $n_1+n_2 = n$, where $n_1:=|J_1|$ and  $n_2:= |J_2|$, the matrix
 \( \left( \begin{smallmatrix} B_{J_1} \\ H_{J_2}D \end{smallmatrix} \right) \) is invertible, and $\lambda$ is the unique solution to the equation\footnote{Given a matrix $A \in \mathbb{R}^{m \times n}$ and an index set $J \subseteq [m]$ corresponding to rows of $A$, we let $A_J$ denote the $|J| \times n$ matrix obtained by selecting only the rows of $A$ indexed by $J$.}
\begin{equation}
\label{eqnPlemEPC1}
\begin{pmatrix}
 B_{J_1} \\
 H_{J_2} D
\end{pmatrix} \lambda =
\begin{pmatrix}
 0 \\
 \mathbbm{1}_{n_2} 
\end{pmatrix}.
\end{equation}
For all $i\in [n_2]$, we have 
\[
e_{n_2,i}^\top H_{J_2} D\lambda =\|D\lambda\|_1=1.
\]
As a consequence, when restricted to the columns in the support of $D\lambda$, the rows of $H_{J_2}$ must all be equal, with entries equal to the sign of $D\lambda$; that is, for $i, j \in [n_2]$ and $k \in \operatorname{supp}(D\lambda)$, we have
\begin{equation}
\label{eqnPlemEPC2}
e_{n_2,i}^\top H_{J_2} e_{n,k}=e_{n_2,j}^\top H_{J_2} e_{n,k}, \quad  \text{ and }\quad e_{n_2,i}^\top H_{J_2} e_{n,k}=\operatorname{sign}(e_{n,k}^\top D\lambda).
\end{equation}
Note that as $D'$ is also a positive matrix with same support as $D$, we have $\operatorname{supp}(D'\lambda)=\operatorname{supp}(D\lambda)$, $\operatorname{sign}(D\lambda)=\operatorname{sign}(D'\lambda)$\footnote{Given a vector $x\in \mathbb{R}^n$, let $\operatorname{sign}(x)$ be the vector in $\mathbb{R}^n$ with $i^{th}$ entry equal to $1$ if $x_i>0$, 0 if $x_i=0$, and -1 if $x_i<0$.}, and the second part of equation \ref{eqnPlemEPC2} remains valid when $D$ is replaced with $D'$, which yields
\begin{equation}
\label{eqnPlemEPC3}
H_{J_2}D'\lambda=\|D'\lambda\|_1 \mathbbm{1}_{n_2},\quad \quad \text{with}\quad \quad \|D'\lambda\|_1>0.
\end{equation}
If we set $\lambda'=\frac{\lambda}{\|D'\lambda\|_1}$, then $B\lambda \leq 0$ (by homogeneity) and $\|D'\lambda'\|_1=1$, and thus $\lambda'\in {\cal D}(B,D')$. We now prove that $\lambda'\in \operatorname{extr}({\cal D}(B,D'))$. Suppose that there exist $\alpha\in (0,1)$ and $\lambda_1,\lambda_2\in {\cal D}(B,D')$ such that
\begin{equation}
\label{eqnPlemEPC4}
\lambda'=\alpha \lambda_1+(1-\alpha)\lambda_2.
\end{equation}
For $i\in \{1,2\}$, as $\lambda_i \in {\cal D}(B,D')$, we must have $\|D'\lambda_i\|_1\leq 1$, and since $\|D'\lambda'\|_1=1$, equations \ref{eqnPlemEPC3} and \ref{eqnPlemEPC4} imply that 
\begin{equation}
\label{eqnPlemEPC5}
\|D'\lambda_i\|_1=1,\quad \quad \text{and}\quad \quad H_{J_2}D'\lambda_i=\mathbbm{1}_{n_2}.
\end{equation}
Also, since by homogeneity $B_{J_1}\lambda'=0$, and, for $i\in \{1,2\}$, $B_{J_1}\lambda_i\leq 0$, equation \ref{eqnPlemEPC4} implies that we must have 
\begin{equation}
\label{eqnPlemEPC6}
B_{J_1} \lambda_i=0.
\end{equation}

As in equation~\ref{eqnPlemEPC2}, equation~\ref{eqnPlemEPC5} implies that, for each \( i \in \{1, 2\} \), the entries of the rows of \( H_{J_2} \) coincide when restricted to the indices of columns corresponding to the support of \( D'\lambda_i \), and these entries are equal to the signs of the corresponding elements in \( D'\lambda_i \). That is, for \( i \in \{1, 2\} \), and for any \( j_1, j_2 \in [n_2] \) and \( k \in \operatorname{supp}(D'\lambda_i) \), we have
\begin{equation}
\label{eqnPlemEPC7}
e_{n_2,j_1}^\top H_{J_2} e_{n,k} = e_{n_2,j_2}^\top H_{J_2} e_{n,k}, \quad \text{and} \quad e_{n_2,j_1}^\top H_{J_2} e_{n,k} = \operatorname{sign}(e_{n,k}^\top D'\lambda_i).
\end{equation}
Since $\operatorname{supp}(D\lambda_i)=\operatorname{supp}(D'\lambda_i)$ and $\operatorname{sign}(D\lambda_i)=\operatorname{sign}(D'\lambda_i)$, equation \ref{eqnPlemEPC7} implies that, for $i\in\{1,2\}$, we have
\begin{equation}
\label{eqnPlemEPC8}
H_{J_2}D\lambda_i=\|D\lambda_i\|_1 \mathbbm{1}_{n_2}.
\end{equation}
As the matrix in equation \ref{eqnPlemEPC1} is invertible, equations \ref{eqnPlemEPC6} and \ref{eqnPlemEPC8} imply that $\lambda_i=\|D\lambda_i\|_1\lambda$. thus $\lambda_1,\ \lambda_2$ and $\lambda'$ are all positive scalar multiples of $\lambda$, with $\|D'\lambda_1\|_1=\|D'\lambda_2\|_1=\|D'\lambda'\|_1=1$, which implies that $\lambda'=\lambda_1=\lambda_2$, and $\lambda'\in \operatorname{extr}({\cal D}(B,D'))\backslash\{0\}$.

We have 
\begin{equation*}
\Bigl|1-\frac{1}{\|D'\lambda\|_1}\Bigr|=\Bigl|1-\frac{\|D\lambda\|_1}{\|D'\lambda\|_1}\Bigr|\leq \frac{\|(D'-D)\lambda\|_1}{\|D'\lambda\|_1}\leq (2/\underline{\sigma})\|D-D'\|
\end{equation*}
where we have used the fact that the smallest diagonal entry in the support of $D'$ is bounded below by $2/\underline{\sigma}$ when $\|D-D'\|\leq \underline{\sigma}/2$. This completes the proof of \ref{eqnlemEPC1}.
\end{proof}

\section {Auxiliary lemmas}

\begin{lemma}
\label{lem2}

Let  
\[
F, F_n: S\times \Delta \subset \mathbb{R}^d \times \mathbb{R}^p \to \mathbb{R}, \quad n\geq 1,
\]
be a sequence of continuous functions defined on a neighborhood \( S\times \Delta \) of the compact set \( S_0\times \Delta_0 \). Suppose that  
\begin{itemize}
    \item \( F_n \) converges uniformly to \( F \) on \( S \times \Delta \),
    \item \( S_n \times \Delta_n \subset \mathbb{R}^d \times \mathbb{R}^p \) are compact sets satisfying  
    \[
    d_H(S_n, S_0) \to 0, \quad d_H(\Delta_n, \Delta_0) \to 0.
    \]
\end{itemize}  
Then, the minimax values satisfy  
\[
\min_{\theta \in S_n} \max_{\lambda \in \Delta_n} F_n(\theta,\lambda) \to  
\min_{\theta \in S_0} \max_{\lambda \in \Delta_0} F(\theta,\lambda).
\]
\end{lemma}

\begin{proof}[\textbf{Proof of Lemma \ref{lem2}}]
For all sufficiently large n's, as $S$ and $\Delta$ are neighborhoods of $S_0$ and $\Delta_0$, we can assume for simplicity that $S_n\times \Delta_n \subset S\times \Delta$ for all $n\geq 1$, and that $S\times \Delta$ is compact. 
Let $\phi, \phi_n: S \to \mathbb{R}$, for $n\geq 1$, be defined by $\phi_n(\theta)=\max_{\lambda \in \Delta_n}F_n(\theta,\lambda)$ and $\phi(\theta)=\max_{\lambda \in \Delta_0}F(\theta,\lambda)$. Then by the theorem of the maximum, $\phi_ n$ and $\phi$ are continuous. 

\underline{Step 1}We first claim that $\phi_n$ converges uniformly to $\phi$ on $S$. Indeed, as $\phi$ is continuous and $S$ is compact, it suffices to show that \(\lim_{n\to \infty} \phi_n(\theta_n)=\phi(\theta^*)\) whenever $\theta_n \to \theta^*$, $\theta_n,\theta^*\in S$. We establish the latter by showing that every subsequence of $\{\phi_n(\theta_n)\}_{n\geq 1}$ has a further subsequence that converges to $\phi(\theta^*)$. For $\theta \in S$, let $\lambda_n(\theta)$ and $\lambda(\theta)$ be such that by \(\lambda_n(\theta)\in \arg\max_{\lambda \in \Delta_n } F_n(\theta,\lambda) \) and \(\lambda(\theta)\in \arg\max_{\lambda \in \Delta_0 } F(\theta,\lambda) \). Then $\phi_n(\theta_n)=F_n(\theta_n,\lambda_n(\theta_n))$. As $S\times \Delta$ is compact, every subsequence of $\{\phi_n(\theta_n)\}_{n\geq 1}$ has a further subsequence, say $\{\phi_{n_k}(\theta_{n_k})\}_{k\geq 1}$, such that $\lambda_{n_k}(\theta_{n_k}) \to \lambda^*$. By uniform convergence of $F_n$ to $F$ and the continuity of $F$, we have $\phi_{n_k}(\theta_{n_k})= F_{n_k}(\theta_{n_k},\lambda_{n_k}(\theta_{n_k}))\to F(\theta^*,\lambda^*)$. As $d_H(\Delta_n,\Delta_0)\to 0$, we have $\lambda^* \in \Delta_0$. We now show that $F(\theta^*,\lambda^*)=\phi(\theta^*)$. Indeed, for each $\lambda \in \Delta_0$, as $d_H(\Delta_{n_k},\Delta_0)\to 0$ there exists $\lambda_k \in \Delta_{n_k}$ such that $\lambda_k \to \lambda$. Then such $\lambda \in \Delta_0$ we get
\[
F(\theta^*,\lambda)=\lim_{n\to \infty} F_{n_k}(\theta_{n_k},\lambda_k)\leq \lim_{n\to \infty} F_{n_k}(\theta_{n_k},\lambda_{n_k})=F(\theta^*,\lambda^*).
\]
Thus $F(\theta^*,\lambda^*)=\phi(\theta^*)$, and we conclude that $\phi_n(\theta_n)\to \phi(\theta^*)$ and thus $\phi_n$ uniformly converges to $\phi$ on S.

\underline{Step 2} We now show that $\min_{\theta \in S_n} \phi_n(\theta) \to \min_{\theta \in S_0} \phi(\theta)$. Let $\phi_n(\theta_n^*)=\min_{\theta \in S_n} \phi_n(\theta)$, where $\theta_n^* \in S_n$. We establish the claim by showing that every subsequence of $\{\phi_n(\theta_n^*)\}$ has a further subsequence that converges to $\min_{\theta \in S_0} \phi(\theta)$. Indeed, given a subsequence of $\{\phi_n(\theta_n^*)\}$, consider a further subsequence, say $\{\phi_{n_k}(\theta_{n_k}^*)\}_{k\geq 1}$, such  that $\theta_{n_k}^*\to \theta^*$. Then the preceding step implies that $\phi_{n_k}(\theta_{n_k}^*)\to \phi(\theta^*)$. Also, as $d_H(S_n,S_0)\to 0$, we have $\theta^* \in S_0$, and given any $\theta \in S_0$, there exists $\theta_{k}\in S_{n_k}$ such that $\theta_k \to \theta$. Hence for $\theta \in S_0$, we have
\[
\phi(\theta)=\lim_{k\to \infty} \phi_{n_k}(\theta_k)\geq \lim_{k\to \infty} \phi_{n_k}(\theta_{n_k}^*)=\phi(\theta^*).
\]
Thus $\phi(\theta^*)=\min_{\theta \in S_0} \phi(\theta)$ and we conclude that $\phi_n(\theta_n^*) \to \min_{\theta \in S_0} \phi(\theta)$.

\end{proof}

The general form of the following lemma, stated with an implicit constant, is known as Hoffman's Theorem \cite{AH}. Here, we present a proof that includes an explicit constant, as understanding its dependence on the inputs \((A, b, c)\) is crucial for our uniformity results. Our proof modifies the argument in \cite{AH} slightly to obtain an explicit Lipschitz constant. 

\begin{lemma}[Hoffman's Theorem]
\label{lemhoff}
Consider the polyhedron \(\mathcal{P} = \{z\in \mathbb{R}^n \mid A_E z = b_E, \ A_I z \leq b_I\}\), where \( A_E \in \mathbb{R}^{m_E \times n} \) and \( A_I \in \mathbb{R}^{m_I \times n} \). Then, for all \( x \in \mathbb{R}^n \), we have
\begin{equation} \label{eq:Hoffeq}
d(x, \mathcal{P}) \leq C \Bigl(\|A_E x - b_E\| + \|(A_I x - b_I)_+\|\Bigr),
\end{equation}
where the constant \( C \) can be chosen as \( C = \bigl(\Lambda(A)\bigr)^{1/2} \), with \( \Lambda(A) \) defined in Definition \ref{deflambda}.  

Here, \( (x)_{+} \) denotes the positive part of \( x \), obtained by replacing each entry \( x_i \) with \( \max\{x_i, 0\} \), and \( \|\cdot\| \) represents the Euclidean norm in the corresponding space.
\end{lemma}

\begin{proof}[\textbf{Proof of Lemma \ref{lemhoff}}]
Suppose \( x \notin \mathcal{P} \), and let \( y \) be the projection of \( x \) onto \( \mathcal{P} \) with respect to the Euclidean norm. Define \( J \subset I \) as the subset of rows of the matrix \( A_I \) that are active at \( y \). The tangent cone of \( \mathcal{P} \) at \( y \) is given by 
\[
\mathcal{C} = \{z \in \mathbb{R}^n \mid A_E z = 0, \ A_J z \leq 0 \}.
\]
Since \( y \) is the closest point in \( \mathcal{P} \) to \( x \), the first-order optimality condition (FOC) implies that \( \langle x - y, z \rangle \leq 0 \) for all \( z \in \mathcal{C} \). By Farkas' Lemma, there exist vectors \( \lambda_E \) and \( \lambda_J \) with \( \lambda_J \geq 0 \) such that
\[
x - y = A_E^T \lambda_E + A_J^T \lambda_J.
\]

Let \( \tilde{E} \subset E \) be the subset of indices corresponding to nonzero entries of \( \lambda_E \), and define \( \tilde{J} \subset J \) similarly for \( \lambda_J \). By an argument similar to the one used in proving Carath\'eodory’s convex hull theorem, we can choose \( \lambda_E \) and \( \lambda_J \) so that the set of rows of $A_E$ and $A_I$ corresponding to the indices in $F:=\tilde{E}\cup \tilde{J}$, \( \{a_i \mid i \in F\} \), is linearly independent. Set \( \lambda_F = (\lambda_{\tilde{E}}', \lambda_{\tilde{J}}')' \) and \( A_F = (A_{\tilde{E}}', A_{\tilde{J}}')' \). Then, we have
\[
x - y = A_F^T \lambda_F,
\]
where the rows of \( A_F \) are linearly independent.\footnote{Alternatively, we can let $F$ be the support of a \underline{vertex solution} of the dual to the linear program $\max\{\langle x-y,z\rangle \mid A_J z\leq 0, \ A_Ez=0\}$. Note that by assumption, the latter LP has value zero.}

Since \( x \notin \mathcal{P} \), it follows that \( x - y \neq 0 \) and \( x - y \notin \mathcal{C} \) (otherwise, the inequality \( \langle x - y, z \rangle \leq 0 \) for all \( z \in \mathcal{C} \) would imply \( \|x - y\| = 0 \), a contradiction). 

Define the polyhedral cone \( K \) by
\[
K = \{z \in \mathbb{R}^{|F|} \mid z_i = 0 \ \forall i \in \tilde{E}, \ z_j \leq 0 \ \forall j \in \tilde{J} \}.
\]
Setting \( w = x - y \), our goal is to bound \( \|w\| \) in terms of \( d(A_F w, K) \). Let \( \Pi_K \) denote the projection matrix onto \( K \), and define
\[
\delta = A_F w - \eta, \quad \text{where} \quad \eta = \Pi_K(A_F w).
\]
We then obtain
\[
\lambda_F^T \delta = \lambda_F^T A_F w - \lambda_F^T \eta = \lambda_F^T A_F A_F^T \lambda_F - \lambda_F^T \eta.
\]

Since \( \lambda_{\tilde{J}} \geq 0 \) and by the definition of \( K \), we have \( \lambda_F^T \eta \leq 0 \), leading to 
\begin{equation}
\label{2eqn1}
\|A_F^T \lambda_F\|^2 \leq |\lambda_F^T \delta| \Rightarrow (\lambda_{\min}(A_F A_F^T))^{1/2} \|\lambda_F\| \|A_F^T \lambda_F\| \leq \|\lambda_F\| \|\delta\|.
\end{equation}
Using the definition of \( K \), it can be easily verified that \( \eta= (0',- (A_{\tilde{J}} w)'_{-})' \). Consequently, we obtain  
\[
\delta = \bigl((A_{\tilde{E}} w)', (A_{\tilde{J}} w)_{+}'\bigr)',
\]
and \( \|\delta\|^2 = \|A_{\tilde{E}} x - b_{\tilde{E}}\|^2 + \|(A_{\tilde{J}} x - b_{\tilde{J}})_{+}\|^2 \). Since  
\[
\|A_{\tilde{E}} x - b_{\tilde{E}}\|^2 + \|(A_{\tilde{J}} x - b_{\tilde{J}})_{+}\|^2 \leq \|A_E x - b_E\|^2 + \|(A_I x - b_I)_{+}\|^2,
\]  
combining this inequality with the final implication in \eqref{2eqn1}, we conclude that 
\[
d(x, \mathcal{P}) \leq \frac{1}{\sqrt{\lambda_{\min}(A_F A_F^T)}} \Bigl(\|A_E x - b_E\| + \|(A_I x - b_I)_{+}\|\Bigr),
\]
which proves the lemma.
\end{proof}

\subsection{Sharper upper bound on Hoffman's constant} \label{secWeakenSSLSLUC4}

The sharp Hoffman bound on a feasible system of inequalities $A \eta \le b$ with an $m \times n$ matrix $A$ is given by the quantity
\begin{equation} \label{eqHpena}
H(A) := \max \left\{\frac{1}{\min_{v \in \mathbbm{R}^{|F|}_+, ||v|| = 1} ||A^T_Fv||} \: \mid \: F \subseteq [m],\ \operatorname{rank}(A_F) = |F| \right\},
\end{equation}
where $||\cdot||$ denotes the Euclidean norm.\footnote{To obtain this expression from Eq. (4) of \citet{PVZ19char}, we use our notation $F$ to denote their set $J$ of rows, and note that the Euclidean norm is self-dual.} The above expression for $H(A)$ is given in \citet{PVZ19char}, and satisfies $d(x,\mathcal{P}) \le  H(A) \cdot ||(Ax - b)_+||$ for all $b$ such that $\{\eta \in \mathbbm{R}^n: A \eta \le b\} \ne \emptyset$, where $\mathcal{P} = \{z \in \mathbbm{R}^n | Az \le b\}$. Proposition 1 of \citet{PVZ19char} shows that this bound is sharp in the sense that, provided that $H(A)>0$, then there exists a $b \in \mathbbm{R}^m$ and $x \in \mathbbm{R}^n$ such that $d(x,\mathcal{P}) = H(A) \cdot ||(Ax - b)_+||,$ i.e. Eq. \eqref{eq:Hoffeq} from this paper holds with equality for $C = H(A)$.\footnote{Sharper bounds on Hoffman's constant are possible, if the bounds are allowed to depend on $b$ as well as $A$.}

The Hoffman constant $\Lambda(A)$ introduced in Section \ref{subsectionAVEAM} must be weakly larger than $H(A)$ for a given matrix $A$. As an illustration, consider the matrix
\begin{equation} \label{eqAallones}
    A = \begin{bmatrix}
    1 & 1\\
    1 & 1-\epsilon
\end{bmatrix}
\end{equation}
For any $\epsilon \in (0,1)$, $r=rank(A)=2$, and the eigenvectors of $A A^T$ are proportional to $\begin{pmatrix}
    -1 \\ \frac{\epsilon \pm \sqrt{4+\epsilon^2}}{2} \end{pmatrix}$, with corresponding eigenvalues $\lambda_j = \frac{1}{2}\{\epsilon^2 - 2\epsilon + 4 \mp (2-\epsilon)\sqrt{\epsilon^2+4}\}$. As $\epsilon \downarrow 0$, the larger eigenvalue approaches 4 and the smaller eigenvalue approaches 0. Let ${\cal A}$ be the set of all $A$ matrices above with $\epsilon \in (0,1)$. For this set of A matrices part (i) of Assumption \ref{assAVEAM4} fails, since $\sup_{A\in {\cal A}}\Lambda(A) = \sup_{\epsilon \in (0,1)} \frac{1}{\frac{1}{2}\{\epsilon^2 - 2\epsilon + 4 - (2-\epsilon)\sqrt{\epsilon^2+4}\}} = \infty$. Meanwhile $H(A)$ remains uniformly bounded:
\begin{align*}
    H(A)^2 &= \max \left\{||A^T_Fv||^2 \mid F \subseteq [m], |F|=r, \ \operatorname{rank}(A_F) = r, v \in \mathbbm{R}^{|F|}_+, ||v|| = 1 \right\}\\
    &=\max \left\{||A^Tv||^2 \mid, v = (t,\sqrt{1-t^2})^T, t \in [0,1]\right\}=||A^T (1,0)^T||^2 = 2
\end{align*}
using that $||A^T(t,\sqrt{1-t^2})^T||^2$ is strictly increasing on $t$ in $[0,1]$ for $\epsilon  \in (0,1)$, and is thus maximized at $t=1$.\footnote{By contrast, if $t$ were not restricted to be positive (i.e. $v$ were not restricted to the positive orthant $\mathbbm{R}_+^{|F|}$ of $\mathbbm{R}^{|F|}$, then the global maximum occurs at a negative value of $t$ and delivers the value $\sqrt{\frac{\epsilon^2 - 2\epsilon + 4 - (2-\epsilon)\sqrt{\epsilon^2+4}}{2}}$ of $||A^Tv||^2$. Note that this quantity goes to zero as $\epsilon \rightarrow 0$, reflecting that $\Lambda(A) \rightarrow \infty$.} Thus $H(A) = 1/\sqrt{2} $ for all $A \in {\cal A}$. 

Thus, replacing (i) of Assumption \ref{assAVEAM4} with the assumption that $\sup_{Q\in {\cal Q}_0}H(A(Q))<\infty$ represents a meaningful relaxation, and with it the conclusion of Lemma \ref{lemhoff} and Theorem \ref{thmAVEAM} remain. 

A sufficient condition for $\sup_{Q\in {\cal Q}_0}H(A(Q))<\infty$ is given the following:
\begin{proposition} \label{propSharpHoffmanPositiveEntries}
Suppose that for some $\rho > 0$, we have that for each $F \subseteq [m]$ there exists an $x_F$ such that $\inf_{Q\in {\cal Q}_0: \operatorname{rank}(A_F(Q))=|F|} A(Q)x_F \ge \rho \mathbbm{1}$. Then $\sup_{Q\in {\cal Q}_0}H(A(Q))<\infty$.
\end{proposition}
\noindent The following are each sufficient conditions for the condition in Proposition \ref{propSharpHoffmanPositiveEntries}:
\begin{enumerate}
    \item{There exists a column $A(Q)_j$ of $A(Q)$ in which all entries have the same sign for all $Q \in {\cal Q}_0$, uniformly in the sense that $(-1)^{d}\inf_{Q\in {\cal Q}_0} \min_k [A(Q)]_{jk} \ge \rho$ for some $d \in \{0,1\}$. This corresponds to $x_F = \pm e_j$.}
    \item{The non-zero entries of $A(Q)$ are uniformly of the same sign, i.e. $(-1)^{d}\inf_{Q\in {\cal Q}_0} \min_{j,k} [A(Q)]_{jk} \ge \rho$ for some $d \in \{0,1\}$. More generally, if the sum along the entries on each row of $A(Q)$ is positive (or negative) uniformly in $A \in {\cal Q}_0$, we can take $x_F$ to be a vector of ones (or -1).}
\end{enumerate}
\noindent Item 2. in Proposition \ref{propSharpHoffmanPositiveEntries} is satisfied for instance in the $A$ matrix given in \eqref{eqAallones} above.

\begin{proof}\textbf{(Proof of Proposition \ref{propSharpHoffmanPositiveEntries})}
    Consider an arbitrary $m \times n$ real matrix $A$. Proposition 1 of \citet{PVZ19equiv} shows that if the system $Ax<0$ is feasible, then $H(A) = 1/R(A)$, where $R(A):=\inf\{ ||\delta||_S \mid (A+\delta)x < 0 \textrm{ is infeasible}\}$, and $||\delta||_S$ denotes the spectral norm of the matrix $A$.\footnote{The quantity $R(A)$ is known as ``Renegar's distance to ill-posedness'' for the matrix $A$.}

    Observe furthermore that $H(A) = \max \{H(A_F): F \subseteq [m],\ \operatorname{rank}(A_F) = |F|\}$. To see this, note that $H(A) \ge H(A_F)$ for any of the sets $F$ appearing in \eqref{eqHpena}: $F \subseteq [m],\ \operatorname{rank}(A_F) = |F|$, because the possible subsets of the rows of $A_F$ are all included in possible subsets of the rows of $A$. Meanwhile,  $H(A) \le \max \{H(A_F): F \subseteq [m],\ \operatorname{rank}(A_F) = |F|\}$, since $ \{H(A_F): F \subseteq [m],\ \operatorname{rank}(A_F) = |F|\} \subseteq \bigcup_{F \subseteq [m],\ \operatorname{rank}(A_F) = |F|}\hspace{.1cm} \{H(A_{F'}): F' \subseteq [F],\ \operatorname{rank}(A_{F'}) = |F'|\}$. Combining, we thus have that $H(A) = \max \{H(A_F): F \subseteq [m],\ \operatorname{rank}(A_F) = |F|\}$. This result allows us to build up $H(A)$ from $R(A_F)$ among collections of rows $F$ such that $A_F$ has full row rank, if those $F$ make $A_Fx<0$ feasible such that $H(A_F) = 1/R(A_F)$.

    By assumption, we have that 
    $\min_{F \subseteq [m]} \inf_{Q\in {\cal Q}_0: \operatorname{rank}(A_F(Q))=|F|} A(Q)x_F \ge \rho \mathbbm{1}$.
    This implies that for each $Q\in {\cal Q}_0$ and $F$ such that $A_F(Q)$ has full row rank, the system $A_F(Q)x > 0$ is feasible (since $A_F(Q) x_F \rho \mathbbm{1} > 0$, $A_F(Q) (-x_F) < 0$). Therefore, $H(A(Q)) = \max \{1/R(A_F(Q)): F \subseteq [m],\ \operatorname{rank}(A_F(Q)) = |F|\}$.

    Now consider the definition of $R(\cdot)$ applied to the matrix $A_F(Q)$. To make $(A_F(Q)+\delta)x < 0$ infeasible, it is necessary for an entry of $\delta$ to be large enough to change the sign of at least one element of $A_F(Q) x_F$ (which recall is elementwise greater than $\rho$), i.e. $\delta_r x_F < -\rho$ for some row vector $\delta_r$ from $\delta$. Using the Cauchy-Schwarz inequality:
    $$|X_F|_2 \cdot |\delta_j|_2 \ge \quad <|\delta_r^T|,|x_F|> \quad  \ge \quad | <\delta_r^T,x_F> | \quad  \ge \quad | <\delta_r^T,x_F> | \quad > \rho$$
    where $|x|$ denotes the element-wise absolute value of a vector $x$. Without loss, we can take Euclidean norm of each $x_F$ to be $1$. Using that the Frobenius norm $||\delta||_F:=\sqrt{\sum_{j,k} \delta_{jk}^2}$ of $\delta$ is at least as large as $|\delta_j|_2 = \sqrt{\sum_{k} \delta_{jk}^2}$, we then have from the above that $||\delta||_F$ must be strictly greater than $\rho$. Finally, using that $||M||_S \cdot \sqrt{\min\{m,n\}} \ge ||M||_F$ for an $m \times n$ matrix $M$, we thus we obtain the bound $R(A_F(Q)) \ge \rho/\sqrt{\min\{|F|,n\}} \ge \rho/\sqrt{\min\{m,n\}}$, where $m$ is the number of rows in the full matrix $A(Q)$

    Now using that $H(A(Q)) = \max \{1/R(A_F(Q)): F \subseteq [m],\ \operatorname{rank}(A_F(Q)) = |F|\}$, we have:
    \begin{equation} \label{eqHAFineq}
        \sup_{Q\in {\cal Q}_0} H(A(Q)) \le \sup_{Q\in {\cal Q}_0} \sqrt{\min\{m,n\}} / \rho = \sqrt{\min\{m,n\}} / \rho < \infty.
    \end{equation}
\end{proof}

\subsection{Sufficient condition for bounded $\Lambda(A)$} \label{app:Suff}
Here we establish an upper bound on $\Lambda(A)$ that is nearly a continuous function of the entries of $A$. This allows us to provide a simple sufficient condition for part (i) of Assumption \ref{assAVEAM4}.

Recall the definition of $\Lambda(A)$ for a $p \times d$ matrix $A$ with $\operatorname{rank}(A)=r$:
\begin{equation*}
\Lambda(A) := \left(\min\bigl\{\lambda_{\min}(A_F A_F^\top) \mid F \subseteq [m],\ |F| = r,\ \operatorname{rank}(A_F) = r \bigr\}\right)^{-1}.
\end{equation*}
Define 
\begin{equation*}
\bar{\Lambda}(A) := \left(\min\bigl\{\frac{\mathrm{det}(A_F A_F^T)}{\mathrm{tr}(A_FA_F^T)^{n-1}} \mid F \subseteq [m],\ |F| = r,\ \operatorname{rank}(A_F) = r \bigr\}\right)^{-1}.
\end{equation*}
$\bar{\Lambda}(A) \ge \Lambda(A)$, as a consequence of Lemma \ref{lemmatracedetbound}, stated below.

Introduce the following set:
$$\mathcal{F}(Q) = \{F \subseteq [m],\ \operatorname{rank}(A_F(Q)) = \operatorname{rank}(A(Q))=|F|\} .$$
The following proposition shows that a \textit{rank stability} condition is sufficient for part (i) of Assumption \ref{assAVEAM4}:
\begin{proposition} \label{prop:suffbound}
Suppose that $\mathcal{F}(Q)$ is the same for all $Q \in \mathcal{Q}_0$, and for all $Q \in \mathcal{Q}_0$,  $\operatorname{vec}(A(Q)) \in \mathcal{A}$ for some compact set $\mathcal{A}$. Then $\sup_{Q \in \mathcal{Q}_0} \Lambda(A) < \infty$.
\end{proposition}
\begin{proof}
    By Lemma \ref{lemmatracedetbound}, it is sufficient to show that $\sup_{Q \in \mathcal{Q}_0} \bar{\Lambda}(A) < \infty$. Given $\mathcal{F}(Q) = \mathcal{F}$ for all $Q \in \mathcal{Q}_0$, this is equivalent to:
    \begin{equation} \label{eq:boundabovezero}
        \inf_{Q \in \mathcal{Q}_0} \frac{\mathrm{det}(A_F(Q) A_F(Q)^T)}{\mathrm{tr}(A_F(Q)A_F(Q)^T)^{n-1}} > 0 \quad \textrm{ for each } F \in \mathcal{F}
    \end{equation}
    Observe that for a given $F \in \mathcal{F}$, $\frac{\mathrm{det}(A_F(Q) A_F(Q)^T)}{\mathrm{tr}(A_F(Q)A_F(Q)^T)^{n-1}}$ is a continuous function of the entries $\operatorname{vec}(A(Q))$ of $A(Q)$ (for $\mathrm{tr}(A_F(Q)A_F(Q)^T) \ne 0$). Meanwhile this quantity is strictly positive because since $A_F(Q)$ has full row rank, the matrix $A_F(Q)A_F(Q)^T$ is positive definite and hence $\mathrm{tr}(A_F(Q)A_F(Q)^T) > 0$ and $\mathrm{det}(A_F(Q)A_F(Q)^T) > 0$. By compactness of $A$, it then follows that the infimum is also strictly positive, thus Eq. \eqref{eq:boundabovezero} is satisfied.
    
\end{proof}
\begin{lemma} \label{lemmatracedetbound}
    Consider a real and non-zero $n \times n$ matrix $A$. The smallest eigenvalue $\lambda_{\min}(AA^T)$ satisfies the lower bound: $\lambda_{\min}(AA^T) \ge \frac{det(AA^\top)}{tr(AA^\top)^{n-1}}(n-1)^{n-1}$, where $\mathrm{det}(M)$ is the determinant of $M$.
\end{lemma}
\begin{proof}
As $AA^T$ is a real symmetric, its eigenvalues $\lambda_i$ are real and non-negative. Order these in descending order so that $\lambda_{\max}(AA^T) = \lambda_1$ and $\lambda_{\min}(AA^T) = \lambda_n$.
Using the AM-GM inequality and positive definiteness, we have 
\[det(AA^\top)^{1/(n-1)}=\lambda_n^{1/(n-1)} \prod_{j\neq n}\lambda_j^{1/(n-1)}\le \lambda_n^{1/(n-1)} \frac{1}{n-1}\sum_{i=1}^{n-1}\lambda_i\le \frac{\lambda_n^{1/(n-1)}}{n-1}tr(AA^\top) 
\]
therefore
\[
\lambda_n\ge \frac{det(AA^\top)}{tr(AA^\top)^{n-1}}(n-1)^{n-1}
\]
\end{proof}

\subsubsection*{Application to Example \ref{exMST}} \label{app:mstbounded}
We now use Proposition \ref{prop:suffbound} to show that part (i) of Assumption \ref{assAVEAM4} is satisfied in the setting of Example \ref{exMST}, detailed in Section \ref{secApplicationATE}. In particular, we show that it is necessary and sufficient that both $p(0)$ and $p(1)$ lie within the interior of the unit interval and that the two are well-separated (i.e. there is a non-zero first stage), uniformly over $\mathcal{Q}_0$. For brevity, we assume that $p(1) > p(0)$, though the argument does not require the sign of the first stage to be known to the econometrician.

Accordingly, let
$$\mathcal{R}_\delta = \{Q: Q\{D=1|Z=0\} \ge \delta, Q\{D=1|Z=1\} \le 1 - \delta, Q\{D=1|Z=1\}-Q\{D=1|Z=0\} \ge \delta\}.$$
That $\mathcal{Q}_0 \subseteq \mathcal{R}_\delta$ for some $\delta > 0$ is sufficient (and necessary) for part (i) of Assumption \ref{assAVEAM4} to hold in this application. To use Proposition \ref{prop:suffbound} to show this, notice first that if $\mathcal{Q}_0 \subseteq \mathcal{R}_\delta$, then $\operatorname{vec}(A(Q))$ the matrix $A(Q)$ defined by \eqref{eqMST} belongs to a compact set in $\mathbbm{R}^{p \times d}$. 

Proposition \ref{prop:suffbound} also requires that the set of $F \subseteq [p]$ such that $A_F(Q)$ has full row rank be the same across all $Q \in \mathcal{Q}_0$. Note that when writing an equality restriction as two inequality restrictions, the rows of $A(Q)$ that are introduced are perfectly linearly dependent. Thus, to evaluate part (i) of Assumption \ref{assAVEAM4}, it is sufficient to consider sets of rows $F$ of the matrix obtained by stacking the coefficients from equality and inequality restrictions, as in Eq. \eqref{eqMST}. Indeed, any three rows in the form given in \eqref{eqMST} are linearly independent given $Q \in \mathcal{R}_\delta$, which can be seen by considering all such triplets case-by-case. Thus we can use Proposition \ref{prop:suffbound} with $\mathcal{F} = \{F \subseteq [p]: |F|=3\}$.
    
\section{Further detail on applications}

\subsection{\cite{HSS22}}
\label{exampHSS}

\cite{HSS22} study inference on a parameter $\theta_0$ defined as a solution to the linear program
\[
\theta_0 \in \arg \max \{ c^\top \theta \mid M\theta \le d \},
\]
where $\theta_0$ denotes any optimizer, and the primitives $(c,M,d)$ may be estimated from data.

Using the KKT conditions, \cite{HSS22} show that a candidate value $\theta$ solves the linear program if and only if there exists a vector $\eta \ge 0$ such that
\[
M\theta \le d, \qquad M^\top \eta = c, \qquad c^\top \theta = d^\top \eta.
\]
This characterization can be expressed as a hypothesis of the form \ref{eqnARP}. Consequently, a confidence region for $\theta_0$ can be constructed by inverting tests of the null hypothesis (evaluated over a grid of $\theta$ values)
\[
H_0: \exists\, \eta \text{ such that } b \le A\eta,
\]
where
\[
A^\top =
\begin{bmatrix}
M^\top & -M^\top & -I & 0 & d^\top & -d^\top
\end{bmatrix}
\quad \text{and} \quad
b^\top =
\begin{bmatrix}
c^\top & -c^\top & 0 & \mu_\theta^\top & c_\theta & -c_\theta
\end{bmatrix},
\]
with $\mu_\theta = M\theta-d$ and $c_\theta = c^\top \theta$.

The procedure proposed in \cite{HSS22} constructs confidence regions via a profile test statistic. However, this approach is computationally demanding, as it requires solving a potentially nonconvex optimization problem over a nonconvex feasible set. In addition, the method becomes impractical when the dimension of $\eta$ (i.e., the number of constraints in the linear program) is large. The approach developed in this paper provides an alternative that remains computationally tractable in high-dimensional settings.

\subsection{Example \ref{exDTE}: bounding the share harmed by treatment} \label{secApShareHarmed}

We begin from the setup in Example \ref{exDTE}, and consider several additional assumptions that increase the size of the matrices $A$ and $b$ in Eq. \eqref{eqnARP}. Our first example adds estimated rows to the matrix $A$, and our second simply adds more rows to $A$ that are known.

\subsubsection{Imposing stochastic increasingness}
The identified set on the parameter $\theta$ is often wide in practice. To narrow it, \citet{FL} propose imposing the assumption that potential outcomes are mutually \textit{stochastically increasing} with one another: that is $P(Y(1) \le k|Y(0) = j)$ and $P(Y(0) \le k|Y(1) = j)$ are both non-increasing in $j$, for all $j,k$. This amounts to adding $2J(J-1)$ additional restrictions on the components of $\eta$, however these constraints are not naturally linear in $\eta$ (due to the ratio form of conditional probabilities). For instance, $P(Y(1) \le k|Y(0) = j) \ge P(Y(1) \le k|Y(0) = j+1)$ can be written as $\frac{\sum_{j' \le j} \eta_{jk}}{\sum_{j'} \eta_{jk}} \ge \frac{\sum_{j' \le j} \eta_{j,k+1}}{\sum_{j'} \eta_{j,k+1}}$. However, given that $\sum_{j'} \eta_{j'k} = P(Y(0)=j) = P_Q(Y=j|D=0)$, we can rewrite mutual stochastic increasingness as $P_Q(Y=k+1|D=0) \cdot \sum_{j' \le j} \eta_{j'k} \ge P_Q(Y=k|D=0) \cdot \sum_{j' \le j} \eta_{j',k+1}$ for each $j$ and $k<J$, and similarly with $Y(1)$ and $Y(0)$ swapped. These can be rewritten as
\begin{subequations}
\begin{align}
\sum_{j',k'} (1[k'=k]-1[k'=k+1])\cdot1[j' \ge j] \cdot P_Q(Y=k'|D=0) \cdot \eta_{j',k'} \ge 0 \label{eqMS1}\\
\sum_{j',k'} (1[j'=j]-1[j'=j+1])\cdot1[k' \ge k] \cdot P_Q(Y=j'|D=1) \cdot \eta_{j',k'} \ge 0 \label{eqMS2}
\end{align}
\end{subequations}
for each $j,k$. This leads to a system with $2J(J-1)$ linear restrictions in $\eta$ that involve estimated components in the corresponding rows of $A$ (and zeros in the corresponding rows of $b$).

\subsubsection{Covariates}

Consider a covariate $X$ that takes discrete values $\{1,2, \dots L\}$, such that $(Y(0),Y(1)) \indep D | X$. Define $\eta$ to know be a $J^2L$ component vector with components $\eta_{jkx}=P(Y(0)=j,Y(1)=k|X=x)$.

Let $M$ be a known $JL \times J^2L$ matrix with entries $M_{j'x',jkx} = 1[j=j',x=x']$, $q_0$ be a $JL-$component vector with components $q^0_{jx}(Q) = P_Q(Y=j|D=0,X=x)$, $N$ is a $JL \times J^2L$ matrix with entries $M_{k'x',jkx} = 1[k=k',x=x']$ while $q_0$ is a $JL-$component vector with components $q^1_{kx}(Q) = P_Q(Y=k|D=1,X=x)$. Let $\theta = \{\theta_x\}{x=1}^L$ where $\theta_x = P(Y(1) \le Y(0)|X=x)$ be a vector of hypothesized values for all of the conditional shares harmed by treatment. Let $Q$ and $C$ be a $L \times J^2L$ matrices with entries $Q_{x',jkx} = 1[x=x']$ and $C_{x',jkx} = 1[j \ge k, x=x']$.

Then we have the system $A\eta = b$ where:
$$A=\begin{bmatrix} M \\ N \\ Q \\ C \end{bmatrix}, \quad \quad b(\theta,Q)=\begin{bmatrix} q^0(Q) \\ q^1(Q) \\ \mathbbm{1}_L \\ \theta \end{bmatrix}$$
where $\mathbbm{1}_L$ is a vector of $L$ ones. The number of columns is $d=LJ^2$ and the number of rows is $p=L(2J+2)$.

Now let us add an assumption that potential outcomes are mutually \textit{stochastically increasing} with $X$: that is $P(Y(0) \le j, Y(1) \le k|X = x)$ and $P(Y(0) \le k|Y(1) = j)$ are both non-increasing in $x$, for all $j,k,x$. This captures the idea that when $X$ increases, so do $Y(0)$ and $Y(1)$. A candidate co-variate X would be the outcome $Y$ measured before the experiment, or some other proxy of the outcome variable. Adding the stochastic increasing-ness assumption, we introduce $(L-1) J^2$ additional constraints of the form $\sum_{j' \le j} \eta_{jkx} \ge \sum_{j' \le j} \eta_{jk,x+1}$, for each $x < L$ and $j,k$. Now we have a system with $p_K=(L-1)J^2+2L$ known rows, and $p_U=2LJ$ rows that need to be estimated. The matrix A is known for a given $\theta$.

The allowance for $p_K$ to grow with the sample size is important in this setting since for moderately sized $J$, $p_K$ may be comparable to or larger than the sample size. For instance, \citet{HDG} report insignificant effects on respiratory function among users of an improved cookstove randomized to villagers in India. It is important to understand whether these null effects are due to most individuals having a small effect or individuals with large positive effects being offset by individuals with large negative effects. Respiratory function is measured as forced expiratory volume (FEV) using a spirometer, reported in hundredths of a liter, taking on $J=304$ distinct values when measured at baseline (before the improved cookstoves are introduced). If we used a binary $X$ variable (L=2), we would have in this case that $p_U=1,216$ while $p_K=92,420$, and the replication data of \citet{HDG} contains $n=10,409$ respondents.

\subsection{Example \ref{exDefiers}: the ATE with more compliers than defiers} \label{secdC}
\citet{dC} introduces the ``more compliers than defiers'' (MCTD) condition that $P(G=C|Y(1)-Y(0)=t) \ge P(G=D|Y_i(1)-Y_i(0)=t)$ for each value $t$, where $D$ indicates the event of being a defier ($D(1) < D(0)$) and $C$ being a complier ($D(1) > D(0)$). He shows that with this assumption replacing the typical LATE assumption of no defiers, an average treatment effect among a subset of the compliers is point identified.

One might instead study the identifying power of the MCTD assumption for the overall average treatment effect. Suppose that we have a discrete outcome with $Y \in \{1, 2, \dots J\}$. Let $\eta_{g,y,y'} = P(G=g,Y(1)=y,Y(0)=y')$. We can write the MCTD condition as $P(G=C,Y(1)-Y(0)=t) \ge P(G=D,Y_i(1)-Y_i(0)=t)$, or equivalently
$$ \sum_{g,y,y'} \{1[g=c,y-y'=t]-1[g=d,y-y'=t]\} \cdot \eta_{g,y,y'} \ge 0 $$
for each $t \in \{-(J-1), \dots J-1\}$.

The identifying content of the data can be written as
$$E[Y D |Z=z] = E[Y(1) \cdot D(z)] = \sum_{g,y,y'} \{y\cdot 1[D_g(z)=1]\} \cdot \eta_{g,y,y'}$$
and similarly for $1-D$, for each $z \in \{0,1\}$.

We can write the ATE as 
$$E[Y(1)-Y(0)] = \sum_{g,y,y'} \{y-y'\} \cdot \eta_{g,y,y'}$$

\subsection{More on monotone IV} \label{secMTSMTR}
Recall the monotone instrumental variables (MIV) assumption of Example \ref{exMP}. \textit{Monotone treatment selection} (MTS) holds as a special case of MIV when $Z=T$. Unlike MIV in general,  MTS imposes that $E[Y(s)|T=t'] \ge E[Y(s)|T=t]$ for $t'\ge t$ and each $s$. To write MTS as a system of linear inequalities, introduce a $L^2$ component vector $\eta$ with components $\eta_{tt'}:=E[Y(t)|T=t']$. Then MTS says that $\eta_{t,j+1} - \eta_{tj} \ge 0$ for each $j = 1 \dots J-1$. The identifying content of the data is that the diagonal elements $\eta_{tt} = E[Y(t)|T=t]$ are identified as $\eta_{tt} = E[Y|T=t]$. The hypothesis that $E[Y(t_2)-Y(t_1)]=\theta$ can be written as $c^T \eta = \theta$, where $c_{tt'} = P(T=t') \cdot \{1[t=t_2]-1[t=t_1]\}$. Testing for a given value of $\theta$ thus amounts to testing for $A(Q) \eta \ge b(Q)$ where just one row of $A(Q)$ is estimated.

Both MTS and the more general MIV assumptions can be combined with an assumption of \textit{monotone treatment response} (MTR), which says that for any $t' \ge t$. $Y(t') \ge Y(t)$ with probability one. This implies that $\eta_{t's}-\eta_{ts} \ge 0$ for any $s$ and $t' \ge t$ (in the case of MIV, it implies $\eta_{t'zs}-\eta_{tzs} \ge 0$ for any $z$ as well). When combined with MTR, MTS yields a bounded identified set even when $Y$ is itself unbounded.

For example in the simplest case of $T \in \{0,1\}$, we let $\eta$ be a 4-component vector with components $\eta_{tt'}:=E[Y(t)|T=t']$. The MTS assumption says that $\eta_{t,1} - \eta_{t0} \ge 0$ for each $t \in \{0,1\}$. The MTR assumption says that $Y(1) \ge Y(0)$ with probability one, which implies $\eta_{1t'}-\eta_{0t'} \ge 0$ for each $t' \in \{0,1\}$. Under the hypothesis that the ATE takes value $\theta$, we obtain the system
    
$$\begin{bmatrix}
    &1 & 0 & 0 & 0\cr
    &0 & 0 & 0 & 1\cr
    &-1 & 1 & 0 & 0\cr
    &0 & 0 & -1 & 1\cr
    &-1 & 0 & 1 & 0\cr
    &0 & -1 & 0 & 1\cr
    &-P(T=0) & -P(T=1) & P(T=0) & P(T=1)
\end{bmatrix} \begin{pmatrix}
    \eta_{00} \\ \eta_{01} \\ \eta_{10} \\ \eta_{11}
\end{pmatrix} \quad \begin{matrix} = \\ = \\ \ge\\ \ge\\ \ge \\ \ge \\ =\end{matrix} \quad \begin{pmatrix} 
E_Q[Y|T=0]\\
E_Q[Y|T=1]\\
0\\
0\\
0\\
0\\
\theta
\end{pmatrix}$$ The first two rows reflect the observable restriction that $\eta_{tt} = E[Y|T=t]$. The following four inequalities reflect the MTS and MTR assumptions, while the final row encodes the hypothesis that $E[Y(1)-Y(0)]=\theta$. This row involves the probabilities $P(T=dt)$, which generally must be estimated.

\citet{MP} note that when MIV and MTR are combined, the sharp analytic bounds that they provide for mean counterfactual outcomes $E[Y(t)]$ lead to bounds on treatment effects that are not in general sharp. Combining MTR with the system given above for MIV allows for nonconservative inference, even without an analytic characterization of the identified set. 


\begin{singlespace}
\bibliographystyle{apalike}
\bibliography{biblio.bib}
\end{singlespace}
\end{appendix}
\end{document}